\documentclass[twocolumn]{aastex61}
\usepackage{graphicx, graphics, amsmath,amssymb, amsfonts}
\usepackage{dcolumn}
\usepackage{bm}
\usepackage{natbib}
\usepackage{enumerate}
\pdfoutput=1

\begin{document}

\title{Nonlinear transverse cascade and sustenance of
MRI-turbulence in Keplerian disks with an azimuthal magnetic field}

\correspondingauthor{G. Mamatsashvili}
\email{g.mamatsashvili@hzdr.de}

\author{D. Gogichaishvili}
\affiliation{Department of Physics, The University of Texas at
Austin, Austin, Texas 78712, USA}

\author{G. Mamatsashvili}
\affiliation{Helmholtz-Zentrum Dresden-Rossendorf, PO Box 510119,
D-01314 Dresden, Germany} \affiliation{Department of Physics,
Faculty of Exact and Natural Sciences, Tbilisi State University,
Tbilisi 0179, Georgia} \affiliation{Abastumani Astrophysical
Observatory, Ilia State University, Tbilisi 0162, Georgia}

\author{W. Horton}
\affiliation{Institute for Fusion Studies, The University of Texas
at Austin, Austin, Texas 78712, USA} \affiliation{Space and
Geophysics Laboratory, The University of Texas at Austin, 10000
Burnet Rd, Austin, Texas 78758, USA}

\author{G. Chagelishvili}
\affiliation{Abastumani Astrophysical Observatory, Ilia State
University, Tbilisi 0162, Georgia}\affiliation{Institute of
Geophysics, Tbilisi State University, Tbilisi 0193, Georgia}

\author{G. Bodo}
\affiliation{INAF/Osservatorio Astrofisico di Torino, Strada
Osservatorio 20, I-10025 Pino Torinese, Italy}

\begin{abstract}
We investigate magnetohydrodynamic turbulence driven by the
magnetorotational instability (MRI) in Keplerian disks with a
nonzero net azimuthal magnetic field using shearing box simulations.
As distinct from most previous studies, we analyze turbulence
dynamics in Fourier (${\bf k}$-) space to understand its sustenance.
The linear growth of MRI with azimuthal field has a transient
character and is anisotropic in Fourier space, leading to anisotropy
of nonlinear processes in Fourier space. As a result, the main
nonlinear process appears to be a new type of angular redistribution
of modes in Fourier space -- the \emph{nonlinear transverse cascade}
-- rather than usual direct/inverse cascade. We demonstrate that the
turbulence is sustained by interplay of the linear transient growth
of MRI (which is the only energy supply for the turbulence) and the
transverse cascade. These two processes operate at large length
scales, comparable to box size and the corresponding small
wavenumber area, called \emph{vital area} in Fourier space is
crucial for the sustenance, while outside the vital area direct
cascade dominates. The interplay of the linear and nonlinear
processes in Fourier space is generally too intertwined for a vivid
schematization. Nevertheless, we reveal the \emph{basic subcycle} of
the sustenance that clearly shows synergy of these processes in the
self-organization of the magnetized flow system. This synergy is
quite robust and persists for the considered different aspect ratios
of the simulation boxes. The spectral characteristics of the
dynamical processes in these boxes are qualitatively similar,
indicating the universality of the sustenance mechanism of the
MRI-turbulence.
\end{abstract}

\section{Introduction}

The problem of the onset and sustenance of turbulence in accretion
disks lies at the basis of understanding different aspects of disk
dynamics and evolution: secular redistribution of angular momentum
yielding observationally obtained accretion rates, dynamo action and
generation of magnetic fields and outflows, possibility of
appearance of coherent structures (e.g., vortices, zonal flows,
pressure bumps) that can form sites for planet formation.
Investigations in this direction acquired new impetus and became
more active since \citet{Balbus_Hawley91} demonstrated the relevance
and significance of the magnetorotational instability (MRI) for
disks. Today the MRI is considered as the most likely cause of
magnetohydrodynamic (MHD) turbulence in disks and hence a driver
agent of the above phenomena. Starting from the 1990s a vast number
of analytical and numerical studies have investigated different
aspects of linear and nonlinear evolution of the MRI in
three-dimensional (3D) Keplerian disks using both local shearing box
and global approaches for different configurations (unstratified and
stratified, incompressible and compressible, with vertical and/or
azimuthal magnetic fields having zero and nonzero net fluxes) at
different domain sizes and resolutions \citep[see e.g.,][for a
review]{Armitage11,Fromang13}.

In this paper, we consider a local model of a disk threaded by a
nonzero net azimuthal/toroidal magnetic field. The linear stability
analysis showed that only non-axisymmetric perturbations can exhibit
the MRI for this orientation of the background field
\citep{Balbus_Hawley92,Ogilvie_Pringle96,
Terquem_Papaloizou96,Papaloizou_Terquem97,Brandenburg_Dintrans06,Salhi_etal12,Shtemler_etal12}.
Such perturbations are, however, sheared by the disk's differential
rotation (shear) and as a result the MRI acquires a transient
nature, while the flow stays exponentially, or spectrally stable.
Nevertheless, as early seminal numerical simulations by
\citet{Hawley_etal95} revealed, the transient MRI in the presence of
an azimuthal field in fact causes transition to MHD turbulence.
However, the transient growth itself, which in this case is the only
available source of energy for turbulence, cannot ensure a long-term
sustenance of the latter without appropriate nonlinear feedback. In
other words, the role of nonlinearity becomes crucial: it lies at
the heart of the sustenance of turbulence. Thus, the transition to
turbulence in the presence of azimuthal field fundamentally differs
from that in the case of the vertical field, where the MRI grows
exponentially forming a channel flow, which, in turn, breaks down
into turbulence due to secondary (parasitic) instabilities
\citep{Goodman_Xu94,Hawley_etal95,Bodo_etal08,Pessah_Goodman09,
Latter_etal09,Pessah10,Longaretti_Lesur10,Murphy_Pessah15}.

The first developments of the MRI in magnetized disks in the 1990s
coincided with the period of the breakthrough of the fluid dynamical
community in understanding the dynamics of spectrally stable (i.e.,
without exponentially growing eigenmodes) hydrodynamic (HD) shear
flows \citep[see
e.g.,][]{Reddy_etal93,Trefethen_etal93,Farrell_Ioannou96,Schmid_Henningson01,Schmid07}.
The nonnormality of these flows, i.e., the nonorthogonality of the
eigenfunctions of classical modal approach, had been demonstrated
and its consequences -- the transient/nonmodal growth of
perturbations and the transition to turbulence were thoroughly
analyzed. There are no exponentially growing modes in such flows and
the turbulence is energetically supported only by the linear
nonmodal growth of perturbations due to the shear flow nonnormality.
Afterwards, the \emph{bypass} concept of the onset and sustenance of
turbulence in spectrally stable shear flows was formulated
\citep[see
e.g.,][]{Gebhardt_Grossmann94,Baggett_etal95,Grossmann00}. According
to this concept, the turbulence is triggered and maintained by a
subtle interplay of shear-induced linear transient growth and
nonlinear processes. These processes appear to be strongly
anisotropic in Fourier (${\bf k}$-) space due to the shear
\citep{Horton_etal10,Mamatsashvili_etal16} in contrast to classical
isotropic and homogeneous forced turbulence without background
shear.

Differentially rotating disks represent special case of shear flows
and hence the effects of nonnormality inevitably play a key role in
their dynamics
\citep[e.g.,][]{Chagelishvili_etal03,Mukhopadhyay_etal05,Zhuravlev_Razdoburdin14,Razdoburdin_Zhuravlev17}.
In particular, in magnetized disks, the nonmodal/transient growth of
the MRI over intermediate (dynamical) times can be actually more
relevant in many situations than its modal growth
\citep{Mamatsashvili_etal13,Squire_Bhattacharjee14}. Since in the
present case of azimuthal field, the MRI exhibits only transient
rather than exponential growth, the resulting turbulence, like in
spectrally stable HD shear flows, is expected to be governed by a
subtle cooperation of this nonmodal growth and nonlinear processes.
As we showed previously \citep{Mamatsashvili_etal14}, this is indeed
the case for an analogous two-dimensional (2D) MHD flow with linear
shear and magnetic field parallel to it and the flow configuration
considered here in fact represents its 3D generalization. So, our
main goal is to investigate the spectral properties and sustaining
dynamics of MHD turbulence driven by the transient amplification of
the MRI in disks with a net nonzero azimuthal field.

The dynamics and statistics of MRI-driven MHD turbulence in
Keplerian disk flows have been commonly analyzed and interpreted in
physical space rather than in Fourier space. This also concerns
studies of disks with nonzero net azimuthal magnetic field. Below we
cite the most relevant ones.
\citet{Hawley_etal95,Guan_etal09,Guan_Gammie11,Nauman_Blackman14,Ross_etal16}
in the shearing box, and
\citet{Fromang_Nelson06,Beckwith_etal11,Flock_etal11,Flock_etal12b,
Sorathia_etal12,Hawley_etal13,Parkin_Bicknell13} in global disk
simulations extensively investigated the dependence of the dynamics
and saturation of the MRI-turbulence without explicit dissipation on
the domain size, resolution and imposed azimuthal field strength.
\citet{Fleming_etal00} in local model and \citet{Flock_etal12a} in
global model addressed the influence of resistivity and established
a critical value of magnetic Reynolds number for the existence of
turbulence. \citet{Simon_Hawley09} and \citet{Meheut_etal15}
included also viscosity together with resistivity and showed that at
fixed field strength the saturation amplitude mainly depends on the
magnetic Prandtl number, that is, the ratio of viscosity to
resistivity, if the latter is larger than unity and the Reynolds
number is high enough. On the other hand, at Prandtl numbers smaller
than unity the turbulence sustenance is more delicate: it appears to
be independent of the Prandtl number and mainly determined by the
magnetic Reynolds number. \citet{Simon_Hawley09} attributed this
behavior to the small-scale resistive dissipation processes
(reconnection), which are thought to be central in the saturation
process.

Part of these papers based on the local approximation
\citep{Hawley_etal95,Fleming_etal00,Nauman_Blackman14,Meheut_etal15}
do present analysis of energy density power spectrum, but in
somewhat restricted manner by considering either averaging over
wavevector angle, i.e., averages over spherical shells of constant
$|{\bf k}|$, or slices along different directions in Fourier space.
However, there are several studies of MRI turbulence also in the
local approximation, but with nonzero net vertical magnetic flux
\citep{Lesur_Longaretti11} and with zero net flux
\citep{Fromang_Papaloizou07,Fromang_etal07,Simon_etal09,Davis_etal10},
which go beyond energy spectrum and describe the dynamics of MRI-
turbulence and associated energy injection (stresses) and nonlinear
transfer processes in Fourier space, but again in a restricted
manner by using shell-averaging procedure and/or reduced
one-dimensional (1D) spectrum along a certain direction in Fourier
space by integrating in the other two. However, as demonstrated by
\citet{Hawley_etal95,Lesur_Longaretti11,Murphy_Pessah15} for
MRI-turbulence (with net vertical field) and by our previous study
of 2D MHD shear flow turbulence in Fourier space
\citep{Mamatsashvili_etal14}, the power spectra and underlying
dynamics are notably anisotropic due to shear, i.e., depend quite
strongly also on the orientation of wavevector ${\bf k}$ in Fourier
space rather than only on its magnitude $|{\bf k}|$. This is in
contrast to a classical isotropic forced turbulence without
background velocity shear, where energy cascade proceeds along ${\bf
k}$ only \citep{Biskamp03}. This shear-induced anisotropy also
differs from the typical anisotropy of classical shearless MHD
turbulence in the presence of a (strong) background magnetic field
\citep{Goldreich_Sridhar95}. It leads to anisotropy of nonlinear
processes and particularly to the nonlinear transverse cascade (see
below) that play a central role in the sustenance of turbulence in
the presence of transient growth. Consequently, the shell-averaging
done in the above studies is misleading, because it completely
leaves out shear-induced spectral anisotropy, which is thus an
essential ingredient of the dynamics of shear MHD turbulence. The
recent works by \citet{Meheut_etal15} and \citet{Murphy_Pessah15}
share a similar point of view, emphasizing the importance of
describing anisotropic shear MRI-turbulence using a full 3D spectral
analysis instead of using spherical shell averaging in Fourier
space, which is applicable only for isotropic turbulence without
shear. Such a generalized treatment is a main goal of this paper. In
particular, \citet{Murphy_Pessah15} employ a new approach that
consists in using invariant maps for characterizing anisotropy of
MRI-driven turbulence in physical space and dissecting the 3D
Fourier spectrum along the most relevant planes, as defined by the
type of anisotropy of the flows.

As for the global disk studies cited above, relatively little
attention is devoted to the dynamics of MRI-turbulence in Fourier
space. This is, however, understandable, since in contrast to the
cartesian shearing box model, global disk geometry makes it harder
to perform Fourier analysis in all three, radial, azimuthal and
meridional directions, so that these studies only consider azimuthal
spectra integrated in other two directions.

Recently, we have numerically studied a cooperative interplay of
linear transient growth and nonlinear processes ensuring the
sustenance of nonlinear perturbations in HD and 2D MHD plane
spectrally stable constant shear flows
\citep{Horton_etal10,Mamatsashvili_etal14,Mamatsashvili_etal16}.
Performing the analysis of dynamical processes in Fourier space, we
showed that the shear-induced spectral anisotropy gives rise to a
new type of nonlinear cascade process that leads to transverse
redistribution of modes in {\bf k}-space, i.e. to a redistribution
over wavevector angles. This process, referred to as \emph{the
nonlinear transverse cascade}, originates ultimately from flow shear
and fundamentally differs from the canonical (direct and inverse)
cascade processes accepted in classical Kolmogorov or
Iroshnikov-Kraichnan (IK) theories of turbulence \citep[see
e.g.,][]{Biskamp03}. The new approach developed in these
studies and the main results can be summarized as follows: \\
-- identifying modes that play a key role in the
sustaining process of the turbulence;\\
-- defining a wavenumber area in Fourier space that is vital in
the sustenance of turbulence;\\
-- defining a range of aspect ratios of the simulation domain for
which the dynamically important modes are fully taken into account;\\
-- revealing the dominance of the nonlinear transverse cascade in
the dynamics;\\
-- showing that the turbulence is sustained by a subtle interplay
between the linear transient (nonmodal) growth and the nonlinear
transverse cascade.

In this paper, with the same spirit and goals in mind, we take the
approach of \citet{Mamatsashvili_etal14} to investigate the dynamics
and sustenance of MHD turbulence driven by the transient growth of
MRI with a net nonzero azimuthal field in a Keplerian disk flow. We
adopt the shearing box model of the disk \citep[see
e.g.,][]{Hawley_etal95}, where the flow is characterized by constant
shear rate, as that considered in that paper, except it is 3D,
including rotation (Coriolis force) and vertical thermal
stratification. To capture the spectral anisotropy of the
MRI-turbulence, we analyze the linear and nonlinear dynamical
processes and their interplay in 3D Fourier space in full
\emph{without} using the above-mentioned procedure of averaging over
spherical shells of constant $k=|{\bf k}|$. So, our study is
intended to be more general than the above-mentioned studies that
also addressed the spectral dynamics of MRI-turbulence. One of our
goals is to demonstrate the realization and efficiency of the
transverse cascade and its role in the turbulence dynamics also in
the 3D case, as we did for 2D MHD shear flow. Although in 3D
perturbation modes are more diverse and, of course, modify the
dynamics, still the essence of the cooperative interplay of linear
(transient) and nonlinear (transverse cascade) processes should be
preserved.

We pay particular attention to the choice of the aspect ratio of the
simulation box, so as to encompass as full as possible the modes
exhibiting the most effective amplification due to the transient
MRI. To this aim, we apply the method of optimal perturbations,
widely used in fluid dynamics for characterizing the nonmodal growth
in spectrally stable shear flows \citep[see
e.g.,][]{Farrell_Ioannou96,Schmid_Henningson01,Zhuravlev_Razdoburdin14},
to the present MRI problem \citep[see
also][]{Squire_Bhattacharjee14}. These are perturbations that
undergo maximal transient growth during the dynamical time. In this
framework, we define areas in Fourier space, where the transient
growth is more effective -- these areas cover small wavenumber
modes. On the other hand, the simulation box includes only a
discrete number of modes and minimum wavenumbers are set by its
size. A dense population of modes in these areas of the effective
growth in ${\bf k}$-space is then achieved by suitably choosing the
box sizes. In particular, we show that simulations with elongated in
the azimuthal direction boxes (i.e., with azimuthal size larger than
radial one), do not fully account for this nonmodal effects, since
the discrete wavenumbers of modes contained in such boxes scarcely
cover the areas of efficient transient growth.

The paper is organized as follows. The physical model and derivation
of dynamical equations in Fourier space is given in Section 2.
Selection of the suitable aspect ratio of the simulation box based
on the optimal growth calculations is made in Section 3. Numerical
simulations of the MRI-turbulence at different aspect ratios of the
simulation box are done in Section 4. In this Section we present
also energy spectra, we determine dynamically active modes and
delineate the vital area of turbulence, where the active modes and
hence the sustaining dynamics are concentrated. The analysis of the
interplay of the linear and nonlinear processes in Fourier space and
the sustaining mechanism of the turbulence is described in Section
5. In this Section we also reveal the basic subcycle of the
sustenance, describe the importance of the magnetic nonlinear term
in the generation and maintenance of the zonal flow, examine the
effect of the box aspect ratio and demonstrate the universality of
the sustaining scheme. A summary and discussion are given in Section
6.

\section{Physical model and equations}

We consider the motion of an incompressible conducting fluid with
constant kinematic viscosity $\nu$, thermal diffusivity $\chi$ and
Ohmic resistivity $\eta$, in the shearing box centered at a radius
$r_0$ and rotating with the disk at angular velocity $\Omega(r_0)$.
Adopting the Boussinesq approximation for vertical thermal
stratification \citep{Balbus_Hawley91,Lesur_Ogilvie10}, the
governing equations of the non-ideal MHD become
\begin{multline}\label{eq:mom}
\frac{\partial {\bf U}}{\partial t}+({\bf U}\cdot \nabla) {\bf
U}=-\frac{1}{\rho}\nabla P +\frac{\left({\bf B}\cdot\nabla
\right){\bf B}}{4\pi \rho} - 2{\bf \Omega}\times{\bf
U}\\+2q{\Omega}^2 x {\bf e}_x - \Lambda N^2 \theta {\bf
~e}_z+\nu\nabla^2 {\bf U},
\end{multline}
\begin{equation}\label{eq:dens}
\frac{\partial \theta}{\partial t}+{\bf U}\cdot \nabla \theta
=\frac{u_z}{\Lambda} +\chi \nabla^2 \theta,
\end{equation}
\begin{equation}\label{eq:ind}
\frac{\partial {\bf B}}{\partial t}=\nabla\times \left( {\bf
U}\times {\bf B}\right)+\eta\nabla^2{\bf B},
\end{equation}
\begin{equation}\label{eq:divv}
\nabla\cdot {\bf U}=0,
\end{equation}
\begin{equation}\label{eq:divb}
\nabla\cdot {\bf B}=0,
\end{equation}
where ${\bf e}_x,~{\bf e}_y,~{\bf e}_z$ are the unit vectors,
respectively, along the radial $(x)$, azimuthal $(y)$ and vertical
$(z)$ directions, $\rho$ is the density, ${\bf U}$ is the velocity,
${\bf B}$ is the magnetic field, $P$ is the total pressure, equal to
the sum of the thermal and magnetic pressures, $\theta \equiv
{\delta \rho}/\rho$ is the perturbation of the density logarithm (or
entropy, since pressure perturbations are neglected in the
Boussinesq approximation). Finally, $N^2$ is the Brunt-V$\ddot{\rm
a}$is$\ddot{\rm a}$l$\ddot{\rm a}$ frequency squared that controls
the stratification. It is assumed to be positive and spatially
constant, equal to $N^2=0.25\Omega^2$, formally corresponding to a
stably stratified (i.e., convectively stable) local model along the
vertical $z$-axis \citep{Lesur_Ogilvie10}. For dimensional
correspondence with the usual Boussinesq approximation, we define a
stratification length $\Lambda \equiv g/N^2$, where $g$ is the
vertical component of the gravity. Note, however, that $\Lambda$
cancels out from the equations if we normalize the density logarithm
by $\Lambda\theta \rightarrow \theta$, which will be used
henceforth. So, here we take into account the effects of thermal
stratification in a simple way. \citet{Bodo_etal12,Bodo_etal13}
studied more sophisticated models of stratified MRI-turbulence in
the shearing box, treating thermal physics self-consistently with
dynamical equations. The shear parameter $q=-d\ln\Omega/d\ln r$ is
set to $q=3/2$ for a Keplerian disk.

Equations (\ref{eq:mom})-(\ref{eq:divb}) have a stationary
equilibrium solution -- an azimuthal flow along the $y$-direction
with linear shear of velocity in the the radial $x$-direction, ${\bf
U}_0=(0,-q\Omega x,0)$, with the total pressure $P_0$, density
$\rho_0$ and threaded by an azimuthal uniform background magnetic
field, ${\bf B}_0=(0,B_{0y},0), B_{0y}>0$. This simple, but
important configuration, which corresponds to a local version of a
Keplerian flow with toroidal field, allows us to grasp the key
effects of the shear on the perturbation dynamics and ultimately on
a resulting turbulent state.

Consider perturbations of the velocity, total pressure and magnetic
field about the equilibrium, ${\bf u}={\bf U}-{\bf U}_0, p=P-P_0,
{\bf b}={\bf B}-{\bf B}_0$. Substituting them into Equations
(\ref{eq:mom})-(\ref{eq:divb}) and rearranging the nonlinear terms
with the help of divergence-free conditions (\ref{eq:divv}) and
(\ref{eq:divb}), we arrive to the system
(\ref{eq:App-ux})-(\ref{eq:App-divperb}) governing the dynamics of
perturbations with arbitrary amplitude that is given in Appendix.
These equations are solved within a box with sizes $(L_x, L_y, L_z)$
and resolutions $(N_x, N_y, N_z)$, respectively, in the
$x,y,z-$directions. We use standard for the shearing box boundary
conditions: shearing-periodic in $x$ and periodic in $y$ and $z$
\citep{Hawley_etal95}. For stratified disks, outflow boundary
conditions in the vertical direction are more appropriate, however,
in the present study, as mentioned above, we adopt a local
approximation in $z$ with spatially constant $N^2$ that justifies
our choice of the periodic boundary conditions in this direction
\citep{Lesur_Ogilvie10}. This does not affect the main dynamical
processes in question.

\subsection{Energy equation}

The perturbation kinetic, thermal and magnetic energy densities are
defined, respectively, as
\[
E_K=\frac{1}{2}\rho_0{\bf u}^2,~~~E_{th}=\frac{1}{2}\rho_0
N^2\theta^2,~~~E_M=\frac{{\bf b}^2}{8\pi}.
\]
From the main Equations (\ref{eq:App-ux})-(\ref{eq:App-divperb}) and
the shearing box boundary conditions, after some algebra, we can
readily derive the evolution equations for the volume-averaged
kinetic, thermal and magnetic energy densities
\begin{multline}\label{eq:ek}
\frac{d}{dt}\langle E_K \rangle=q\Omega \left\langle\rho_0
u_xu_y\right\rangle-N^2\left\langle \rho_0\theta u_z
\right\rangle+\frac{1}{4\pi}\left\langle {\bf B}_0{\bf u}\otimes
\nabla {\bf b}\right\rangle\\-\frac{1}{4\pi}\left\langle {\bf b}{\bf
b}\otimes \nabla{\bf u}\right\rangle-\rho_0\nu\langle
\left(\nabla{\bf u}\right)^2\rangle,
\end{multline}
\begin{equation}\label{eq:eth}
\frac{d}{dt}\langle E_{th} \rangle=N^2\left\langle \rho_0\theta u_z
\right\rangle-\rho_0N^2\chi\langle \left(\nabla
\theta\right)^2\rangle,
\end{equation}
\begin{multline}\label{eq:em}
\frac{d}{dt}\langle E_M \rangle=q\Omega\left\langle
-\frac{b_xb_y}{4\pi}\right\rangle+\frac{1}{4\pi}\left\langle {\bf
B}_0{\bf b}\otimes \nabla {\bf
u}\right\rangle\\+\frac{1}{4\pi}\left\langle {\bf b}{\bf b}\otimes
\nabla{\bf u}\right\rangle-\frac{\eta}{4\pi}\langle \left(\nabla{\bf
b}\right)^2\rangle,
\end{multline}
where the angle brackets denote an average over the box. Adding up
Equations (\ref{eq:ek})-(\ref{eq:em}), the cross terms of linear
origin on the right hand side (rhs), proportional to $N^2$ and ${\bf
B}_0$ (which describe kinetic-thermal and kinetic-magnetic energy
exchange, respectively) and the nonlinear terms cancel out because
of the boundary conditions. As a result, we obtain the equation for
the total energy density $E=E_K+E_{th}+E_M$,
\begin{multline}\label{eq:etot}
\frac{d\langle E \rangle}{dt}=q\Omega\left\langle
\rho_0u_xu_y-\frac{b_xb_y}{4\pi}\right\rangle\\-\rho_0\nu\langle
\left(\nabla {\bf u}\right)^2\rangle-\rho_0N^2\chi\langle
\left(\nabla
\theta\right)^2\rangle-\frac{\eta}{4\pi}\langle\left(\nabla {\bf
b}\right)^2\rangle.
\end{multline}
The first term on the rhs of Equation (\ref{eq:etot}) is the flow
shear, $q\Omega$, multiplied by the volume-averaged total stress.
The total stress is a sum of the Reynolds, $\rho_0u_xu_y$, and
Maxwell, $-b_xb_y/4\pi$, stresses that describe, respectively,
exchange of kinetic and magnetic energies between perturbations and
the background flow in Equations (\ref{eq:ek}) and (\ref{eq:em}).
Note that they originate from the linear terms proportional to shear
in Equations (\ref{eq:App-uy}) and (\ref{eq:App-by}). The stresses
also determine the rate of angular momentum transport
\citep[e.g.,][]{Hawley_etal95,Balbus03} and thus are one of the
important diagnostics of turbulence. The negative definite second,
third and fourth terms describe energy dissipation due to viscosity,
thermal diffusivity and resistivity, respectively. Note that a net
contribution from the nonlinear terms has canceled out in the total
energy evolution Equation (\ref{eq:etot}) after averaging over the
box. Thus, only Reynolds and Maxwell stresses can supply
perturbations with energy, extracting it from the background flow
due to the shear. In the case of the MRI-turbulence studied below,
these stresses ensure energy injection into turbulent fluctuations.
The nonlinear terms, not directly tapping into the flow energy and
therefore not changing the total perturbation energy, act only to
redistribute energy among different wavenumbers as well as among
components of velocity and magnetic field (see below). In the
absence of shear ($q=0$), the contribution from the Reynolds and
Maxwell stresses disappears in Equation (\ref{eq:etot}) and hence
the total perturbation energy cannot grow, gradually decaying due to
dissipation.

\subsection{Spectral representation of the equations}

Before proceeding further, we normalize the variables by taking
$\Omega^{-1}$ as the unit of time, the disk scale height $H$ as the
unit of length, $\Omega H$ as the unit of velocity, $\Omega
H\sqrt{4\pi \rho_0}$ as the unit of magnetic field and
$\rho_0\Omega^2H^2$ as the unit of pressure and energy. Viscosity,
thermal diffusivity and resistivity are measured, respectively, by
Reynolds number, ${\rm Re}$, P\'eclet number, ${\rm Pe}$, and
magnetic Reynolds number, ${\rm Rm}$, defined as
\[
{\rm Re}= \frac{\Omega H^2}{\nu},~~~{\rm Pe}=\frac{\Omega
H^2}{\chi},~~~{\rm Rm}=\frac{\Omega H^2}{\eta}.
\]
All the simulations share the same ${\rm Re}={\rm Pe}={\rm Rm}=3200$
(i.e., the magnetic Prandtl number ${\rm Pm}={\rm Rm}/{\rm Re}=1$).
The strength of the imposed background uniform azimuthal magnetic
field is measured by a parameter $\beta=2\Omega^2H^2/v_A^2$, which
we fix to $\beta=200$, where $v_A=B_{0y}/(4\pi \rho_0)^{1/2}$ is the
corresponding Alfv\'en speed. In the incompressible case, this
parameter is a proxy of the usual plasma $\beta$ parameter
\citep{Longaretti_Lesur10}, since the sound speed in thin disks is
$c_s\sim \Omega H$. In this non-dimensional units, the mean field
becomes $B_{0y}=\sqrt{2/\beta}=0.1.$

Our primary interest lies in the spectral aspect of the dynamics, so
we start with decomposing the perturbations $f\equiv ({\bf
u},p,\theta,{\bf b})$ into spatial Fourier harmonics/modes
\begin{equation}\label{eq:fourier}
f({\bf r},t)=\int \bar{f}({\bf k},t)\exp\left({\rm i}{\bf
k}\cdot{\bf r} \right)d^3{\bf k}
\end{equation}
where $\bar{f}\equiv (\bar{\bf u}, \bar{p}, \bar{\theta},\bar{\bf
b})$ denotes the corresponding Fourier transforms. Substituting
decomposition (\ref{eq:fourier}) into perturbation Equations
(\ref{eq:App-ux})-(\ref{eq:App-divperb}), taking into account the
above normalization and eliminating the pressure (see derivation in
Appendix), we obtain the following evolution equations for the
quadratic forms of the spectral velocity, logarithmic density
(entropy) and magnetic field:
\begin{equation}\label{eq:uxk}
\frac{\partial}{\partial t}\frac{|\bar{u}_x|^2}{2} = -
qk_y\frac{\partial}{\partial k_x}\frac{|\bar{u}_x|^2}{2} + {\cal
H}_x + {\cal I}_x^{(u\theta)} + {\cal I}_x^{(ub)}+{\cal
D}_x^{(u)}+{\cal N}_x^{(u)},
\end{equation}
\begin{equation}\label{eq:uyk}
\frac{\partial}{\partial t} \frac{|\bar{u}_y|^2}{2} = -
qk_y\frac{\partial }{\partial k_x}\frac{|\bar{u}_y|^2}{2} + {\cal
H}_y + {\cal I}_y^{(u\theta)} + {\cal I}_y^{(ub)}+{\cal
D}_y^{(u)}+{\cal N}_y^{(u)},
\end{equation}
\begin{equation}\label{eq:uzk}
\frac{\partial}{\partial t} \frac{|\bar{u}_z|^2}{2} = -
qk_y\frac{\partial}{\partial k_x}\frac{|\bar{u}_z|^2}{2} + {\cal
H}_z + {\cal I}_z^{(u\theta)} + {\cal I}_z^{(ub)}+{\cal
D}_z^{(u)}+{\cal N}_z^{(u)},
\end{equation}
\begin{equation}\label{eq:thetak}
\frac{\partial}{\partial t} \frac{|\bar{\theta}|^2}{2}= -
qk_y\frac{\partial}{\partial k_x} \frac{|\bar{\theta}|^2}{2}+{\cal
I}^{(\theta u)} + {\cal D}^{(\theta)} + {\cal N}^{(\theta)},
\end{equation}
\begin{equation}\label{eq:bxk}
\frac{\partial}{\partial t}\frac{|\bar{b}_x|^2}{2} = -
qk_y\frac{\partial}{\partial k_x}\frac{|\bar{b}_x|^2}{2} + {\cal
I}_x^{(bu)}+{\cal D}_x^{(b)}+{\cal N}_x^{(b)},
\end{equation}
\begin{equation}\label{eq:byk}
\frac{\partial}{\partial t}\frac{|\bar{b}_y|^2}{2} = -
qk_y\frac{\partial}{\partial k_x} \frac{|\bar{b}_y|^2}{2} + {\cal
M}+{\cal I}_y^{(bu)} + {\cal D}_y^{(b)}+{\cal N}_y^{(b)},
\end{equation}
\begin{equation}\label{eq:bzk}
\frac{\partial}{\partial t} \frac{|\bar{b}_z|^2}{2} = -
qk_y\frac{\partial}{\partial k_x}\frac{|\bar{b}_z|^2}{2} + {\cal
I}_z^{(bu)}+{\cal D}_z^{(b)}+{\cal N}_z^{(b)}.
\end{equation}
These seven dynamical equations in Fourier space, which are the
basis for the subsequent analysis, describe processes of linear,
$~{\cal H}_i({\bf k},t)$, ${\cal I}_i^{(u\theta)}({\bf k},t)$,
${\cal I}^{(\theta u)}({\bf k},t)$, ${\cal I}_i^{(ub)}({\bf k},t)$,
${\cal I}_i^{(bu)}({\bf k},t)$, ${\cal M}({\bf k},t)$, and
nonlinear, $~{\cal N}_i^{(u)}({\bf k},t)$, ${\cal N}^{(\theta)}({\bf
k},t)$, ${\cal N}_i^{(b)}({\bf k},t)$, origin, where the index
$i=x,y,z$ henceforth. ${\cal D}_i^{(u)}({\bf k},t)$, ${\cal
D}^{(\theta)}({\bf k},t)$, ${\cal D}_i^{(b)}({\bf k},t)$ describe
the effects of viscous, thermal and resistive dissipation as a
function of wavenumber and are negative definite. These terms come
from the respective linear and nonlinear terms in main Equations
(\ref{eq:App-ux})-(\ref{eq:App-bz}) and their explicit expressions
are derived in Appendix. In the turbulent regime, these basic linear
and nonlinear processes are subtly intertwined, so before embarking
on calculating and analyzing these terms from the simulation data,
we first describe them in more detail below. Equations
(\ref{eq:uxk})-(\ref{eq:bzk}) serve as a mathematical basis for our
main goal -- understanding the character of the interplay of the
dynamical processes sustaining the MRI-turbulence. Since we consider
a finite box in physical space, the perturbation dynamics also
depends on the smallest wavenumber available in the box (see Section
3), which is set by its sizes $L_x, L_y, L_z$ and is a free
parameter in the shearing box.

To get a general feeling, as in
\citet{Simon_etal09,Lesur_Longaretti11}, we derive also equations
for the spectral kinetic energy, ${\cal E}_K =
(|\bar{u}_x|^2+|\bar{u}_y|^2+|\bar{u}_z|^2)/2$, by combining
Equations (\ref{eq:uxk})-(\ref{eq:uzk}),
\begin{equation}\label{eq:ekspec}
\frac{\partial {\cal E}_K}{\partial t} = -qk_y \frac{\partial {\cal
E}_K}{\partial k_x} + {\cal H} + {\cal I}^{(u\theta)} + {\cal
I}^{(ub)} + {\cal D}^{(u)} + {\cal N}^{(u)},
\end{equation}
where
\[
{\cal H} = \sum_i {\cal
H}_i=\frac{q}{2}(\bar{u}_x\bar{u}_y^{\ast}+\bar{u}_x^{\ast}\bar{u}_y),
\]
\[
{\cal I}^{(u\theta)}=\sum_i {\cal I}_i^{(u\theta)},~~~~{\cal
I}^{(ub)}=\sum_i {\cal I}_i^{(ub)},
\]
\[
{\cal D}^{(u)}=\sum_i {\cal D}_i^{(u)}=-\frac{2k^2}{\rm Re}{\cal
E}_K,~~~{\cal N}^{(u)}=\sum_i {\cal N}^{(u)}_i
\]
and for the spectral magnetic energy, ${\cal
E}_M=(|\bar{b}_x|^2+|\bar{b}_y|^2+|\bar{b}_z|^2)/2$, by combining
Equations (\ref{eq:bxk})-(\ref{eq:bzk}),
\begin{equation}\label{eq:emspec}
\frac{\partial {\cal E}_M}{\partial t} = -qk_y\frac{\partial {\cal
E}_M}{\partial k_x} + {\cal M}+{\cal I}^{(bu)}+{\cal D}^{(b)}+{\cal
N}^{(b)},
\end{equation}
where
\[
{\cal
M}=-\frac{q}{2}(\bar{b}_x\bar{b}_y^{\ast}+\bar{b}_x^{\ast}\bar{b}_y),~~~~{\cal
I}^{(bu)}=\sum_i {\cal I}_i^{(bu)}=-{\cal I}^{(ub)},
\]
\[
{\cal D}^{(u)}=\sum_i {\cal D}_i^{(u)}=-\frac{2k^2}{\rm Rm}{\cal
E}_M,~~~{\cal N}^{(b)}=\sum_i {\cal N}^{(b)}_i.
\]
The equation of the thermal energy, ${\cal E}_{th}=N^2|\theta|^2/2$,
is straightforward to derive by multiplying Equation
(\ref{eq:thetak}) just by $N^2$, so we do not write it here.
Besides, we will see below that the thermal energy is much less than
the magnetic and kinetic energies, so the thermal processes have a
minor contribution in forming the final picture of the turbulence.
Similarly, we get the equation for the total spectral energy of
perturbations, ${\cal E}={\cal E}_K+{\cal E}_{th}+{\cal E}_M$,
\begin{multline}\label{eq:etotspec}
\frac{\partial {\cal E}}{\partial t} = -qk_y\frac{\partial {\cal
E}}{\partial k_x} + {\cal H}+{\cal M}+{\cal D}^{(u)}+N^2{\cal
D}^{(\theta)}\\+{\cal D}^{(b)}+{\cal N}^{(u)}+N^2{\cal
N}^{(\theta)}+{\cal N}^{(b)}.
\end{multline}

One can distinguish six basic processes, five of linear and one of
nonlinear origin, in Equations (\ref{eq:uxk}) and (\ref{eq:bzk})
(and therefore in energy Equations \ref{eq:ekspec} and
\ref{eq:emspec}) that underlie the perturbation dynamics:
\begin{enumerate}
\item
The first terms on the rhs of Equations
(\ref{eq:uxk})-(\ref{eq:bzk}), $-qk_y\partial (.)/\partial k_x$,
describe the linear ``drift'' of the related quadratic forms
parallel to the $k_x$-axis with the normalized velocity $qk_y$.
These terms are of linear origin, arising from the convective
derivative on the lhs of the main Equations
(\ref{eq:App-ux})-(\ref{eq:App-bz}) and therefore correspond to the
advection by the background flow. In other words, background shear
makes the spectral quantities (Fourier transforms) drift in ${\bf
k}-$space, non-axisymmetric harmonics with $k_y>0$ and $k_y<0$
travel, respectively, along and opposite the $k_x-$axis at a speed
$|qk_y|$, whereas the ones with $k_y=0$ are not advected by the
flow. This drift in Fourier space is equivalent to the time-varying
radial wavenumber, $k_x(t)=k_x(0)+q\Omega k_yt$, in the linear
analysis of non-axisymmetric shearing waves in magnetized disks
\citep[e.g.,][]{Balbus_Hawley92,Johnson07,Pessah_Chan12}. In the
energy Equations (\ref{eq:ekspec}) and (\ref{eq:emspec}), the
spectral energy drift, of course, does not change the total kinetic
and magnetic energies, since $\int d^3{\bf k}\partial (k_y{\cal
E}_{K,M})/\partial k_x=0$.
\item
The second rhs terms of Equations (\ref{eq:uxk})-(\ref{eq:uzk}),
${\cal H}_i$, and Equation (\ref{eq:byk}), ${\cal M}$, are also of
linear origin associated with the shear (Equations
\ref{eq:App-Hx}-\ref{eq:App-Hz} and \ref{eq:App-M}), i.e., originate
from the linear terms proportional to the shear parameter in
Equations (\ref{eq:App-uy}) and (\ref{eq:App-by}). They describe the
interaction between the flow and individual Fourier modes, where the
velocity components $|\bar{u}_i|^2$ and the azimuthal field
perturbation $|\bar{b}_y|^2$ can grow, respectively, due to ${\cal
H}_i$ and ${\cal M}$, at the expense of the flow. In the present
case, such amplification is due to the linear azimuthal MRI fed by
the shear. In the presence of the mean azimuthal field, only
non-axisymmetric modes exhibit the MRI and since they also undergo
the drift in ${\bf k}-$space, their amplification acquires a
transient nature
\citep{Balbus_Hawley92,Papaloizou_Terquem97,Brandenburg_Dintrans06,Salhi_etal12,Shtemler_etal12}.
From the expressions (\ref{eq:App-Hx})-(\ref{eq:App-Hz}) and
(\ref{eq:App-M}), we can see that ${\cal H}_i$ and ${\cal M}$ are
related to the volume-averaged nondimensional Reynolds and Maxwell
stresses entering energy Equations (\ref{eq:ek}) and (\ref{eq:em})
through
\[
q\langle u_xu_y\rangle=\int {\cal H} d^3{\bf k},~~~~~q\langle
-b_xb_y\rangle=\int {\cal M}d^3{\bf k},
\]
where ${\cal H}=\sum_i {\cal H}_i$, and hence represent,
respectively, the spectra of the Reynolds and Maxwell stresses,
acting as the source, or injection of kinetic and magnetic energies
for perturbation modes at each wavenumber (see Equations
\ref{eq:ekspec} and \ref{eq:emspec}) \citep[see
also][]{Fromang_Papaloizou07,Simon_etal09,Davis_etal10,Lesur_Longaretti11}.
\item
the cross terms, ${\cal I}_i^{(u\theta)}$ and ${\cal I}^{(\theta
u)}$ (Equations \ref{eq:App-Iuthi} and \ref{eq:App-Ithu}) describe,
respectively, the effect of the thermal process on the $i$-component
of the velocity, $\bar{u}_i$, and the effect of the $z$-component of
the velocity on the logarithmic density (entropy) for each mode.
These terms are also of linear origin, related to the
Brunt-V$\ddot{\rm a}$is$\ddot{\rm a}$l$\ddot{\rm a}$ frequency
squared $N^2$, and come from the corresponding linear terms in
Equations (\ref{eq:App-uz}) and (\ref{eq:App-theta}). They are not a
source of new energy, as $\sum_i {\cal I}_i^{(u\theta)}+N^2{\cal
I}^{(\theta u)}=0$, but rather characterize exchange between kinetic
and thermal energies (Equation \ref{eq:thetak} and \ref{eq:ekspec}),
so they cancel out in the total spectral energy Equation
(\ref{eq:etotspec}).
\item
the second type of cross terms, ${\cal I}_i^{(ub)}$ and ${\cal
I}_i^{(bu)}$ (Equations \ref{eq:App-Iubi} and \ref{eq:App-Ibui}),
describe, respectively, the influence of the $i$-component of the
magnetic field, $\bar{b}_i$, on the same component of the velocity,
$\bar{u}_i$, and vice versa for each mode. These terms are of linear
origin too, proportional to the mean field $B_{0y}$, and originate
from the corresponding terms in Equations
(\ref{eq:App-ux})-(\ref{eq:App-uz}) and
(\ref{eq:App-bx})-(\ref{eq:App-bz}). From the definition it follows
that ${\cal I}_i^{(ub)}= - {\cal I}_i^{(bu)}$ and hence these terms
also do not generate new energy for perturbations, but rather
exchange between kinetic and magnetic energies (Equations
\ref{eq:ekspec} and \ref{eq:emspec}). They also cancel out in the
total spectral energy equation.
\item
The terms ${\cal D}_i^{(u)}$, ${\cal D}^{(\theta)}$ and ${\cal
D}_i^{(b)}$ (Equations \ref{eq:App-Dui}, \ref{eq:App-Dth} and
\ref{eq:App-Dbi}) describe, respectively, dissipation of velocity,
logarithmic density (entropy) and magnetic field for each
wavenumber. They are obviously of linear origin and negative
definite. Comparing these dissipation terms with the
energy-supplying terms ${\cal H}_i$ and ${\cal M}$, we see that the
dissipation is at work at large wavenumbers $k \gtrsim k_D \equiv
{\rm min}(\sqrt{\rm Re}, \sqrt{\rm Pe}, \sqrt{\rm Rm})$.
\item
The terms ${\cal N}_i^{(u)}$, $\cal N^{(\theta)}$ and ${\cal
N}_i^{(b)}$ (Equations \ref{eq:App-Nui}, \ref{eq:App-Nth1} and
\ref{eq:App-Nbi}) originate from the nonlinear terms in main
Equations (\ref{eq:App-ux})-(\ref{eq:App-bz}) and therefore describe
redistributions, or transfers/cascades of the squared amplitudes,
respectively, of the $i$-component of the velocity, $|\bar{u}_i|^2$,
entropy, $|\bar{\theta}|^2$, and the $i$-component of the magnetic
field, $|\bar{b}_i|^2$, over wavenumbers in $\textbf{k}-$space as
well as among each other via nonlinear triad interactions.
Similarly, the above-defined ${\cal N}^{(u)}$, $N^2{\cal
N}^{(\theta)}$, ${\cal N}^{(b)}$ describe nonlinear transfers of
kinetic, thermal and magnetic energies, respectively. It follows
from the definition of these terms that their sum integrated over an
entire Fourier space is zero,
\begin{equation}\label{eq:sum}
\int [{\cal N}^{(u)}({\bf k},t)+N^2{\cal N}^{(\theta)}({\bf
k},t)+{\cal N}^{(b)}({\bf k},t)]d^3{\bf k}=0,
\end{equation}
which is, in fact, a direct consequence of cancelation of the
nonlinear terms in the total energy Equation (\ref{eq:etot}) in
physical space. This implies that the main effect of nonlinearity is
only to redistribute (scatter) energy (drawn from the background
flow by Reynolds and Maxwell stresses) of the kinetic, thermal and
magnetic components over wavenumbers and among each other, while
leaving the total spectral energy summed over all wavenumbers
unchanged. The nonlinear transfer functions (${\cal N}^{(u)}$,
${\cal N}^{(\theta)}$, ${\cal N}^{(b)}$) play a central role in MHD
turbulence theory -- they determine cascades of energies in
$\textbf{k}$-space, leading to the development of their specific
spectra
\citep[e.g.,][]{Verma04,Alexakis_etal07,Teaca_etal09,Sundar_etal17}.
These transfer functions are one of the main focus of the present
analysis. One of our main goals is to explore how they operate in
the presence of the azimuthal field MRI in disks and ultimately of
the shear. Specifically, below we will show that, like in 2D HD and
MHD shear flows we studied before
\citep{Horton_etal10,Mamatsashvili_etal14}, energy spectra,
energy-injection as well as nonlinear transfers are also anisotropic
in the quasi-steady MRI-turbulence, resulting in the redistribution
of power among wavevector angles in $\textbf{k}-$space, i.e., the
\emph{nonlinear transverse cascade}.
\end{enumerate}

Having described all the terms in spectral equations, we now turn to
the total spectral energy Equation (\ref{eq:etotspec}). Each mode
drifting parallel to the $k_x-$axis, go through a dynamically
important region in Fourier space, which we call the \emph{vital
area}, where \emph{energy-supplying} linear terms, ${\cal H}$ and
${\cal M}$, and \emph{redistributing} nonlinear terms, ${\cal
N}^{(u)}$, ${\cal N}^{(\theta)}$, ${\cal N}^{(b)}$ operate. The net
effect of the nonlinear terms in the total spectral energy budget
over all wavenumbers is zero according to Equation (\ref{eq:sum}).
Thus, the only source for the total perturbation energy is the
integral over an entire ${\bf k}$-space $\int ({\cal H}+{\cal
M})d^3{\bf k}$ that extracts energy from a vast reservoir of shear
flow and injects it into perturbations. Since the terms ${\cal H}$
and ${\cal M}$, as noted above, are of linear origin, the energy
extraction and perturbation growth mechanisms (the azimuthal MRI)
are essentially linear by nature. The role of nonlinearity is to
continually provide, or regenerate those modes in ${\bf k}$-space
that are able to undergo the transient MRI, drawing on mean flow
energy, and in this way feed the nonlinear state over long times.
This scenario of a sustained state, based on a subtle cooperation
between linear and nonlinear processes, is a keystone of the bypass
concept of turbulence in spectrally stable HD shear flows
\citep{Gebhardt_Grossmann94,Baggett_etal95,Grossmann00,Chapman02}.

\section{Optimization of the box aspect ratio -- linear analysis}

It is well known from numerical simulations of MRI-turbulence that
its dynamics (saturation) generally depends on the aspect ratio
($L_y/L_x$, $L_x/L_z$) of a computational box
\citep[e.g.,][]{Hawley_etal95,Bodo_etal08,Guan_etal09,Johansen_etal09,Shi_etal16}.
In order to understand this dependence and hence appropriately
select the aspect ratio in simulations, in our opinion, one should
take into account as fully as possible the nonmodal growth of the
MRI during intermediate (dynamical) timescales, because it can
ultimately play an important role in the turbulence dynamics
\citep{Squire_Bhattacharjee14}. However, this is often overlooked in
numerical studies. So, in this Section, we identify the aspect
ratios of the preselected boxes that better take into account the
linear transient growth process.

In fluid dynamics, the linear transient growth of perturbations in
shear flows is usually quantified using the formalism of optimal
perturbations
\citep{Farrell_Ioannou96,Schmid_Henningson01,Schmid07}. This
approach has already been successfully applied to (magnetized) disk
flows
\citep{Mukhopadhyay_etal05,Zhuravlev_Razdoburdin14,Squire_Bhattacharjee14,Razdoburdin_Zhuravlev17}.
Such perturbations yield maximum linear nonmodal growth during
finite times and therefore are responsible for most of the energy
extraction from the background flow. So, in this framework, we
quantify the linear nonmodal optimal amplification of the azimuthal
MRI as a function of mode wavenumbers for the same parameters
adopted in the simulations.

In the shearing box, the radial wavenumber of each non-axisymmetric
perturbation mode (shearing wave) changes linearly with time due to
shear, $k_x(t)=k_x(0)+q\Omega k_yt$. The maximum possible
amplification of the total energy ${\cal E}={\cal E}_K+{\cal
E}_{th}+{\cal E}_M$ of a shearing wave, with an initial wavenumber
${\bf k}(0)=(k_x(0),k_y,k_z)$ by a specific (dynamical) time $t_d$
is given by
\begin{equation}\label{eq:gf}
G({\bf k}(t_d))=\max_{\bar{f}(0)}\frac{{\cal E}({\bf k}(t_d))}{{\cal
E}({\bf k}(0))},
\end{equation}
where the maximum is taken over all initial conditions $\bar{f}(0)$
with a given energy ${\cal E}({\bf k}(0))$. The final state at $t_d$
and the corresponding energy ${\cal E}({\bf k}(t_d))$ are found from
the initial state at $t=0$ by integrating the linearized version of
spectral Equations (\ref{eq:App-uxk})-(\ref{eq:App-divbk}) in time
for each shearing wave and finding a propagator matrix connecting
the initial and final states. Then, expression (\ref{eq:gf}) is
usually calculated by means of the singular value decomposition of
the propagator matrix. The square of the largest singular value then
gives the optimal growth factor $G$ for this set of wavenumbers. The
corresponding initial conditions, leading to this highest growth at
$t_d$ are called optimal perturbations. (A reader interested in the
details of these calculations is referred to
\citet{Squire_Bhattacharjee14}, where the formalism of optimal
growth and optimal perturbations in MRI-active disks, which is
adopted here, is described to a greater extent.) Reference time,
during which to calculate the nonmodal growth, is generally
arbitrary. We choose it equal to the characteristic ($e$-folding)
time of the most unstable MRI mode,
$t_d=1/\gamma_{max}=1.33\Omega^{-1}$, where
$\gamma_{max}=0.75\Omega$ is its growth rate
\citep{Balbus03,Ogilvie_Pringle96}, since it is effectively a
dynamical time as well.

\begin{figure}
\includegraphics[width=\columnwidth]{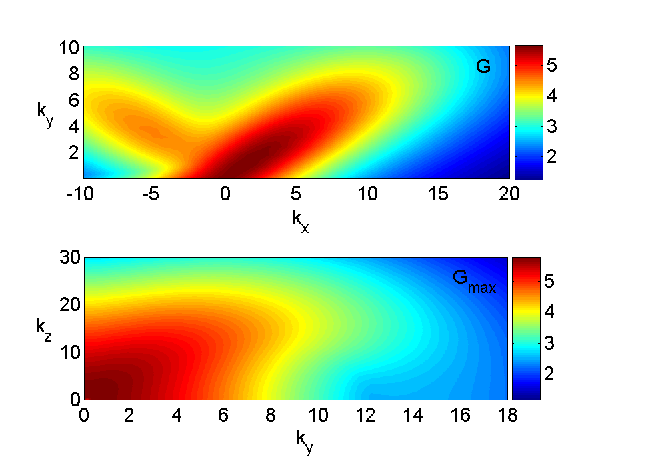}
\caption{(upper panel) Optimal nonmodal growth factor, $G$, in
$(k_x,k_y)$-plane at $t_d=1.33$ and $k_z/2\pi=1$ (which is the same
as $k_z=1$ in new mode number notations used in the next Sections).
(lower panel) Maximized over initial $k_x(0)$ growth factor,
$G_{max}$, as a function of $k_y$ and $k_z$.}
\label{fig:nonmodalgrowth}
\end{figure}

Figure \ref{fig:nonmodalgrowth} shows $G$ in $(k_x,k_y)-$plane at
fixed $k_z$ as well as its value maximized over the initial
$k_x(0)$, $G_{max}=\max_{k_x(0)}G$, represented as a function of
$k_y,k_z$. Because of the $k_x$-drift, the optimal mode with some
initial radial wavenumber $k_x(0)$, at $t_d$ will have the
wavenumber $k_x(t_d)=k_x(0)+qk_yt_d$. In the top panel, $G$ is
represented as a function of this final wavenumber $k_x(t_d)$.
Because of the shear, the typical distribution at fixed $k_z$ is
inclined towards the $k_x$-axis, having larger values on the
$k_x/k_y>0$ side (red region). The most effective nonmodal MRI
amplification occurs at smaller wavenumbers, in the areas marked by
dark red in both $(k_x,k_y)$ and $(k_y,k_z)$-planes in Figure
\ref{fig:nonmodalgrowth}. Thus, the growth of the MRI during the
dynamical time appears to favor smaller $k_z$ \citep[see
also][]{Squire_Bhattacharjee14}, as opposed to the transient growth
of the azimuthal MRI often calculated over times much longer than
the dynamical time, which is the more effective the larger is $k_z$
\citep{Balbus_Hawley92,Papaloizou_Terquem97}. Obviously, the growth
over such long timescales is irrelevant for the nonlinear
(turbulence) dynamics.
\begin{figure}
\includegraphics[width=\columnwidth]{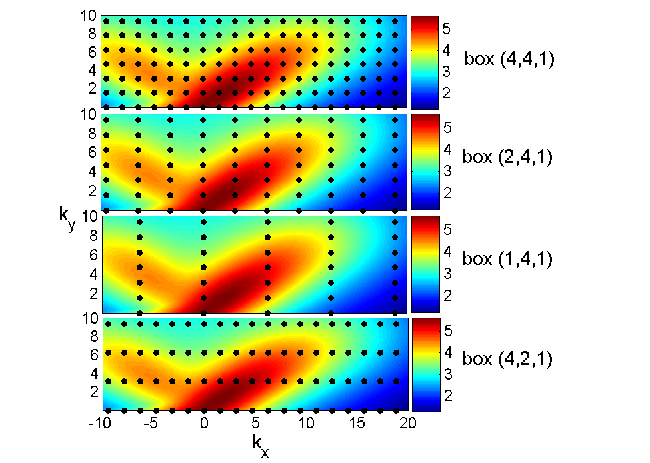}
\caption{Discrete modes (black dots) contained in each simulation
box superimposed on the distribution of $G$ in $(k_x,k_y)$-plane
from the upper panel of Figure 1. From all the selected boxes, the
box $(4,4,1)$ contains most of the effectively amplified modes.}
\label{fig:nonmodalgrowthmodes}
\end{figure}

In the simulation box, however, the wavenumber spectrum is
inherently discrete, with smallest wavenumbers being defined by the
box size $(L_x, L_y, L_z)$ as $k_{i,min}=2\pi/L_i$, while other
wavenumbers being multiples of them. We take $L_z=1$ (i.e., $L_z=H$
in dimensional units) and mainly consider four aspect ratios
$(L_x,L_y,L_z)=(4,4,1),~(2,4,1),~(1,4,1),~(4,2,1)$. Figure
\ref{fig:nonmodalgrowthmodes} shows the modes (black dots) in each
box superimposed on the map of $G$ in $(k_x,k_y)$-plane from Figure
\ref{fig:nonmodalgrowth} for the first vertical harmonics with
$k_{z,min}$, or equivalently $k_z=1$ (in new notations used below).
We see that from among these four boxes, the box $(4,4,1)$ contains
the largest possible number of modes in the area of the effective
transient growth and therefore best accounts for the role of the
nonmodal effects in the energy exchange processes in the case of
turbulence. Of course, further increasing $L_x$ and $L_y$ leads to
larger number of modes in the area of effective growth, however, as
also seen from Figure \ref{fig:nonmodalgrowthmodes}, already for the
box $(4,4,1)$ this area appears to be sufficiently well populated
with modes, i.e., enough resolution (measured in terms of $\Delta
k_i=2\pi/L_i$) is achieved in Fourier space to adequately capture
the nonmodal effects. To ascertain this, we also carried out a
simulation for the box $(8,8,1)$ and found that the ratio of the
number of the active modes (i.e., the number in the growth area) to
the total number of modes in this larger box is almost the same as
for the box $(4,4,1)$. Consequently, these boxes should give
qualitatively similar dynamical pictures in Fourier space. For this
reason, below we choose the box $(4,4,1)$ as fiducial and present
only some results for other boxes for comparison at the end of
Section 5.

\section{Simulations and general characteristics}

The main Equations (\ref{eq:App-ux})-(\ref{eq:App-divperb}) are
solved using the pseudo-spectral code SNOOPY
\citep{Lesur_Longaretti07}. It is a general-purpose code, solving HD
and MHD equations, including shear, rotation, stratification and
several other physical effects in the shearing box model. Fourier
transforms are computed using the FFTW library, taking also into
account the drift of radial wavenumber $k_x(t)$ in ${\bf k}$-space
due to shear in order to comply with the shearing-periodic boundary
conditions. Nonlinear terms are computed using a pseudo-spectral
algorithm \citep{Canuto_etal88}, and antialiasing is enforced using
the $2/3$-rule. Time integration is done by a standard explicit
third-order Runge-Kutta scheme, except for viscous and resistive
terms, which are integrated using an implicit scheme. The code has
been extensively used in the shearing box studies of disk turbulence
\citep[e.g.,][]{Lesur_Ogilvie10,Lesur_Longaretti11,Herault_etal11,Meheut_etal15,Murphy_Pessah15,Riols_etal17}.

\begin{table*}[t]
\caption{Simulation characteristics: box size, number of grid
points, volume- and time-averaged values (denoted by double
brackets) of the perturbed kinetic, $E_K$, magnetic, $E_M$, and
thermal, $E_{th}$, energy densities as well as the rms values of the
magnetic field components and the Reynolds, $u_xu_y$, and Maxwell,
$-b_xb_y$, stresses in the fully developed turbulence.}
\begin{ruledtabular}
\begin{tabular}{ccccccccccccc}
$(L_x,L_y,L_z)$ & $(N_x, N_y, N_z)$ & $\langle\langle E_K
\rangle\rangle$ & $\langle\langle E_M \rangle\rangle$ &
$\langle\langle E_{th} \rangle\rangle$ & $\langle\langle b_x^2
\rangle\rangle^{1/2}$ & $\langle\langle b_y^2 \rangle\rangle^{1/2}$
& $\langle\langle b_z^2 \rangle\rangle^{1/2}$ &
$\langle\langle u_xu_y \rangle\rangle$ & $\langle\langle -b_xb_y \rangle\rangle$ \\
\hline
$(8,8,1)$ & $(512,512,64)$ & 0.0173 & 0.0422 & 0.0022 & 0.101 & 0.266 & 0.06 &  0.0037 & 0.0198\\
$(4,4,1)$ & $(256,256,64)$ & 0.0125 & 0.03   & 0.0019 & 0.086 & 0.224 & 0.05 &  0.0028 & 0.0146\\
$(2,4,1)$ & $(128,256,64)$ & 0.0116 & 0.0298 & 0.0019 & 0.085 & 0.223 & 0.05 &  0.0028 & 0.0144\\
$(1,4,1)$ & $(64,256,64)$  & 0.0111 & 0.0295 & 0.0018 & 0.085 & 0.222 & 0.05 &  0.0027 & 0.0143\\
$(4,2,1)$ & $(256,128,64)$ & 0.0056 & 0.012  & 0.0011 & 0.053 & 0.14  & 0.03 &  0.0013 & 0.0059\\
\end{tabular}
\end{ruledtabular}
\end{table*}

We carry out simulations for boxes with different radial and
azimuthal sizes $(L_x,L_y,L_z)=(4,4,1)$, $(2,4,1)$, $(1,4,1)$,
$(4,2,1)$, $(8,8,1)$ and resolution of $64$ grid points per scale
height $H=1$ (Table 1). The numerical resolution adopted ensures
that the dissipation wavenumber, $k_D$, is smaller than the maximum
wavenumber, $k_{i,max}=2\pi N_i/3L_i$, in the box (taking into
account the 2/3-rule). The initial conditions consist of small
amplitude random noise perturbations of velocity on top of the
Keplerian shear flow. A subsequent evolution is followed up to
$t_f=630$ (about 100 orbits). The wavenumbers $k_x,k_y,k_z$ are
normalized, respectively, by the grid cell sizes of Fourier space,
$\Delta k_x=2\pi/L_x, \Delta k_y=2\pi/L_y$ and $\Delta
k_z=2\pi/L_z$, that is, $(k_x/\Delta k_x, k_y/\Delta k_y, k_z/\Delta
k_z)\rightarrow (k_x,k_y,k_z)$. As a result, the normalized
azimuthal and vertical wavenumbers are integers $k_y,k_z=0,\pm 1,
\pm 2, ...$, while $k_x$, although changes with time due to drift,
is integer at discrete moments $t_n=nL_y/(q|k_y|L_x)$, where $n$ is
a positive integer.

\begin{figure}
\includegraphics[width=\columnwidth]{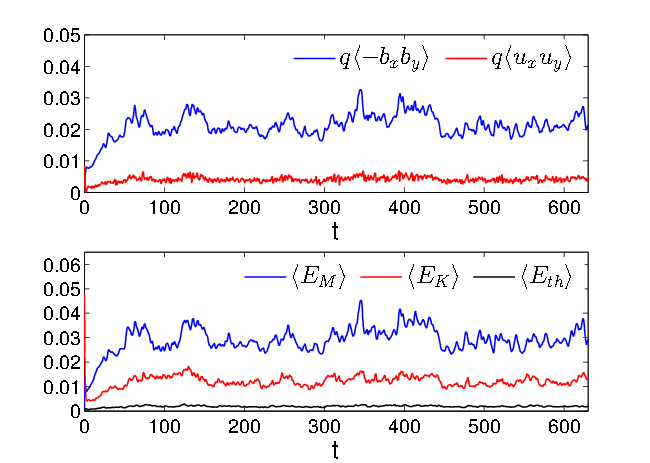}
\includegraphics[width=\columnwidth]{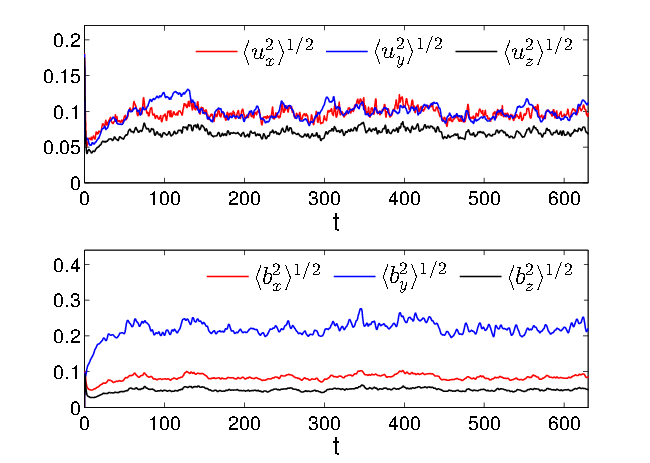}
\caption{Evolution of volume-averaged Reynolds and Maxwell stresses
(top row), kinetic, thermal and magnetic energy densities (second
row), rms of velocity (third row) and magnetic field (bottom row)
components for the fiducial box $(4,4,1)$. Turbulence sets in after
several orbits, with the magnetic energy dominating kinetic and
thermal energies, and the Maxwell stress the Reynolds one. The
azimuthal component of the turbulent magnetic field is larger than
the other two ones due to the shear. It is also about twice larger
than the mean field $B_{0y}=0.1$ (Table 1).}\label{fig:time
evolution}
\end{figure}
\begin{figure*}
\includegraphics[width=0.32\textwidth]{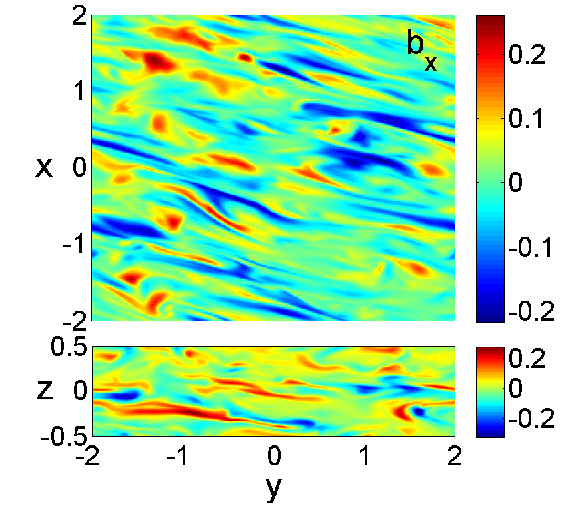}
\includegraphics[width=0.32\textwidth]{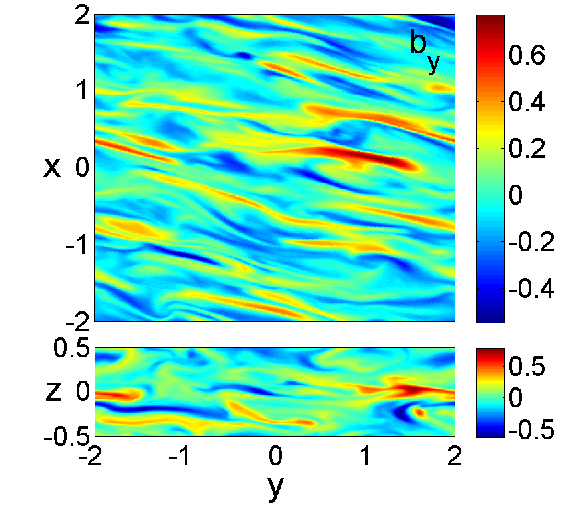}
\includegraphics[width=0.32\textwidth]{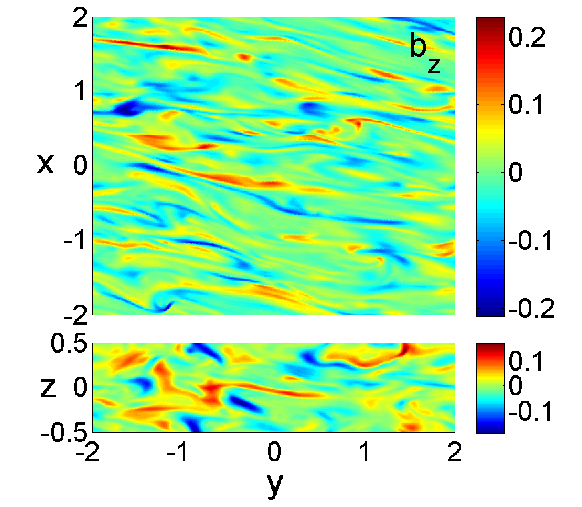}
\caption{Typical structure of the magnetic field in the fully
developed quasi-steady turbulent state at $t=550$ for the box
$(4,4,1)$. Shown are the sections in $(y,x)$ and
$(y,z)$-planes.}\label{fig:physical space}
\end{figure*}

In all the boxes, initially imposed small perturbations start to
grow as a result of the nonmodal MRI amplification of the
constituent Fourier modes. Then, after several orbits, the
perturbation amplitude becomes high enough, reaching the nonlinear
regime and eventually the flow settles down into a quasi-steady
sustained MHD turbulence. Figure \ref{fig:time evolution} shows the
time-development of the volume-averaged perturbed kinetic, $\langle
E_K\rangle$, thermal, $\langle E_{th}\rangle$, and magnetic,
$\langle E_M \rangle$, energy densities as well as the Reynolds,
$\langle u_xu_y \rangle$, and Maxwell $-\langle b_xb_y \rangle$
stresses for the fiducial box $(4,4,1)$. For completeness, in this
figure, we also show the evolution of the rms values of the
turbulent velocity and magnetic field components. The magnetic
energy dominates the kinetic and thermal ones, with the latter being
much smaller than the former two, while the Maxwell stress is about
5 times larger than the Reynolds one. This indicates that the
magnetic field perturbations are primarily responsible for energy
extraction from the mean flow by the Maxwell stress, transporting
angular momentum outward and sustaining turbulence. In contrast to
the 2D plane case \citep{Mamatsashvili_etal14}, the Reynolds stress
in this 3D case is positive and also contributes to the outward
transport. The temporal behavior of the volume-averaged kinetic and
magnetic energy densities and stresses is consistent with analogous
studies of MRI-turbulence in disks with a net azimuthal field
\citep{Hawley_etal95,Guan_etal09,Simon_Hawley09,Meheut_etal15}. For
all the models, the time- and volume-averaged quantities over the
whole quasi-steady state, between $t=100$ and the end of the run at
$t_f$, are listed in Table 1. For the fiducial model, the ratios of
the magnetic energy to kinetic and thermal ones are $\langle\langle
E_M \rangle\rangle/\langle\langle E_K\rangle\rangle = 2.4$ and
$\langle\langle E_M \rangle\rangle/\langle\langle
E_{th}\rangle\rangle = 15.8$, respectively, and the ratio of the
Maxwell stress to the Reynolds stress is $\langle\langle -b_xb_y
\rangle\rangle/\langle\langle u_xu_y\rangle\rangle = 5.21$. For
other boxes, similar ratios hold between magnetic and hydrodynamic
quantities, as can be read off from Table 1, with the magnetic
energy and stresses being always dominant over respective
hydrodynamic ones. Interestingly, for all boxes in the quasi-steady
turbulent state, $\langle E_M \rangle$ and $\langle-b_xb_y\rangle$
closely follow each other at all times, with the ratio being nearly
constant, $\langle E_M \rangle/\langle-b_xb_y\rangle \approx 2$
\citep[see also][]{Hawley_etal95,Guan_etal09}. From Table 1, we can
also see how the level (intensity) of the turbulence varies with the
radial and azimuthal sizes of the boxes. For fixed $L_y=4$, the
saturated values of the energies and stresses increase with $L_x$,
but only very little, so they can be considered as nearly unchanged,
especially after $L_x>1$. By contrast, at fixed $L_x=4$, these
quantities are more sensitive to the azimuthal size $L_y$,
increasing more than twice with the increase of the latter from
$L_y=2$ to $L_y=4$. However, after $L_y=4$ the increase of the
turbulence strength with the box size is slower, as evident from the
box $(8,8,1)$. This type of dependence of the azimuthal
MRI-turbulence characteristics on the horizontal sizes of the
simulation box is consistent with that of \cite{Guan_etal09}.

The structure of the turbulent magnetic field in the fully developed
quasi-steady turbulence in physical space is presented in Figure
\ref{fig:physical space}. It is chaotic and stretched along the
$y$-axis due to the shear, with $b_y$ achieving higher values than
$b_x$ and $b_z$. At this moment, the rms values of these components
are, $\langle b_x^2\rangle^{1/2}=0.079$, $\langle
b_z^2\rangle^{1/2}=0.044$, while $\langle b_y^2\rangle^{1/2}=0.2$
and is twice larger than the background field $B_{0y}=0.1$. These
values, as expected, are consistent with the bottom panel of Figure
\ref{fig:time evolution}. So, the turbulent field satisfies $\langle
b_z^2\rangle^{1/2}< \langle b_x^2\rangle^{1/2} < B_{0y}< \langle
b_y^2\rangle^{1/2}$, which in fact holds throughout the evolution
for all models (Table 1).

\subsection{Analysis in Fourier space -- an overview}

A deeper insight into the nature of the turbulence driven by the
azimuthal MRI can be gained by performing analysis in Fourier space.
So, following
\citet{Horton_etal10,Mamatsashvili_etal14,Mamatsashvili_etal16}, we
examine in detail the specific spectra and sustaining dynamics of
the quasi-steady turbulent state by explicitly calculating and
visualizing the individual linear and nonlinear terms in spectral
Equations (\ref{eq:uxk})-(\ref{eq:bzk}), which have been classified
and described in Section 2, based on the simulation data. These
equations govern the evolution of the quadratic forms (squared
amplitudes) of Fourier transforms of velocity, thermal and magnetic
field perturbations and are more informative than Equations
(\ref{eq:ekspec}) and (\ref{eq:emspec}) for spectral kinetic and
magnetic energies. In the latter equations a lot of essential
information is averaged and lost. Therefore, energy equations alone
are insufficient for understanding intertwined linear and nonlinear
processes that underlie the sustaining dynamics of the turbulence.
For this reason, we rely largely on Equations
(\ref{eq:uxk})-(\ref{eq:bzk}), enabling us to form a complete
picture of the turbulence dynamics. So, we divide our analysis in
Fourier space into several steps:
\begin{enumerate}[I.]
\item
Three-dimensionality, of course, complicates the analysis.
Therefore, initially, we find out which vertical wavenumbers are
important by integrating the spectral energies and stresses in
$(k_x,k_y)$-plane (Figure \ref{fig:integrated in plane}). As will be
evident from such analysis, mostly the lower vertical harmonics,
$|k_z|=0,1,2$, (i.e., with vertical scales comparable to the box
size $L_z$) engage in the turbulence maintaining process.
\item
Next, concentrating on these modes with lower vertical wavenumber,
we present the spectral magnetic energy in $(k_x,k_y)$-plane (Figure
\ref{fig:spectral energy}) and identify the energy-carrying modes in
this plane (Figure \ref{fig:modes}). From these modes, we delineate
a narrower set of dynamically important active ones, which are
central in the sustenance process. Based on this, we identify a
region in Fourier space -- \emph{the vital area} -- where the basic
linear and nonlinear processes for these modes operate. Despite a
limited extent of the vital area, the number of the dynamically
important modes within it appears to be quite large and they are
distributed anisotropically in Fourier space.
\item
Integrating in $(k_x,k_y)$-plane the quadratic forms of the spectral
velocity and magnetic field components (${|\bar{u}_i|^2}$ and
${|\bar{b}_i|^2}$) as well as the corresponding linear and nonlinear
terms on the rhs of Equations \ref{eq:uxk}-\ref{eq:bzk}), we obtain
a first idea about the importance of each of them in the dynamics as
a function of $k_z$ (Figure \ref{fig:integrated in plane-ub}). Note
that the action of the linear drift terms vanishes after the
integration. Nevertheless, the universality and importance of the
linear drift is obvious in any case.
\item
Finally, we analyze the interplay of these processes/terms that
determines the turbulence dynamics (Figures
\ref{fig:spectral-bx}-\ref{fig:spectral-uz}). As a result, we
construct the turbulence sustaining picture/mechanism by revealing
the transverse nature of the nonlinear processes -- the nonlinear
transverse cascade -- and demonstrating its key role in the
sustenance.
\end{enumerate}

\citet{Fromang_Papaloizou07,
Simon_etal09,Davis_etal10,Lesur_Longaretti11} took a similar
approach of representing the MHD equations in Fourier space and
analyzing individual linear and nonlinear (transfer) terms in the
dynamics of MRI-turbulence. They derived evolution equations for the
kinetic and magnetic energy spectra, which are similar to our
Equations (\ref{eq:ekspec})-(\ref{eq:emspec}) except for notation
and mean field direction. As mentioned above, we do not make the
shell-averages in Fourier space, as done in these studies, that
completely wipes out spectral anisotropy due to the shear crucial to
the turbulence dynamics.

Since our analysis primarily focuses on the spectral aspect of the
dynamics, the SNOOPY code, being of spectral type, is particularly
convenient for this purpose, as it allows us to directly extract
Fourier transforms. From now on we consider the evolution after the
quasi-steady turbulence has set in, so all the spectral
quantities/terms in Equations (\ref{eq:uxk})-(\ref{eq:bzk}) are
averaged in time over an entire saturated turbulent state between
$t=200$ and the end of the run. Below we concentrate on the fiducial
box $(4,4,1)$. Comparison of the spectral dynamics in other boxes
and the effects of the box aspect ratio will be presented in the
next Section.

\begin{figure}
\includegraphics[width=\columnwidth]{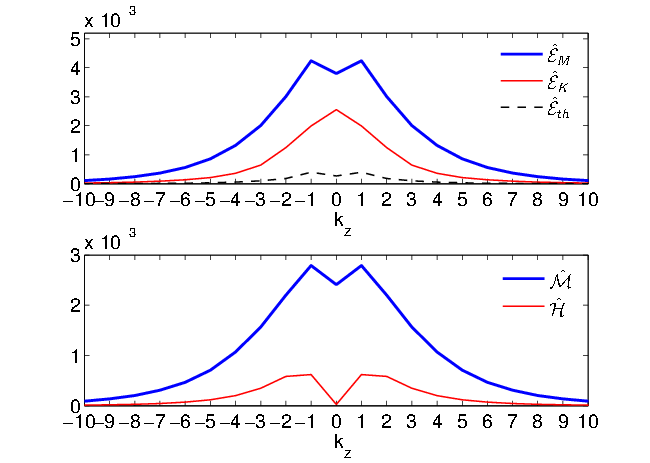}
\caption{Integrated in $(k_x,k_y)$-plane time-averaged kinetic,
$\hat{\cal E}_K$, magnetic, $\hat{\cal E}_M$ and thermal $\hat{\cal
E}_{th}$ energy spectra (upper panel) as well as Reynolds,
$\hat{\cal H}$, and Maxwell, $\hat{\cal M}$, stresses (lower panel)
as a function of $k_z$ for the box $(4,4,1)$.}\label{fig:integrated
in plane}
\end{figure}
\begin{figure*}
\includegraphics[width=0.32\textwidth]{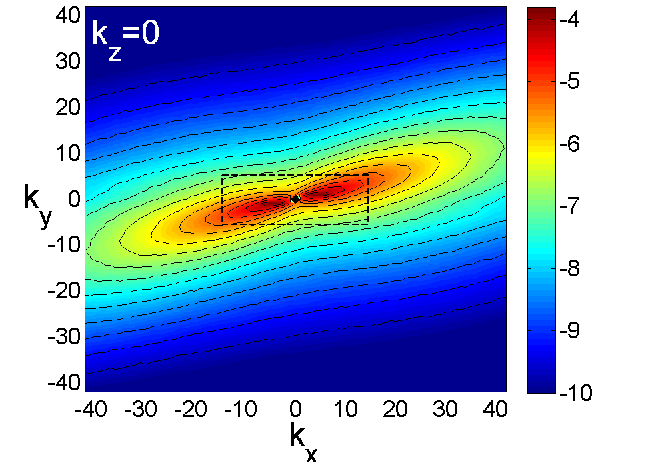}
\includegraphics[width=0.32\textwidth]{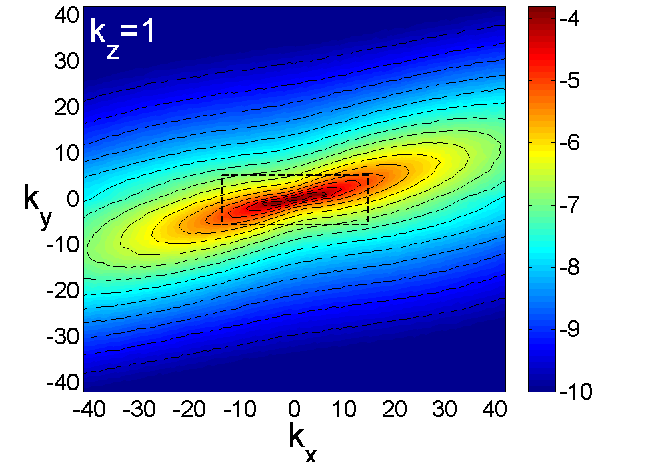}
\includegraphics[width=0.32\textwidth]{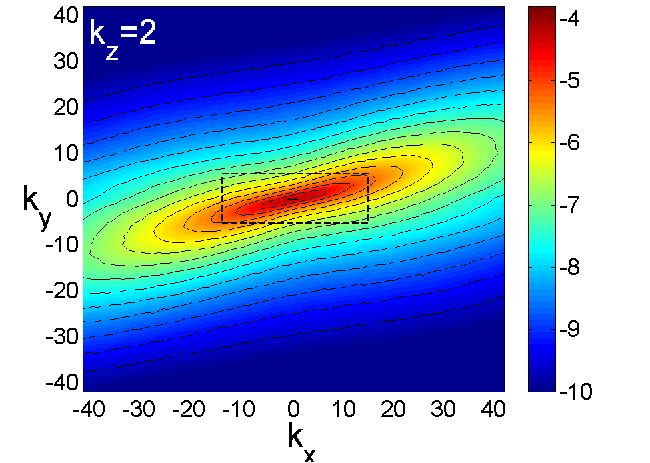}
\caption{Logarithm of the spectral magnetic energy, $log_{10}{\cal
E}_M$, in $(k_x,k_y)$-plane at $k_z=0,1,2$ for the box $(4,4,1)$.
The spectra is strongly anisotropic due to the shear, having larger
power on the $k_x/k_y>0$ side at a given $k_y$. Dashed rectangles
delineate the vital area of turbulence, where the sustaining process
is concentrated (see Figures
\ref{fig:spectral-bx}-\ref{fig:spectral-uz}).}\label{fig:spectral
energy}
\end{figure*}
\begin{figure}
\includegraphics[width=\columnwidth]{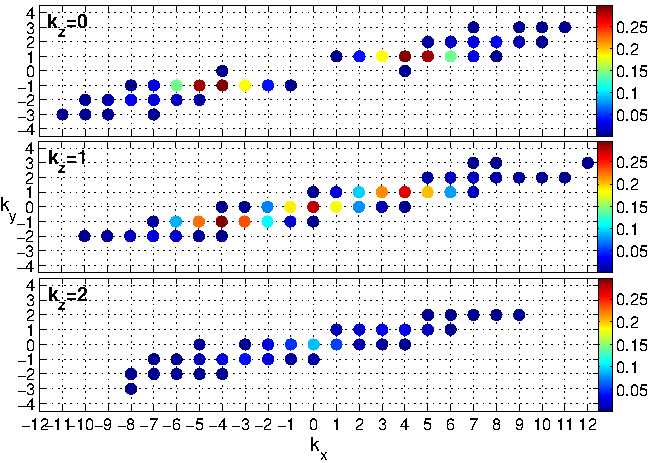}
\caption{Distribution of the active modes in {\textbf{k}}-space at
$k_z=0,1,2$ for the box $(4,4,1)$. The color dots represent the
modes whose magnetic energy, ${\cal E}_M$, grows more than $50 \%$
of the maximum spectral magnetic energy, ${\cal E}_{M,max}$, and the
colors indicate the fraction of time each mode contains this higher
energy during the quasi-steady state until the end of the
simulation.}\label{fig:modes}
\end{figure}

\subsection{Energy spectra, active modes and the vital area}

Figure \ref{fig:integrated in plane} shows the time-averaged spectra
of the kinetic, magnetic and thermal energies as well as the
Reynolds and Maxwell stresses integrated in $(k_x,k_y)$-plane,
$\hat{\cal E}_{K,M,th}(k_z)=\int {\cal E}_{K,M,th}dk_xdk_y$ and
$(\hat{\cal H}(k_z),\hat{\cal M}(k_z))=\int ({\cal H},{\cal
M})dk_xdk_y$ as a function of $k_z$. The magnetic energy is the
largest and the thermal energy the smallest, while the Maxwell
stress dominates the Reynolds one, at all $k_z$. All the three
energy spectra and stresses reach a maximum at small $|k_z|$ -- the
magnetic and thermal energies as well as the stresses at $|k_z|= 1$,
while the kinetic energy at $k_z=0$ -- and rapidly decrease with
increasing $|k_z|$. As a result, in particular, the magnetic energy
injection into turbulence due to the Maxwell stress takes place
mostly at small $k_z$, which is consistent with our linear optimal
growth calculations (Section 3) and also with
\citet{Squire_Bhattacharjee14}, but is in contrast to the accepted
view that the purely azimuthal field MRI is stronger at high $k_z$
\citep{Balbus_Hawley92,Hawley_etal95}. The main reason for this
difference, as mentioned above, is that the latter is usually
calculated over much longer times (spanning from tens to hundred
dynamical times), following the evolution of the shearing waves from
initial tightly leading to final tightly trailing orientation,
whereas the optimal growth is usually calculated over a finite
(dynamical) time, which seems more appropriate in the case of
turbulence. Thus, the large-scale modes with the first few $k_z$
contain most of the energy and hence play a dynamically important
role.

\begin{figure*}
\centering
\includegraphics[width=\columnwidth]{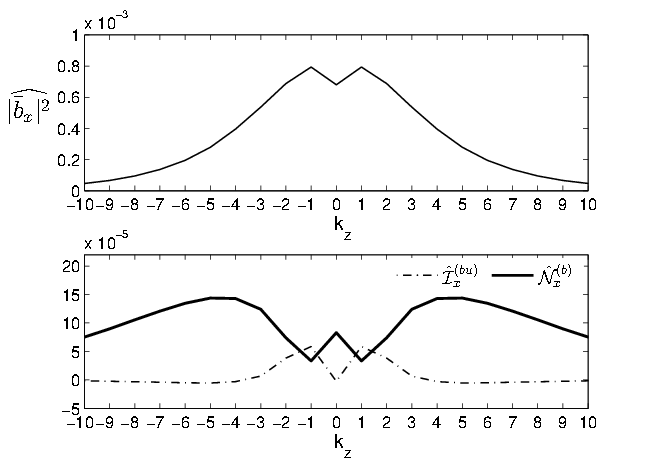}
\includegraphics[width=\columnwidth]{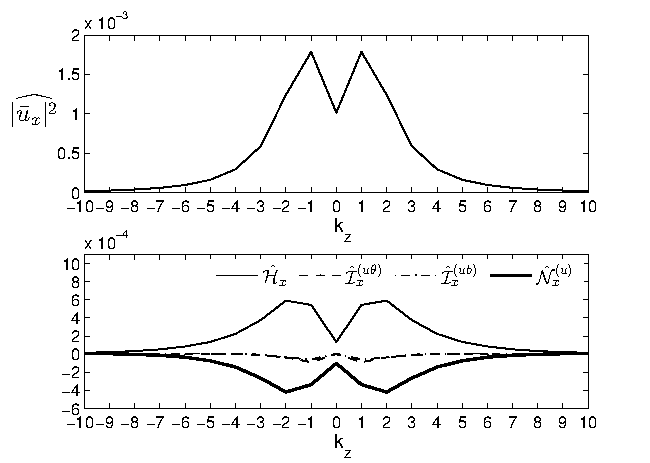}
\includegraphics[width=\columnwidth]{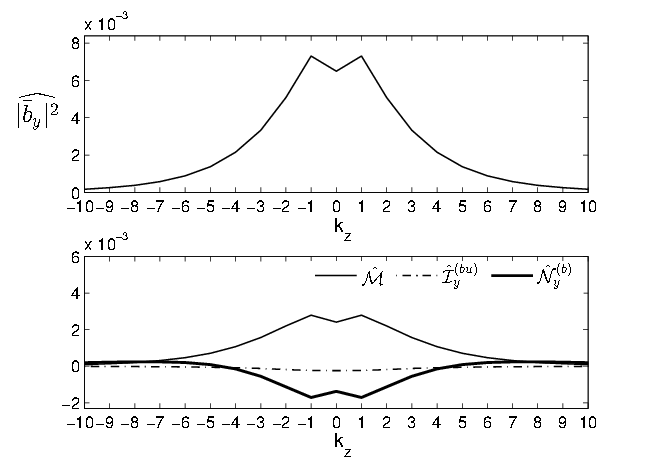}
\includegraphics[width=\columnwidth]{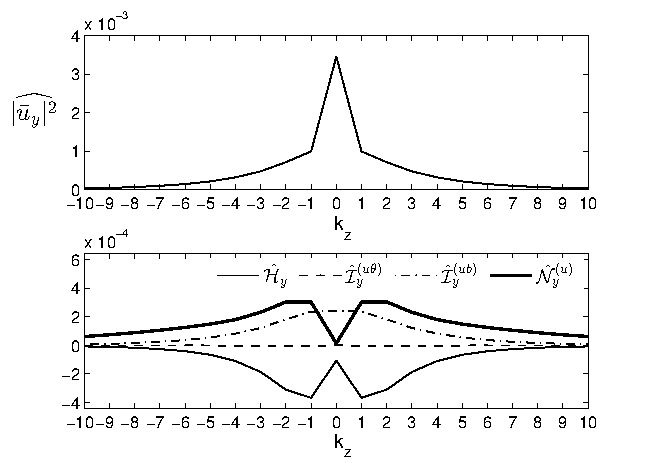}
\includegraphics[width=\columnwidth]{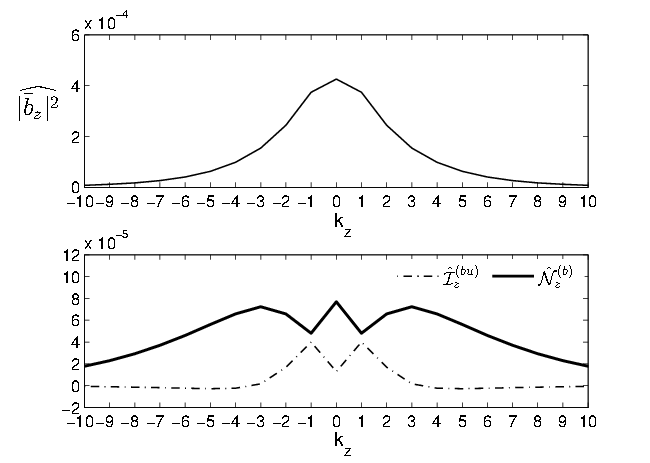}
\includegraphics[width=\columnwidth]{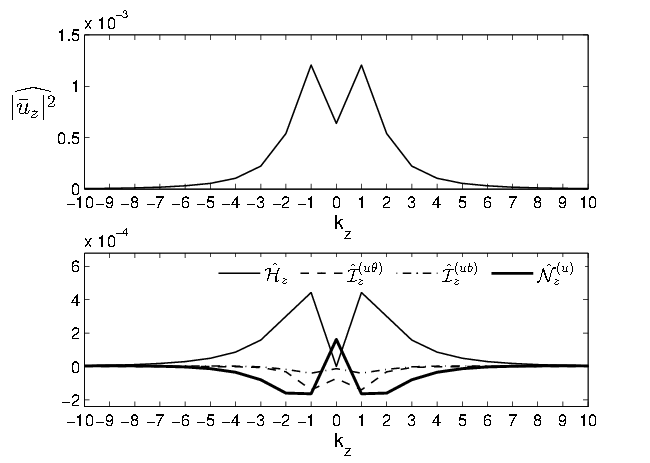}
\caption{Integrated in $(k_x,k_y)$-plane the quadratic forms of the
spectral velocity and magnetic field components together with the
corresponding linear and nonlinear terms from corresponding
Equations (\ref{eq:uxk})-(\ref{eq:bzk}) as a function of
$k_z$.}\label{fig:integrated in plane-ub}
\end{figure*}

To have a fuller picture of the energy spectra, in Figure
\ref{fig:spectral energy} we present sections of ${\cal E}_M$ in
$(k_x,k_y)$-plane again at first three vertical wavenumbers
$k_z=0,1,2$, for which it is higher (see Figure \ref{fig:integrated
in plane}). The spectrum is highly anisotropic due to the shear with
the same elliptical shape and inclination towards the $k_x-$axis
irrespective of $k_z$. This indicates that modes with $k_x/k_y>0$
have more energy than those with $k_x/k_y<0$ at fixed $k_y$. The
kinetic energy spectrum shares similar properties and is not shown
here. A similar anisotropic spectrum was already reported in the
shearing-box simulations of MRI-turbulence with a nonzero net
vertical field
\citep{Hawley_etal95,Lesur_Longaretti11,Murphy_Pessah15}. This
energy spectrum, which clearly differs from a typical turbulent
spectrum in the classical case of forced MHD turbulence without
shear \citep{Biskamp03}, arises as a consequence of a specific
anisotropy of the linear and nonlinear terms of Equations
(\ref{eq:uxk})-(\ref{eq:bzk}) in ${\bf k}-$space. These new features
are not common to shearless MHD turbulence and hence it is not
surprising that Kolmogorov or IK theory cannot adequately describe
shear flow turbulence.

Having described the energy spectrum, we now look at how
energy-carrying modes, most actively participating in the dynamics,
are distributed in $(k_x,k_y)$-plane. We refer to modes whose
magnetic energy reaches values higher than 50\% of the maximum
spectral magnetic energy as active modes. Although this definition
is somewhat arbitrary, it gives an idea on where the dynamically
important modes are located in Fourier space. Figure \ref{fig:modes}
shows these modes in $(k_x,k_y)$-plane at $k_z=0,1,2$ with color
dots. They are obtained by following the evolution of all the modes
in the box during an entire quasi-steady state and selecting those
modes whose magnetic energy becomes higher than the above threshold.
The color of each mode indicates the fraction of time, from the
onset of the quasi-steady state till the end of the simulation,
during which it contains this higher energy. We have also checked
that Figure \ref{fig:modes} is not qualitatively affected upon
changing the 50\% threshold to either 20\% or 70\%. Like the energy
spectrum, the active modes with different duration of ``activity''
are distributed quite anisotropically in $(k_x,k_y)$-plane,
occupying a broader range of radial wavenumbers $|k_x|\lesssim 12$
than that of azimuthal ones $|k_y|\lesssim 3$. This main,
energy-containing area in {\bf k}-space represents \emph{the vital
area} of turbulence. Essentially, the active modes in the vital area
take part in the sustaining dynamics of turbulence. The other modes
with larger wavenumbers lie outside the vital area and always have
energies and stresses less than 50\% of the maximum value,
therefore, not playing as much a role in the energy-exchange process
between the background flow and turbulence. Note that the total
number of the active modes (color dots) in Figure \ref{fig:modes} is
equal to $114$, implying that the dynamics of the MRI-turbulence,
strictly speaking, cannot be reduced to low-order models of the
sustaining processes, involving only a small number of active modes
\citep[e.g.,][]{Herault_etal11,Riols_etal17}.

\subsection{Vertical spectra of the dynamical terms}

Having identified the vital area, we now examine the significance of
each of the linear and nonlinear terms in this area first along the
vertical $k_z$-direction in Fourier space. For this purpose, we
integrate in $(k_x,k_y)$-plane the quadratic forms of the spectral
velocity and magnetic field components as well as the rhs terms of
Equations (\ref{eq:uxk})-(\ref{eq:uzk}) and
(\ref{eq:bxk})-(\ref{eq:bzk}), as we have done for the spectral
energies and stresses above. We do not apply this procedure to the
linear drift term (which vanishes after such integration) and
dissipation terms, as their action is well known. The results are
presented in Figure \ref{fig:integrated in plane-ub} (the spectral
quantities integrated in $(k_x,k_y)$-plane are all denoted by hats),
which shows that:
\begin{enumerate}[$\bullet$]
\item
The dynamics of $\widehat{|\bar{b}_x|^2}$ is governed by $\hat{\cal
I}_x^{(bu)}$ and $\hat{\cal N}^{(b)}_x$, which are both positive and
therefore act as a source for the radial field at all $k_z$.
\item
The dynamics of $\widehat{|\bar{b}_y|^2}$ is governed by $\hat{\cal
M}$ and $\hat{\cal N}^{(b)}_y$, the action of $\hat{\cal
I}_y^{(bu)}$ is negligible compared with these terms. The effect of
$\hat{\cal M}$ is positive for all $k_z$, reaching  a maximum, as we
have seen before, at $|k_z|=1$. This implies that the energy
injection into turbulence from the background flow due to the MRI
occurs over a range of length scales, preventing the development of
the proper inertial range in the classical sense \citep[see
also][]{Lesur_Longaretti11}. On the other hand, $\hat{\cal
N}^{(b)}_y$ is negative and hence acts as a sink for low/active
$k_z$, but positive at large $|k_z|$. So, the nonlinear term
transfers the azimuthal field component from these wavenumbers to
large $|k_z|$ as well as (which is more important) to other
components.
\item
The dynamics of $\widehat{|\bar{b}_z|^2}$ is governed by $\hat{\cal
I}_z^{(bu)}$ and $\hat{\cal N}^{(b)}_z$, which are both positive,
with the latter being larger than the former at all $k_z$. Note that
$\widehat{|\bar{b}_z|^2}$ is smaller compared to the other two
components, while $\widehat{|\bar{b}_y|^2}$ is the largest.
\item
The dynamics of $\widehat{|\bar{u}_x|^2}$ is governed by $\hat{\cal
H}_x$ and $\hat{\cal N}^{(u)}_x$, the action of the exchange terms,
$\hat{\cal I}_x^{(u\theta)}$ and $\hat{\cal I}_x^{(ub)}$, are
negligible compared to these terms. The effect of $\hat{\cal H}_x$
is positive for all $k_z$, acting as the only source for
$\bar{u}_x$. By contrast, $\hat{\cal N}^{(u)}_x$ is negative (sink),
opposing $\hat{\cal H}_x$, with a similar dependence of its absolute
value on $k_z$. So, the nonlinear term transfers the radial velocity
to other components.
\item
The dynamics of $\widehat{|\bar{u}_y|^2}$ is governed by $\hat{\cal
H}_y$, $\hat{\cal I}_y^{(ub)}$ and $\hat{\cal N}^{(u)}_y$, the
action of $\hat{\cal I}_y^{(u\theta)}$ is negligible. The effects of
$\hat{\cal N}^{(u)}_y$ and $\hat{\cal I}_y^{(ub)}$ are positive for
all $k_z$, while $\hat{\cal H}_y$ is negative. Special attention
deserves the sharp peak of $\widehat{|\bar{u}_y|^2}$ at $k_z=0$.
This peak is related to the formation of the zonal flow with
$|k_x|=1$ and $k_y=0$ in the MRI-turbulence \citep{Johansen_etal09},
which will be analyzed below.
\item
The dynamics of $\widehat{|\bar{u}_z|^2}$ is governed by $\hat{\cal
H}_z$, $\hat{\cal I}_z^{(u\theta)}$ and $\hat{\cal N}^{(u)}_z$, the
action of $\hat{\cal I}_z^{(ub)}$ is negligible.
$\widehat{|\bar{u}_z|^2}$ is the only term that explicitly depends
on the thermal processes. Note also that $\hat{\cal N}^{(u)}_z$ is
negative at $|k_z|\geq 1$, but becomes positive at $k_z=0$, implying
inverse transfer towards small $k_z$. We do not go into the details
of this dependence, as $\widehat{|\bar{u}_z|^2}$ is anyway smaller
compared to the other components. Besides, the thermal processes do
not play a major role in the overall dynamics, since their energy is
much smaller than the magnetic end kinetic energies (see also Figure
\ref{fig:time evolution}).
\end{enumerate}
It is seen from Figure \ref{fig:integrated in plane-ub} that all the
dynamical terms primarily operate at small vertical wavenumbers
$|k_z|=0,1,2$. Some of them ($\hat{\cal N}_x^{(b)}$ and $\hat{\cal
N}_z^{(b)}$) may extend up to $|k_z|=3-6$, but eventually decay at
large $|k_z|$. Similarly, the spectra of the velocity and magnetic
field have relatively large values also at small $|k_z|$. So,
$|k_z|=2$ can be viewed as an upper vertical boundary of the vital
area in Fourier space.

\begin{figure*}[t!]
\centering
\includegraphics[width=0.32\textwidth, height=0.2\textwidth]{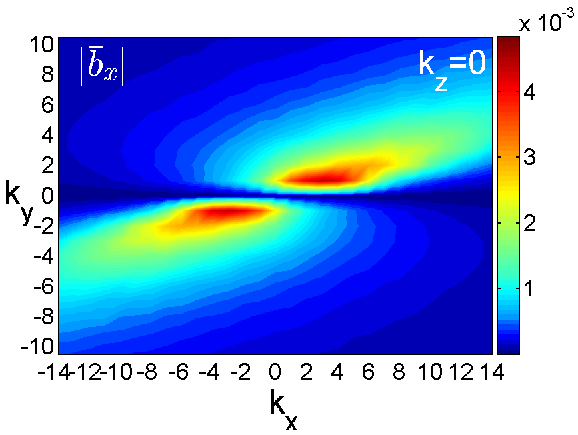}
\includegraphics[width=0.32\textwidth, height=0.2\textwidth]{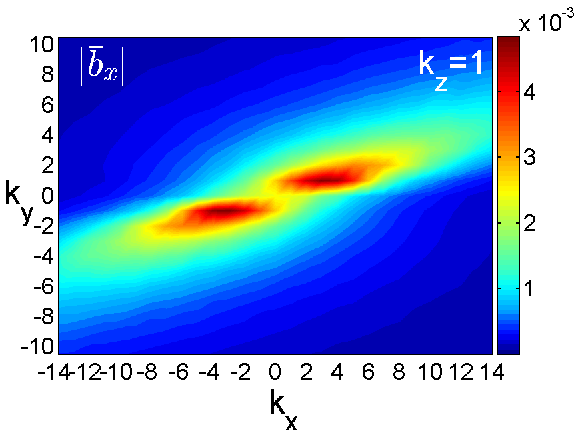}
\includegraphics[width=0.32\textwidth, height=0.2\textwidth]{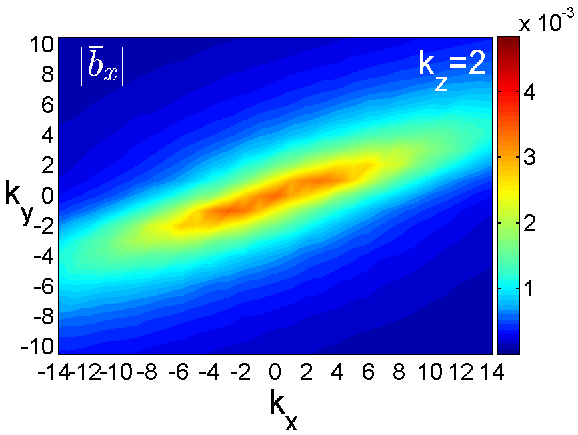}
\includegraphics[width=0.32\textwidth, height=0.2\textwidth]{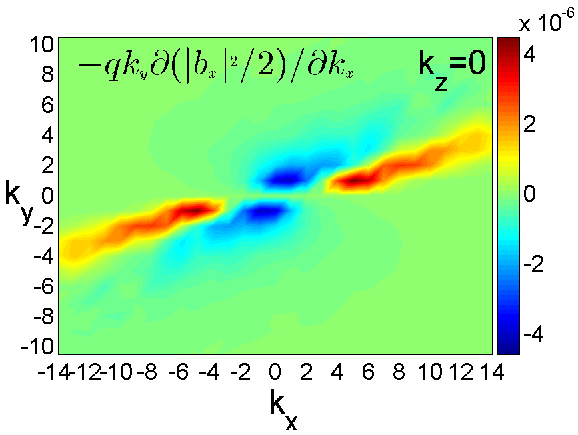}
\includegraphics[width=0.32\textwidth, height=0.2\textwidth]{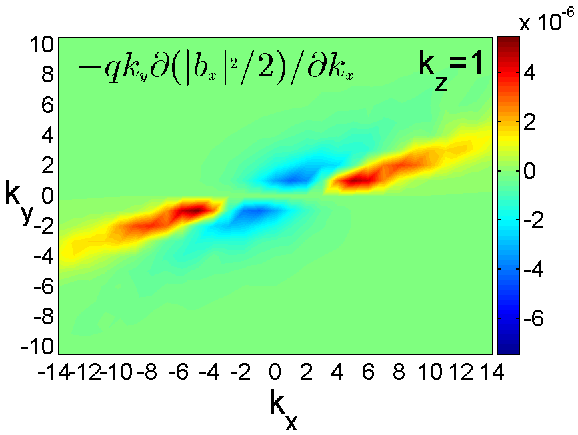}
\includegraphics[width=0.32\textwidth, height=0.2\textwidth]{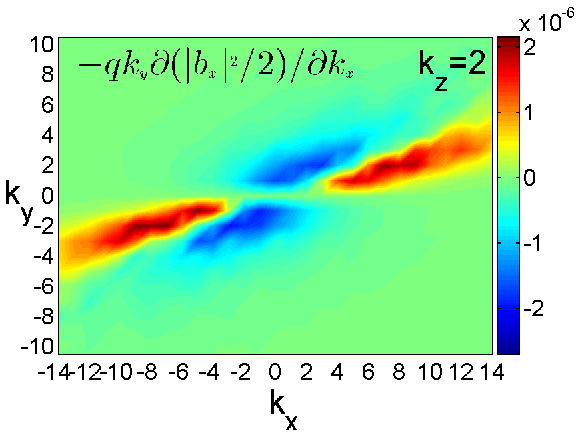}
\includegraphics[width=0.32\textwidth, height=0.2\textwidth]{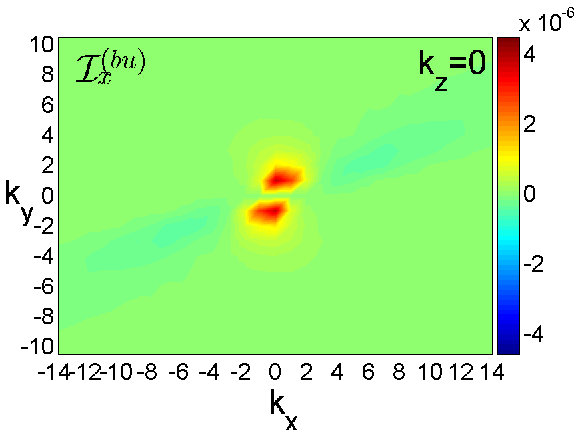}
\includegraphics[width=0.32\textwidth, height=0.2\textwidth]{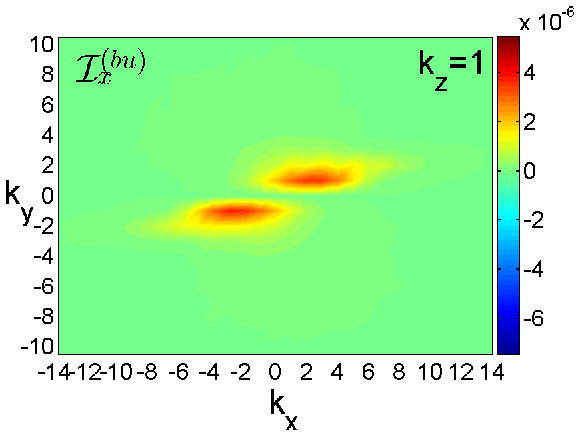}
\includegraphics[width=0.32\textwidth, height=0.2\textwidth]{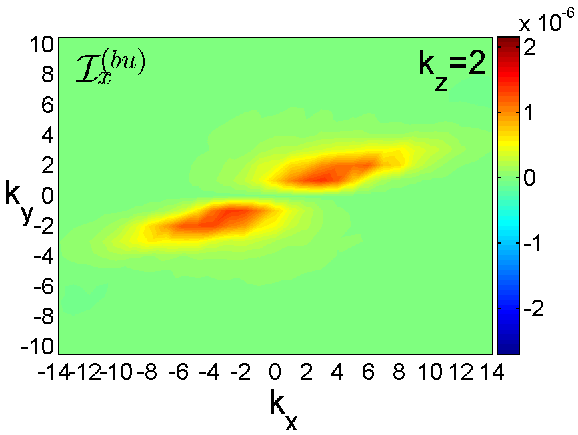}
\includegraphics[width=0.32\textwidth, height=0.2\textwidth]{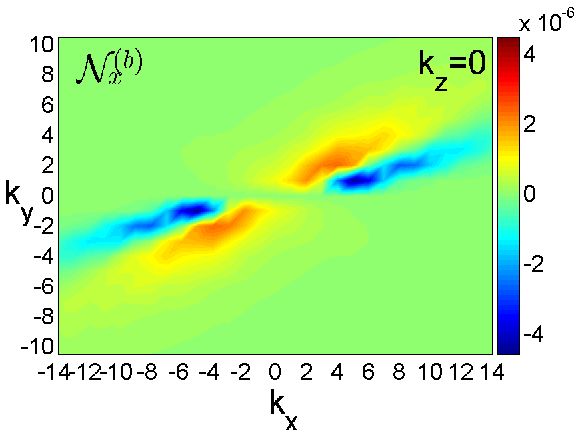}
\includegraphics[width=0.32\textwidth, height=0.2\textwidth]{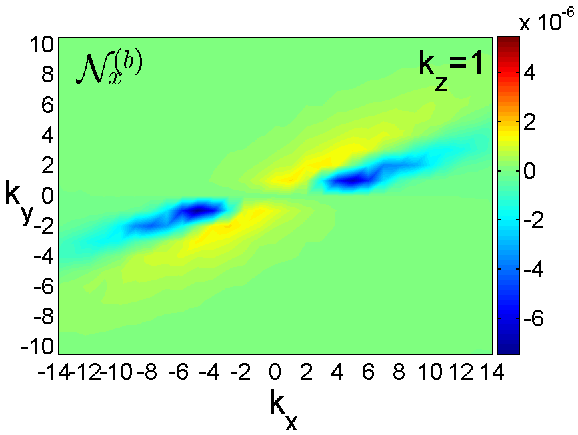}
\includegraphics[width=0.32\textwidth, height=0.2\textwidth]{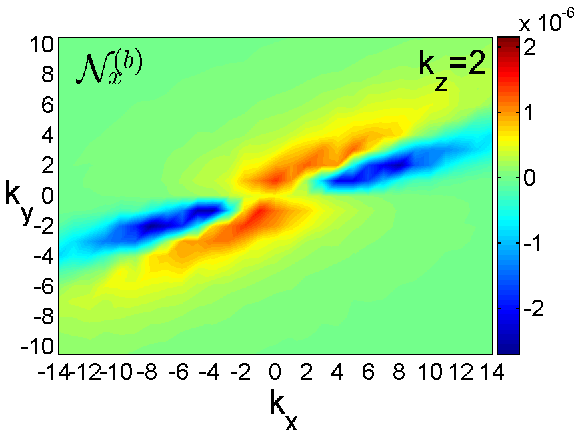}
\caption{Spectrum of the radial field, $|\bar{b}_x|$, and the maps
of the corresponding linear and nonlinear terms, governing its
dynamics (Equation \ref{eq:bxk}), in $(k_x,k_y)$-plane at $k_z=0
(left),~1 (middle),~2 (right)$. The spectrum as well as the action
of these terms are highly anisotropic, (i.e., depend on the
wavevector azimuthal angle) due to the shear. These terms are
appreciable and primarily operate in the vital area $|k_x|\lesssim
12, |k_y|\lesssim 3$. The red and yellow (blue and dark blue)
regions in each panel correspond to wavenumbers where respective
dynamical terms are positive (negative) and hence act as a source
(sink) for $|\bar{b}_x|^2$. In light green regions, outside the
vital area, these terms are small, although, as we checked, preserve
the same anisotropic shape. In particular, the nonlinear transfer
term ${\cal N}_x^{(b)}$ transversely redistributes $|\bar{b}_x|^2$
from the blue and dark blue regions, where ${\cal N}_x^{(b)}<0$, to
the red and yellow regions, where ${\cal N}_x^{(b)}>0$. These
regions exhibit considerable variations with the azimuthal angle of
the wavevector and also depend on $k_z$.}\label{fig:spectral-bx}
\end{figure*}

\begin{figure*}[t!]
\centering
\includegraphics[width=0.32\textwidth, height=0.2\textwidth]{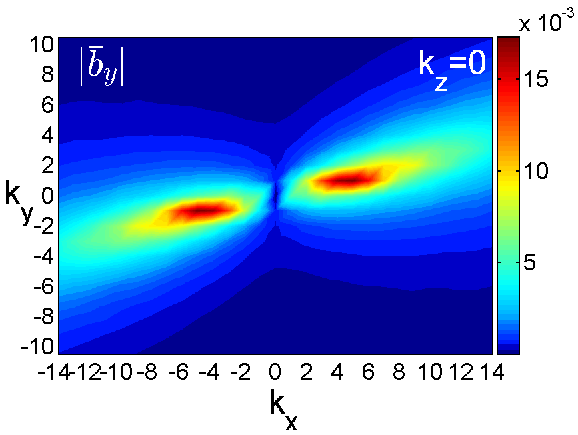}
\includegraphics[width=0.32\textwidth, height=0.2\textwidth]{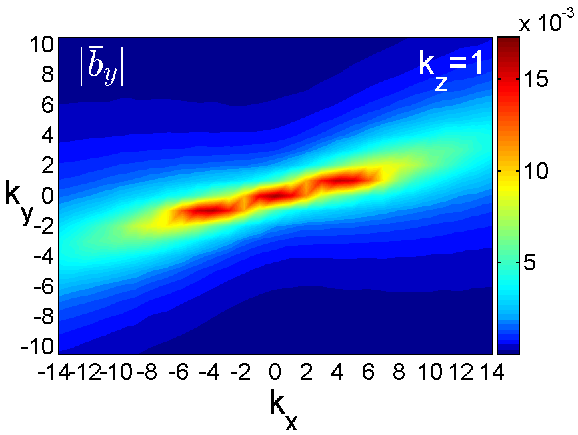}
\includegraphics[width=0.32\textwidth, height=0.2\textwidth]{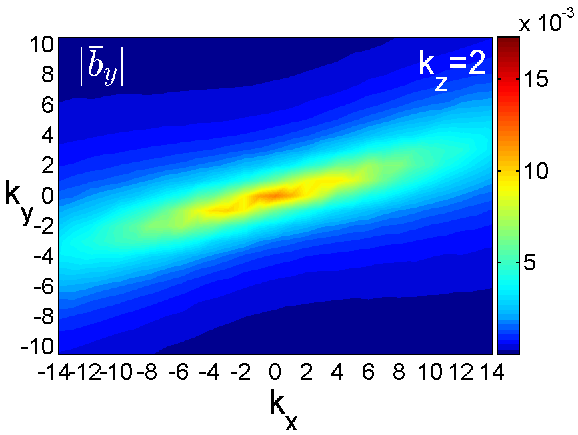}
\includegraphics[width=0.32\textwidth, height=0.2\textwidth]{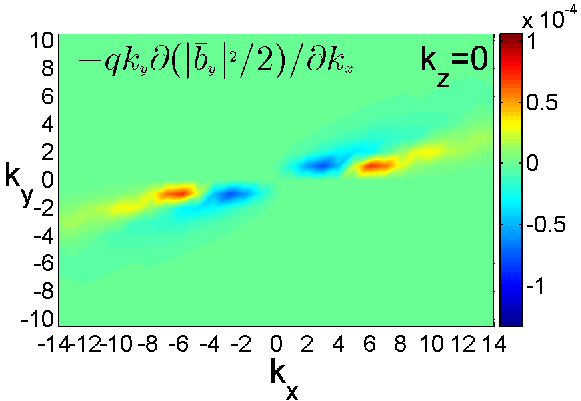}
\includegraphics[width=0.32\textwidth, height=0.2\textwidth]{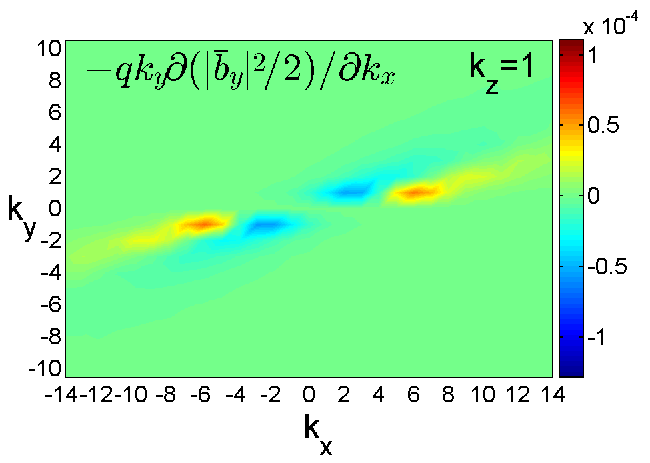}
\includegraphics[width=0.32\textwidth, height=0.2\textwidth]{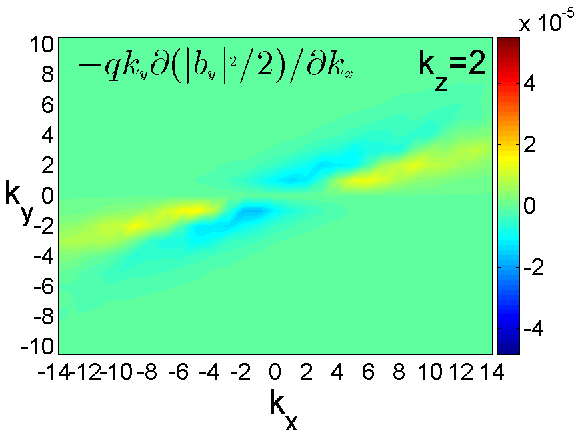}
\includegraphics[width=0.32\textwidth, height=0.2\textwidth]{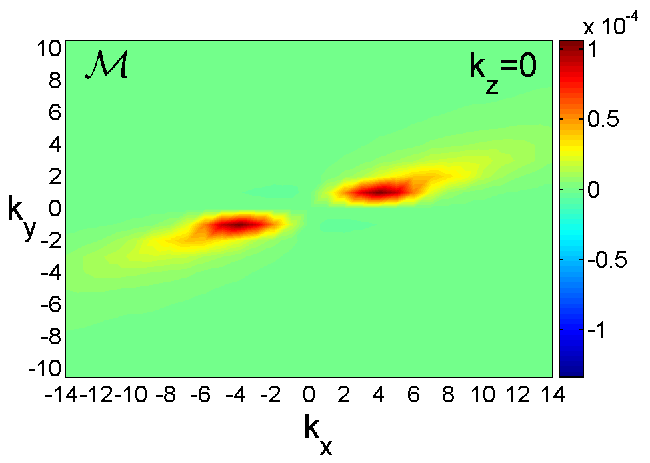}
\includegraphics[width=0.32\textwidth, height=0.2\textwidth]{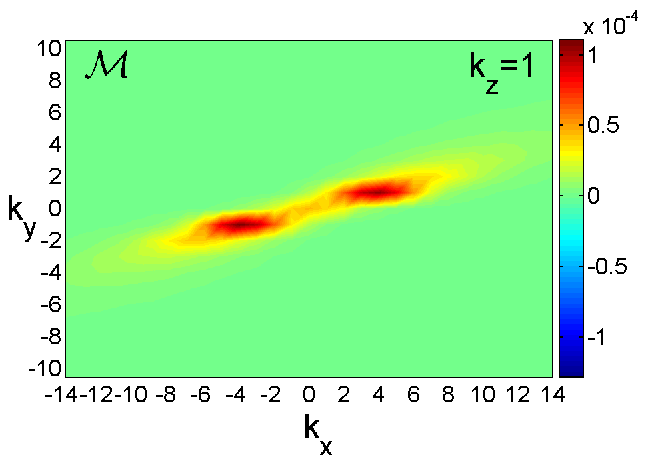}
\includegraphics[width=0.32\textwidth, height=0.2\textwidth]{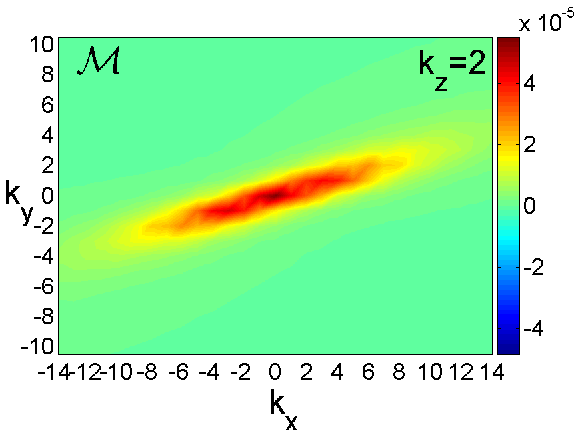}
\includegraphics[width=0.32\textwidth, height=0.2\textwidth]{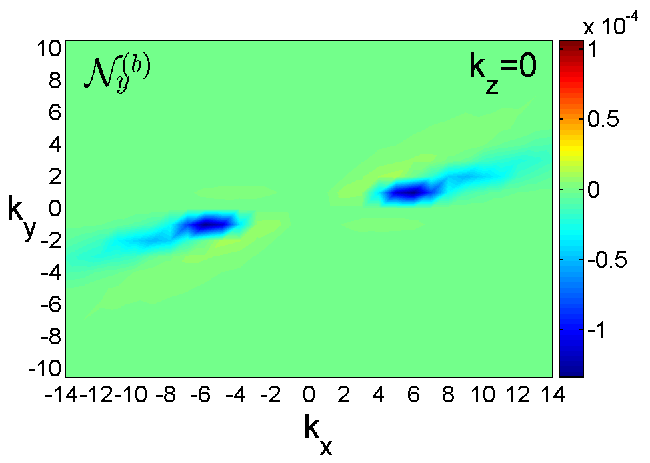}
\includegraphics[width=0.32\textwidth, height=0.2\textwidth]{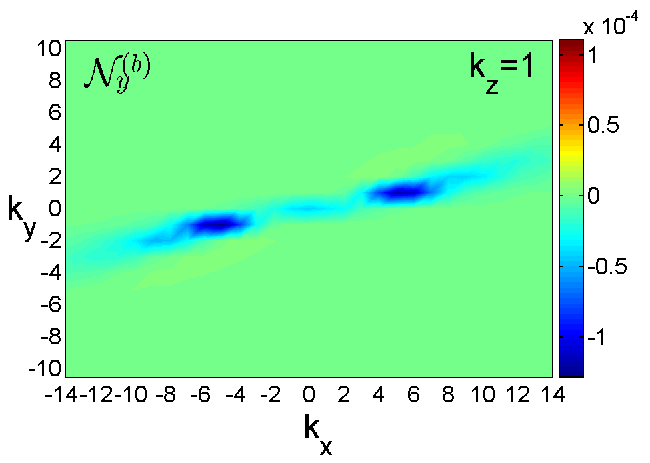}
\includegraphics[width=0.32\textwidth, height=0.2\textwidth]{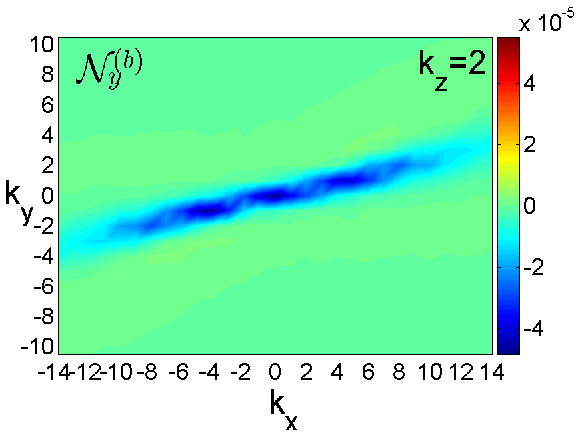}
\caption{Same as in Figure \ref{fig:spectral-bx}, but for the
azimuthal field $\bar{b}_y$ with the corresponding dynamical terms
from Equation (\ref{eq:byk}). The dynamics of this component is
primarily determined by the combined action of the drift, the
Maxwell stress ${\cal M}$, which is positive (energy injection) and
the nonlinear term ${\cal N}_y^{(b)}$, which is negative (sink) in
the vital area. The linear exchange term ${\cal I}_y^{(bu)}$ is
negligible compared with the above terms and is not shown
here.}\label{fig:spectral-by}
\end{figure*}

\begin{figure*}[t!]
\centering
\includegraphics[width=0.32\textwidth, height=0.2\textwidth]{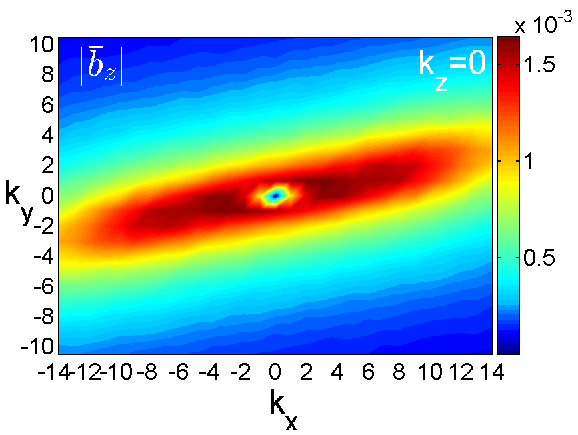}
\includegraphics[width=0.32\textwidth, height=0.2\textwidth]{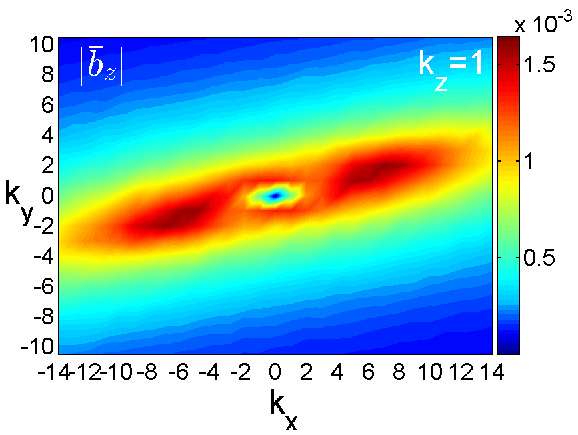}
\includegraphics[width=0.32\textwidth, height=0.2\textwidth]{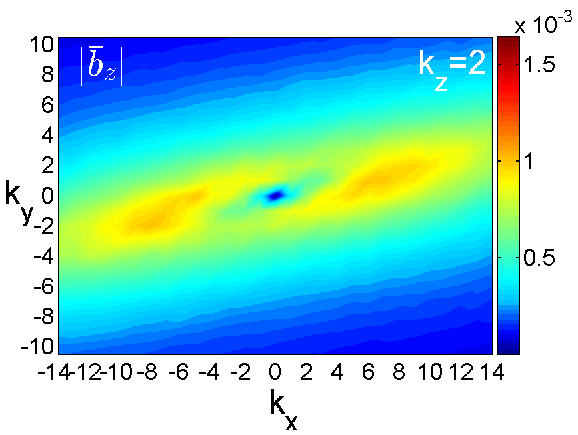}
\includegraphics[width=0.32\textwidth, height=0.2\textwidth]{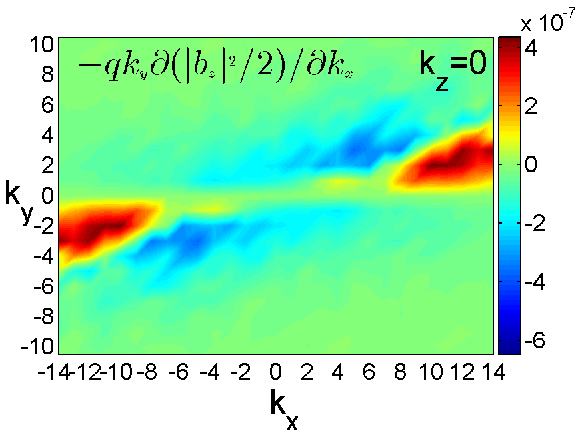}
\includegraphics[width=0.32\textwidth, height=0.2\textwidth]{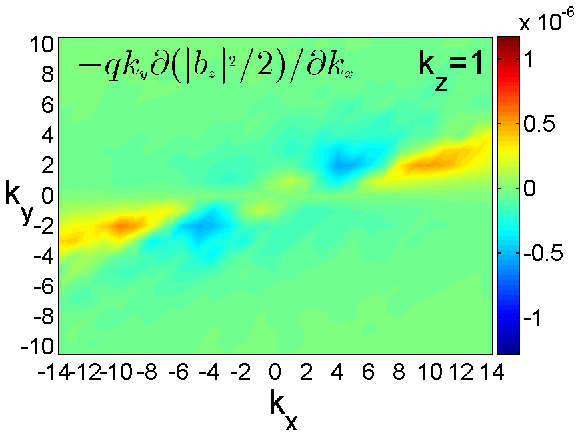}
\includegraphics[width=0.32\textwidth, height=0.2\textwidth]{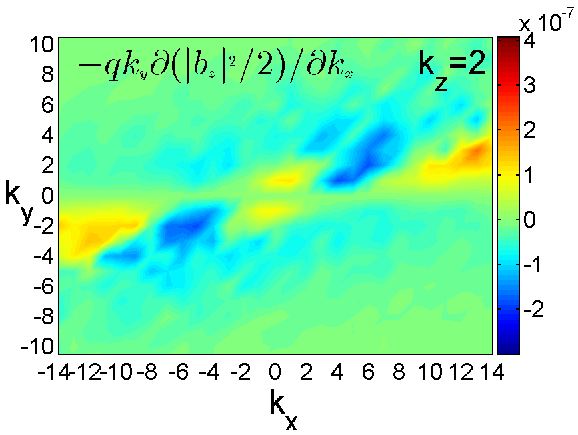}
\includegraphics[width=0.32\textwidth, height=0.2\textwidth]{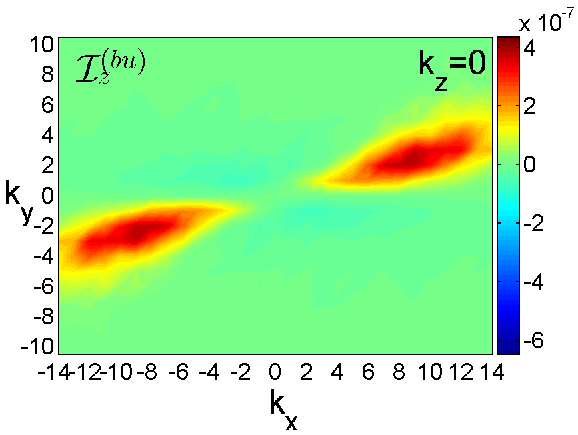}
\includegraphics[width=0.32\textwidth, height=0.2\textwidth]{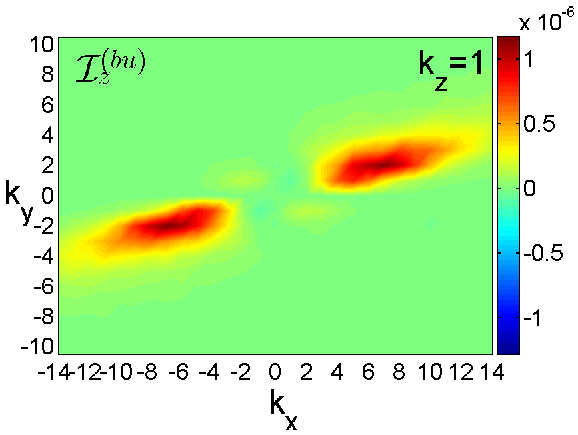}
\includegraphics[width=0.32\textwidth, height=0.2\textwidth]{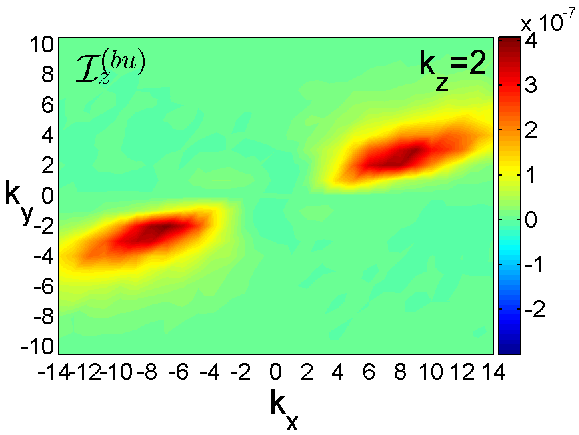}
\includegraphics[width=0.32\textwidth, height=0.2\textwidth]{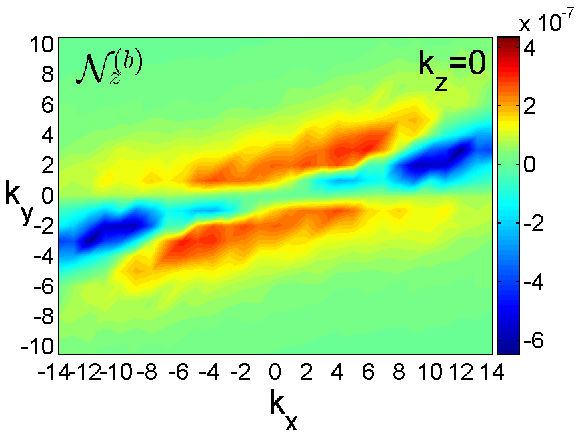}
\includegraphics[width=0.32\textwidth, height=0.2\textwidth]{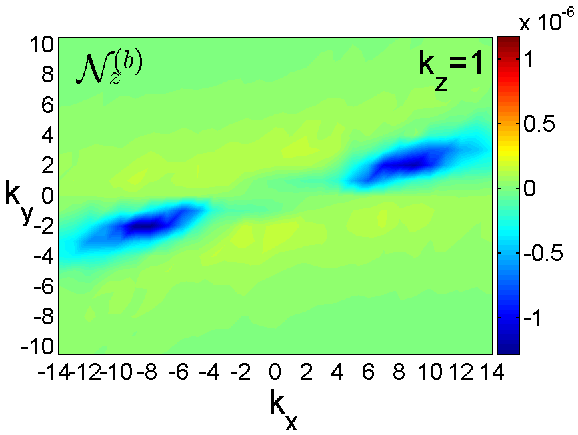}
\includegraphics[width=0.32\textwidth, height=0.2\textwidth]{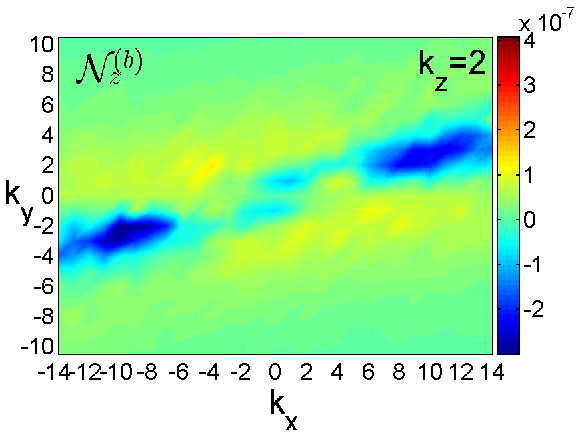}
\caption{Same as in Figure \ref{fig:spectral-bx}, but for
$\bar{b}_z$ with the corresponding dynamical terms from Equation
(\ref{eq:bzk}). The transverse character of the nonlinear
redistribution, ${\cal N}_z^{(b)}$, is also evident. $|\bar{b}_z|$
is small in comparison with $|\bar{b}_x|$ and
$|\bar{b}_y|$.}\label{fig:spectral-bz}
\end{figure*}

\begin{figure*}[t!]
\centering
\includegraphics[width=0.32\textwidth, height=0.2\textwidth]{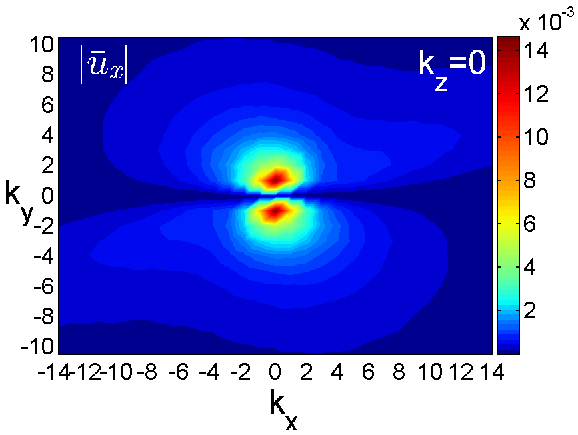}
\includegraphics[width=0.32\textwidth, height=0.2\textwidth]{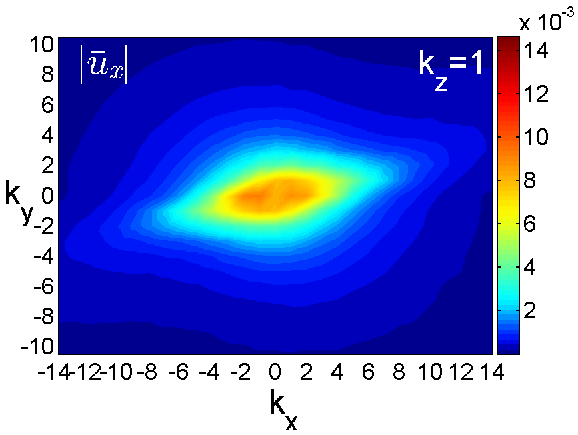}
\includegraphics[width=0.32\textwidth, height=0.2\textwidth]{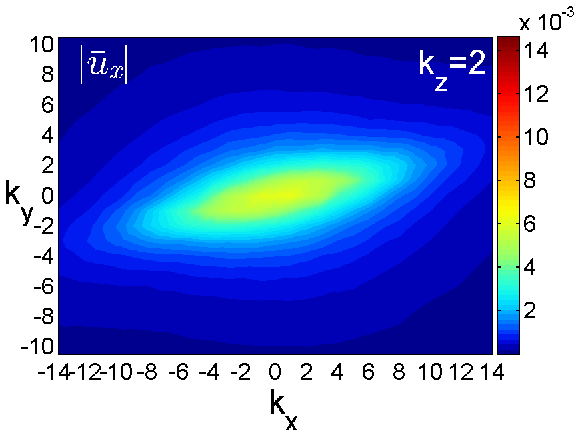}
\includegraphics[width=0.32\textwidth, height=0.2\textwidth]{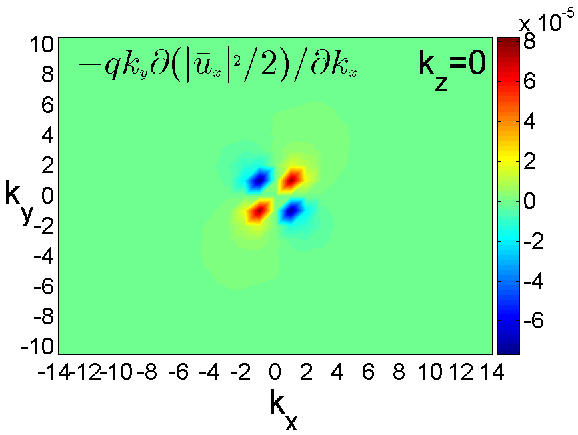}
\includegraphics[width=0.32\textwidth, height=0.2\textwidth]{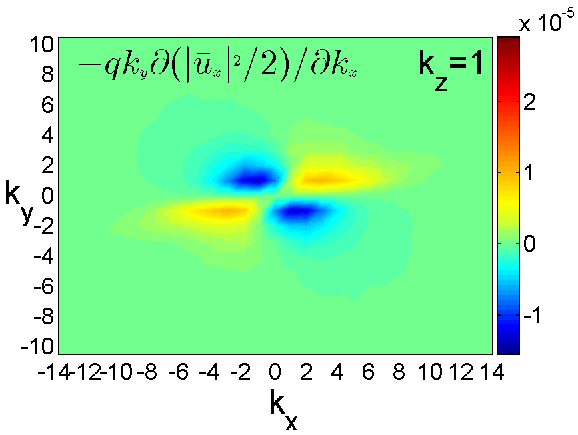}
\includegraphics[width=0.32\textwidth, height=0.2\textwidth]{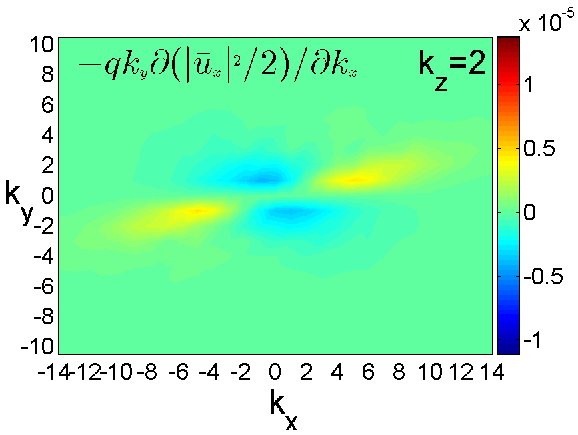}
\includegraphics[width=0.32\textwidth, height=0.2\textwidth]{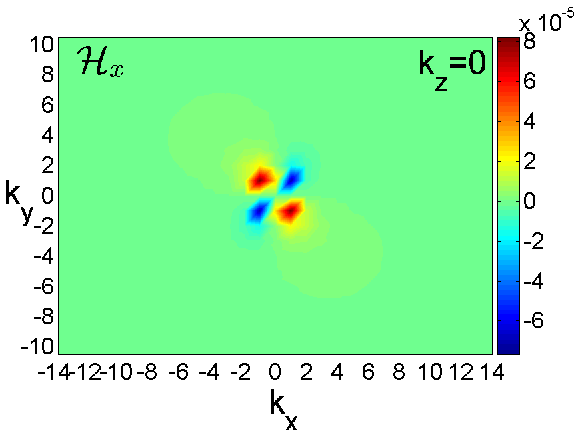}
\includegraphics[width=0.32\textwidth, height=0.2\textwidth]{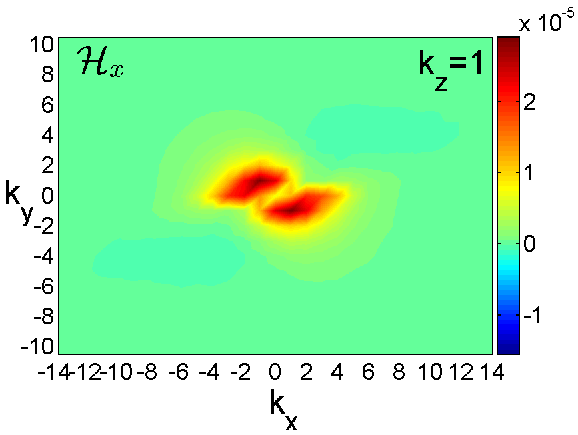}
\includegraphics[width=0.32\textwidth, height=0.2\textwidth]{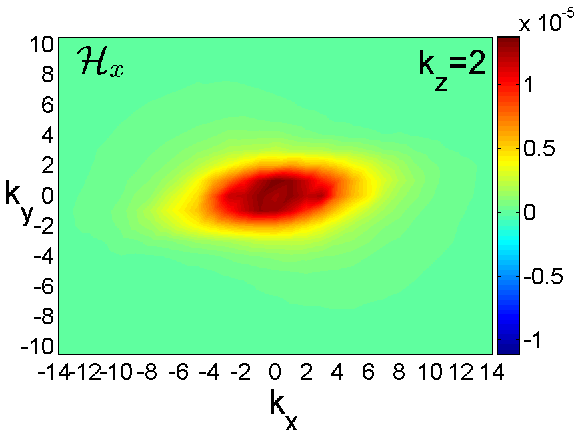}
\includegraphics[width=0.32\textwidth, height=0.2\textwidth]{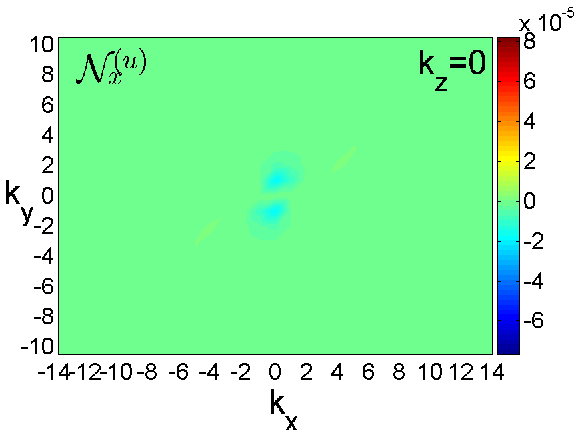}
\includegraphics[width=0.32\textwidth, height=0.2\textwidth]{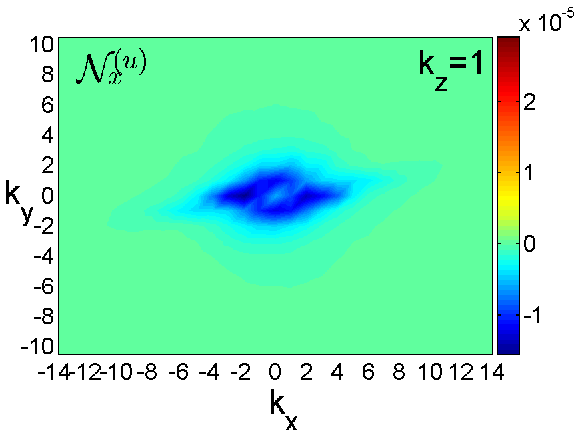}
\includegraphics[width=0.32\textwidth, height=0.2\textwidth]{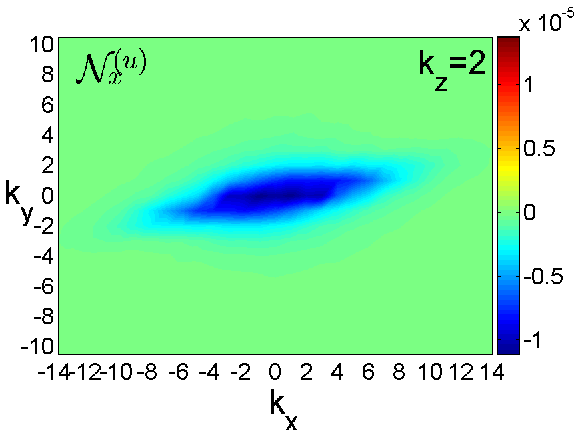}
\caption{Spectra of $|\bar{u}_x|$ and the maps of the corresponding
linear and nonlinear terms governing its dynamics (Equation
\ref{eq:uxk}) in $(k_x, k_y)$-plane at $k_z = 0(left), 1(middle),
2(right)$. The dynamics of this velocity component is primarily
determined by ${\cal H}_x$ (source) and ${\cal N}_x^{(u)}$ (sink),
the linear exchange terms, ${\cal I}_x^{(u\theta)}$ and ${\cal
I}_x^{(ub)}$, are negligible compared with the above terms and are
not shown here.}\label{fig:spectral-ux}
\end{figure*}

\begin{figure*}[t!]
\centering
\includegraphics[width=0.32\textwidth, height=0.2\textwidth]{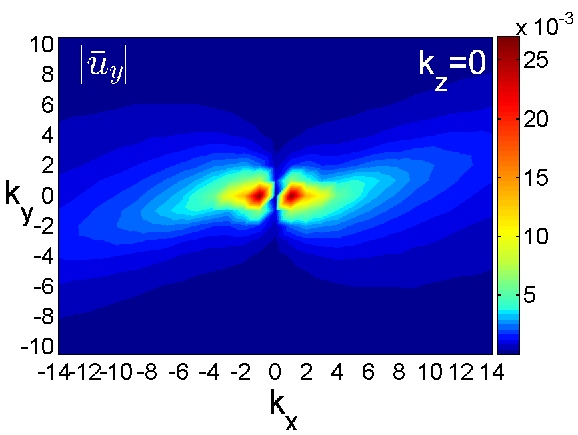}
\includegraphics[width=0.32\textwidth, height=0.2\textwidth]{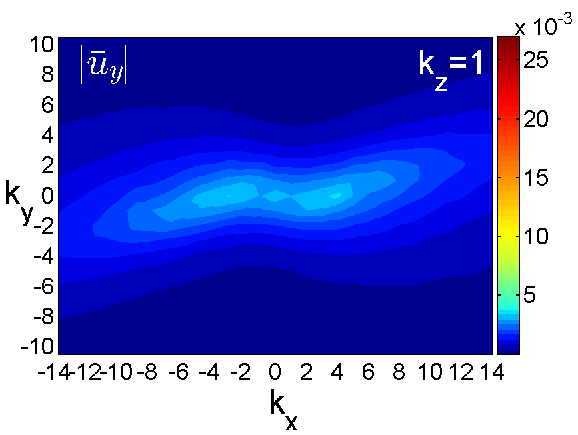}
\includegraphics[width=0.32\textwidth, height=0.2\textwidth]{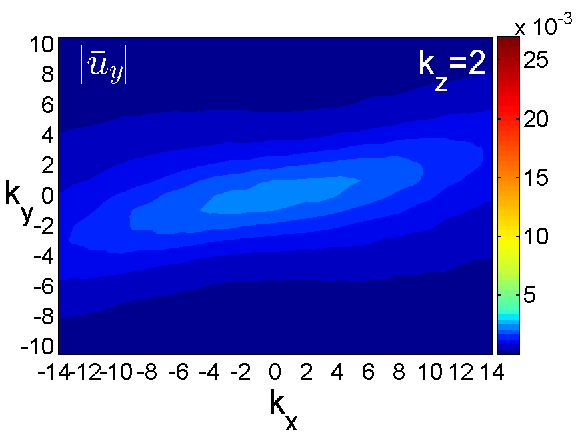}
\includegraphics[width=0.32\textwidth, height=0.2\textwidth]{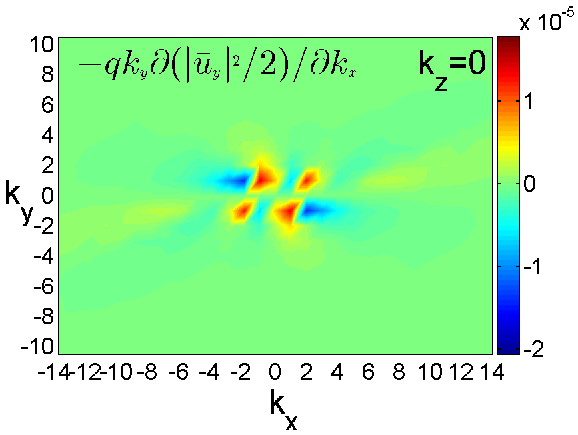}
\includegraphics[width=0.32\textwidth, height=0.2\textwidth]{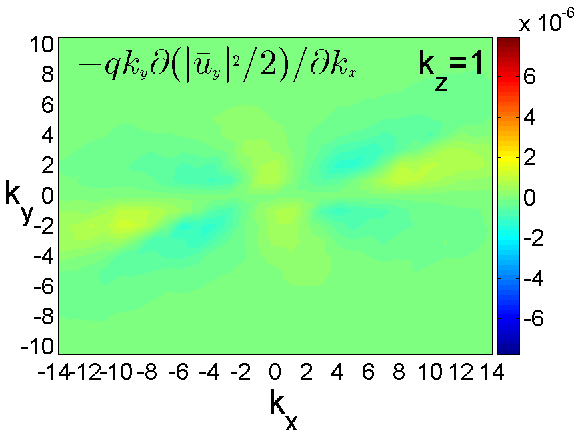}
\includegraphics[width=0.32\textwidth, height=0.2\textwidth]{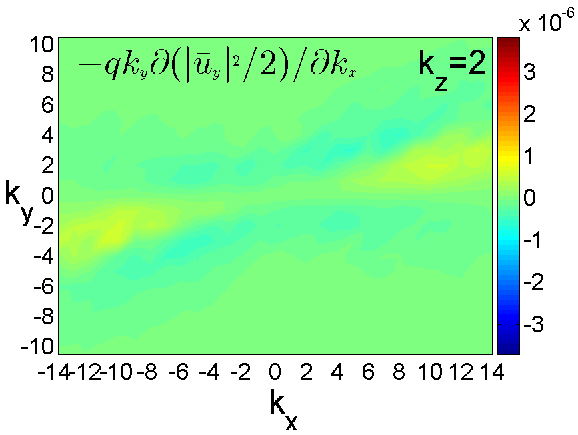}
\includegraphics[width=0.32\textwidth, height=0.2\textwidth]{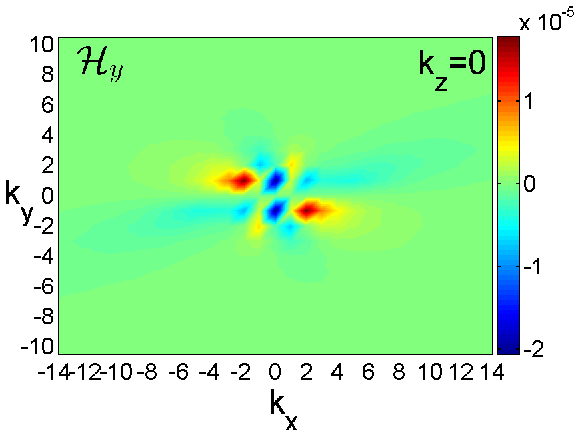}
\includegraphics[width=0.32\textwidth, height=0.2\textwidth]{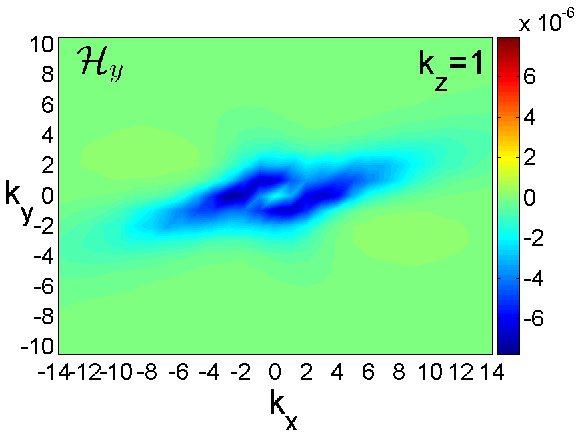}
\includegraphics[width=0.32\textwidth, height=0.2\textwidth]{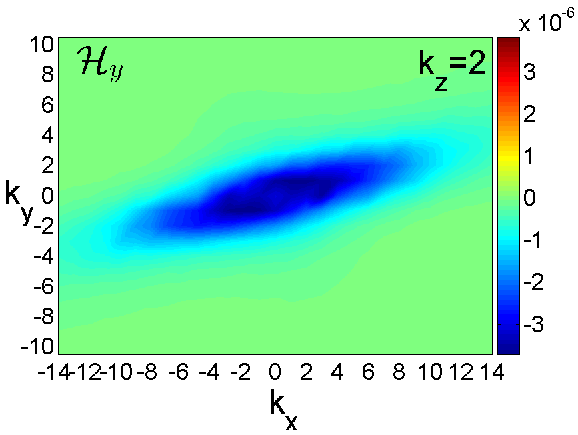}
\includegraphics[width=0.32\textwidth, height=0.2\textwidth]{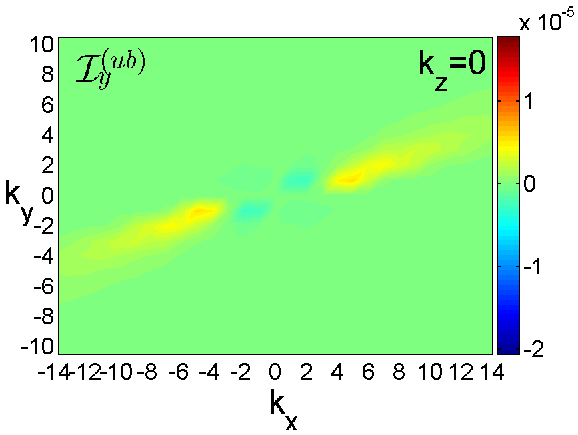}
\includegraphics[width=0.32\textwidth, height=0.2\textwidth]{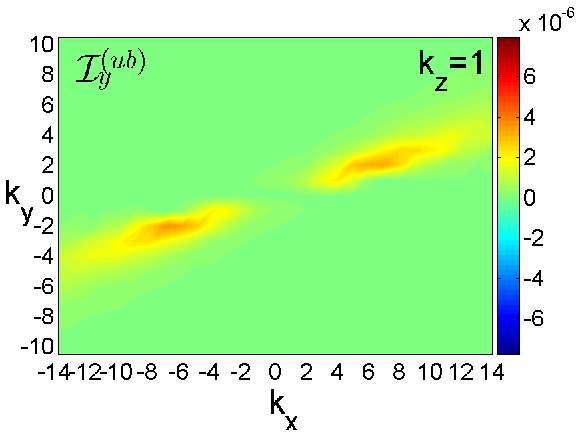}
\includegraphics[width=0.32\textwidth, height=0.2\textwidth]{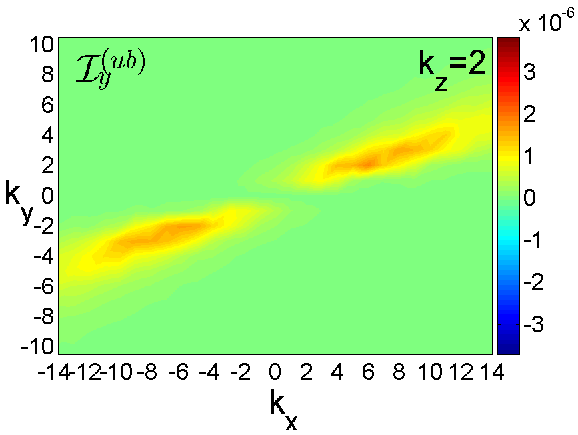}
\includegraphics[width=0.32\textwidth, height=0.2\textwidth]{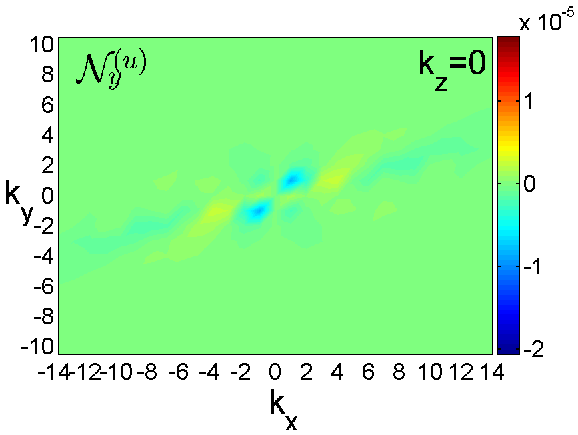}
\includegraphics[width=0.32\textwidth, height=0.2\textwidth]{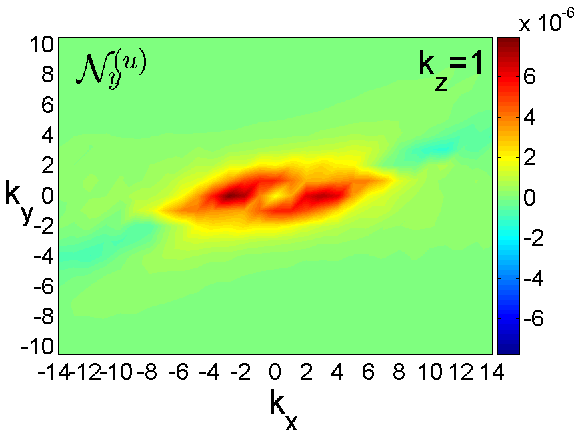}
\includegraphics[width=0.32\textwidth, height=0.2\textwidth]{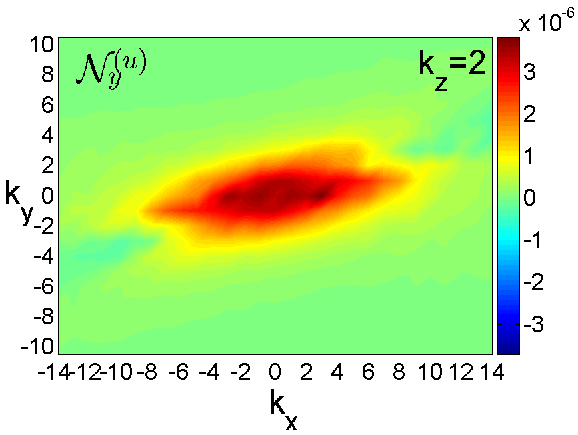}
\caption{Same as in Figure \ref{fig:spectral-ux}, but for
$\bar{u}_y$ with the corresponding dynamical terms from Equation
(\ref{eq:uyk}). The influence of the thermal process, ${\cal
I}_y^{(u\theta)}$, is negligible and not shown here. The spectrum of
$|\bar{u}_y|$ reaches a maximum at $k_x=\pm 1, k_y=k_z=0$, which
corresponds to the zonal flow in physical
space.}\label{fig:spectral-uy}
\end{figure*}

\begin{figure*}[t!]
\centering
\includegraphics[width=0.32\textwidth, height=0.2\textwidth]{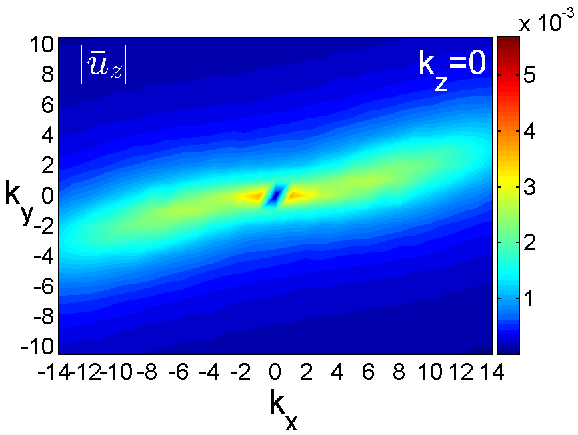}
\includegraphics[width=0.32\textwidth, height=0.2\textwidth]{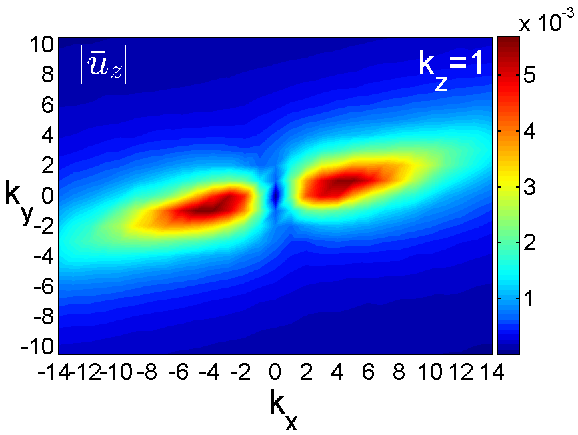}
\includegraphics[width=0.32\textwidth, height=0.2\textwidth]{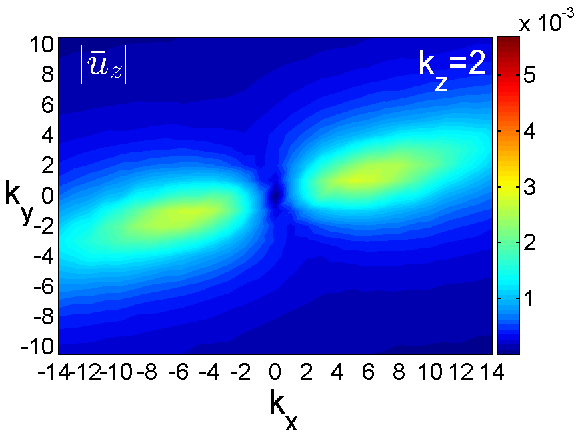}
\includegraphics[width=0.32\textwidth, height=0.2\textwidth]{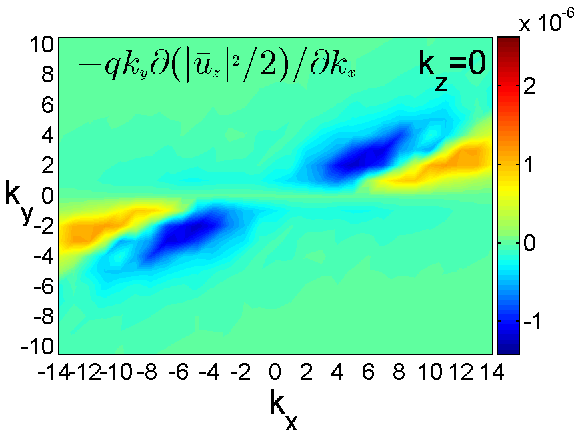}
\includegraphics[width=0.32\textwidth, height=0.2\textwidth]{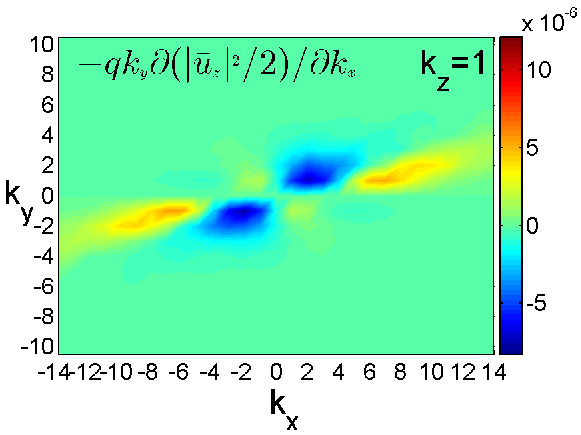}
\includegraphics[width=0.32\textwidth, height=0.2\textwidth]{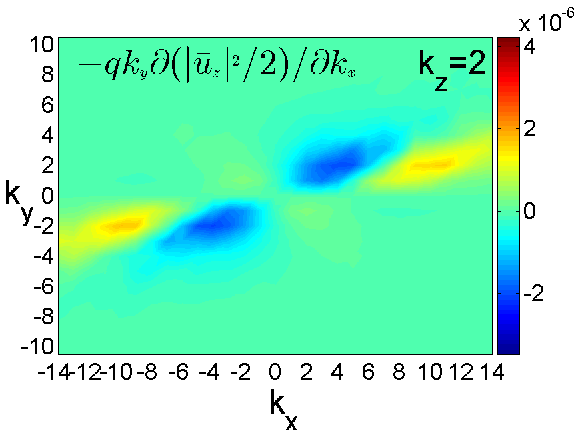}
\includegraphics[width=0.32\textwidth, height=0.2\textwidth]{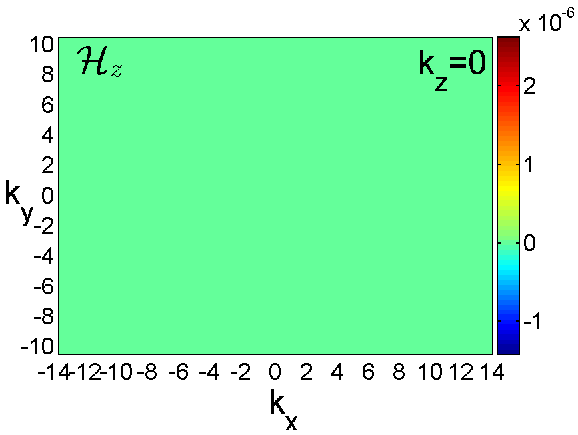}
\includegraphics[width=0.32\textwidth, height=0.2\textwidth]{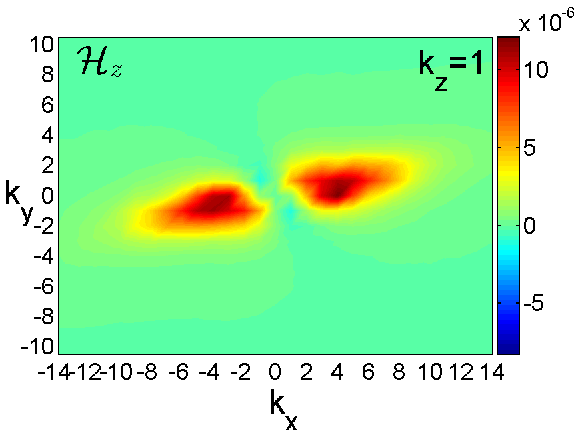}
\includegraphics[width=0.32\textwidth, height=0.2\textwidth]{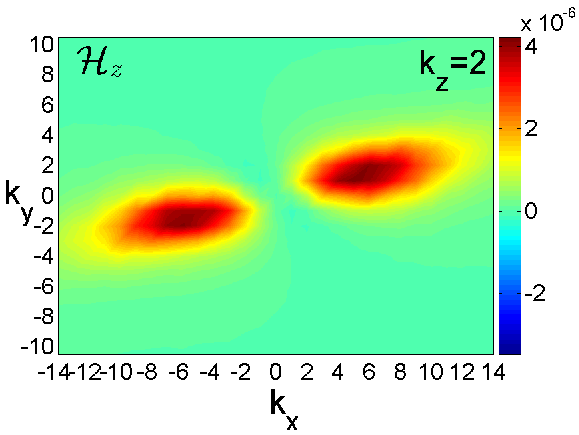}
\includegraphics[width=0.32\textwidth, height=0.2\textwidth]{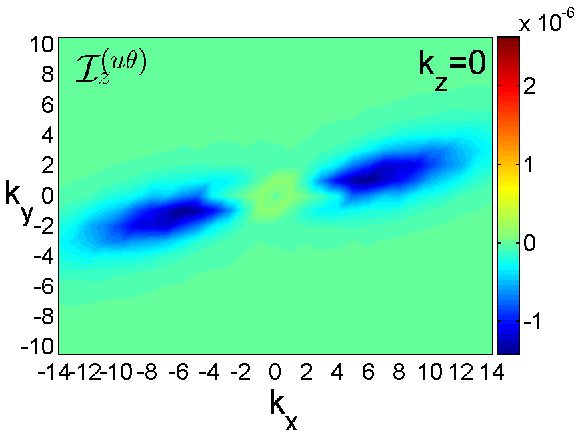}
\includegraphics[width=0.32\textwidth, height=0.2\textwidth]{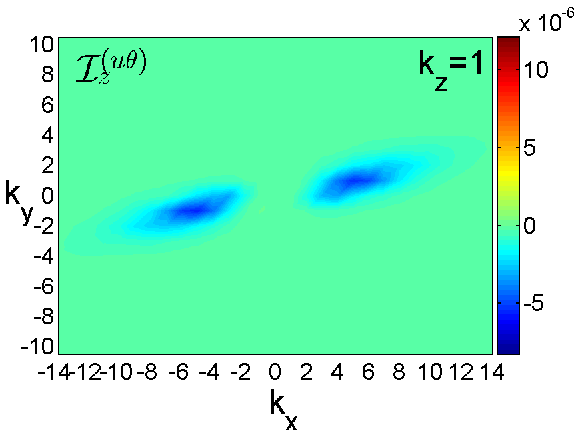}
\includegraphics[width=0.32\textwidth, height=0.2\textwidth]{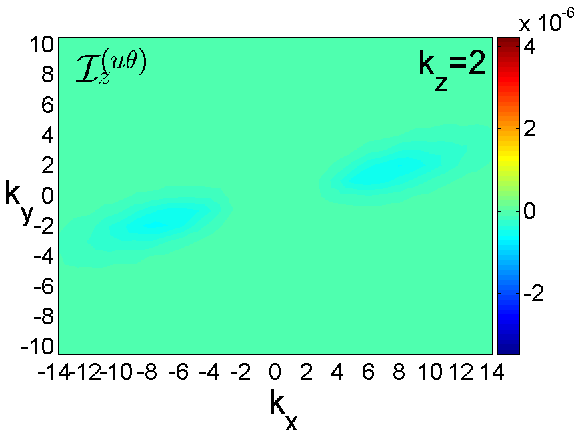}
\includegraphics[width=0.32\textwidth, height=0.2\textwidth]{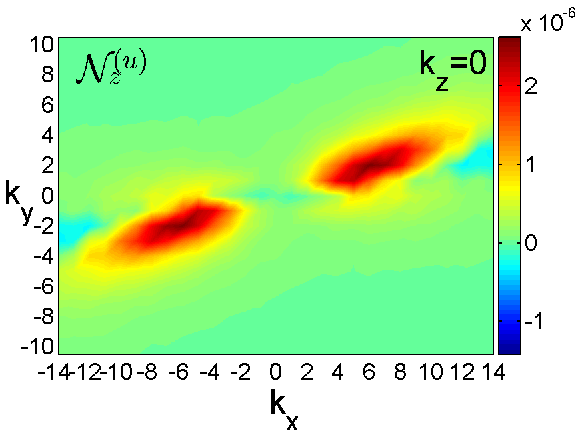}
\includegraphics[width=0.32\textwidth, height=0.2\textwidth]{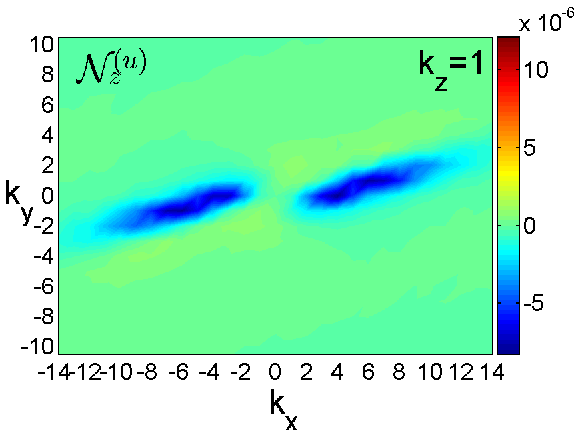}
\includegraphics[width=0.32\textwidth, height=0.2\textwidth]{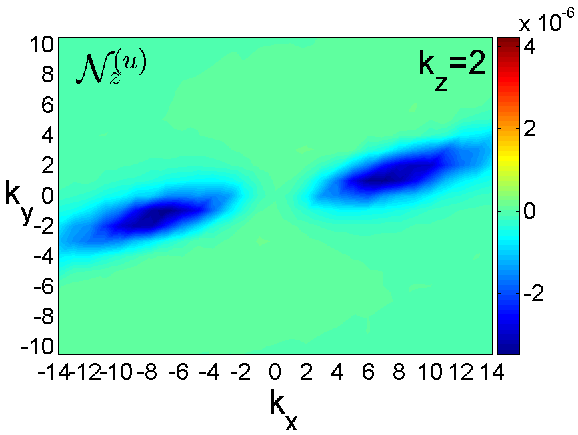}
\caption{Same as in Figure \ref{fig:spectral-ux}, but for
$\bar{u}_z$ with the corresponding dynamical terms from Equation
(\ref{eq:uzk}). The influence of the linear magnetic exchange term
${\cal I}_z^{(ub)}$ is negligible and is not shown here. The
nonlinear term ${\cal N}_z^{(u)}$ transfers $|\bar{u}_z|^2$ towards
small $k_z$ (as it also seen in the corresponding panel of Figure
\ref{fig:integrated in plane-ub}).}\label{fig:spectral-uz}
\end{figure*}

\section{Interplay of the linear and nonlinear processes in the
sustenance of the turbulence}

We have seen above that the sustaining dynamics of turbulence is
primarily concentrated at small vertical wavenumbers, so now we
present the distribution of the time-averaged amplitudes of the
spectral quantities $\bar{u}_i$, $\bar{b}_i$ as well as the linear
($k_x$-drift, ${\cal H}_i, {\cal I}_i, {\cal M}$) and nonlinear
(${\cal N}_i$) dynamical terms in $(k_x,k_y)$-plane again at
$k_z=0,1,2$ in Figures \ref{fig:spectral-bx}-\ref{fig:spectral-uz}
(as noted before, we omit here the thermal processes,
$\bar{\theta}$, which play a minor role). These figures give quite a
detailed information and insight about all the linear and nonlinear
processes involved in Equations (\ref{eq:uxk})-(\ref{eq:bzk}) and
allow us to properly understand their interplay leading to the
turbulence sustenance. We start the analysis of this interplay with
a general outline of the figures. We do not show here the viscous
(${\cal D}_i^{(u)}$) and resistive ${\cal D}_i^{(b)}$ terms, since
their action is quite simple -- they are always negative and reduce
the corresponding quantities, thereby opposing the sustenance
process. They increase with $k$, but in the vital area are too small
to have any influence on the dynamics.

A first glance at the plots makes it clear that all the spectra of
the physical quantities and processes are highly anisotropic due to
the shear, i.e., strongly depend on the azimuthal angle in
$(k_x,k_y)$-planes as well as vary with $k_z$, with a similar type
of anisotropy and inclination towards the $k_x$-axis, as the energy
spectrum in Figure \ref{fig:spectral energy}. For the nonlinear
processes represented by ${\cal N}_i^{(u)}$ and ${\cal N}_i^{(b)}$
(bottom row in Figures \ref{fig:spectral-bx}-\ref{fig:spectral-uz}),
this anisotropy can not be put within the framework of commonly
considered forms of nonlinear -- direct and inverse -- cascades,
since its main manifestation is the transverse (among wavevector
angles) nonlinear redistribution of modes in $(k_x,k_y)$-plane as
well as among different $k_z$. In these figures, the nonlinear terms
transfer the corresponding quadratic forms of the velocity and
magnetic field components transversely away from the regions where
they are negative (${\cal N}_i^{(u)}<0, {\cal N}_i^{(b)}<0$, blue
and dark blue) towards the regions where they are positive (${\cal
N}_i^{(u)}>0, {\cal N}_i^{(b)}>0$, yellow and red). These regions
display quite a strong angular variation in $(k_x,k_y)$-planes.

Similarly, the terms of linear origin ${\cal H}_i, {\cal I}_i, {\cal
M}$ are strongly anisotropic in $(k_x,k_y)$-plane. For the
corresponding quantity, they act as a source when positive (red and
yellow regions) and as a sink when negative (blue and dark blue
regions). The linear exchange of energy with the background shear
flow (which is the central energy supply for turbulence) involves
all the components of the velocity perturbation through ${\cal H}_i$
terms in Equations (\ref{eq:uxk})-(\ref{eq:uzk}) and only the
azimuthal $y$-component of the magnetic field perturbation through
the Maxwell stress term, ${\cal M}$, in Equation (\ref{eq:byk}).
However, the other quadratic forms can grow due to the linear
exchange, ${\cal I}_i$, and nonlinear, ${\cal N}_i$, terms. The
growth of the quadratic forms and energy extraction from the flow as
a result of the operation of all these linear terms essentially
constitutes the azimuthal MRI in the flow.

The linear drift parallel to the $k_x$-axis is equally important for
all the physical quantities. The plots depicting the drift (second
row in Figures \ref{fig:spectral-bx}-\ref{fig:spectral-uz}), show
that this process transfers modes with velocity $|qk_y|$ along
$k_x$-axis at $k_y>0$ and in the opposite direction at $k_y<0$.
Namely, the drift gives the linear growth of individual harmonics a
transient nature, as it sweeps them through the vital area in
$\textbf{k}$-space. One has to note that the dynamics of
axisymmetric modes with $k_y=0$ should be analyzed separately, as
the drift does not affect them. Consequently, the drift can not
limit the duration of their amplification and if there is any, even
weak, linear or nonlinear source of growth at $k_y=0$, these
harmonics can reach high amplitudes.

Let us turn to the analysis of the route ensuring the turbulence
sustenance. First of all, we point out that it should primarily rely
on magnetic perturbations, as the Maxwell stress is mainly
responsible for energy supply for turbulence. From Figure
\ref{fig:spectral-bx}, it is seen that the linear exchange term
${\cal I}_x^{(bu)}$ and the nonlinear term ${\cal N}^{(b)}_x$ make
comparable contributions to the generation and maintenance of the
radial field component $|\bar{b}_x|$. This is also consistent with
the related plots in Figure \ref{fig:integrated in plane-ub}. The
exchange term takes energy from the radial velocity $\bar{u}_x$ and
gives to $\bar{b}_x$. The distribution of ${\cal N}^{(b)}_x$ clearly
demonstrates transversal transfer of $|\bar{b}_x|^2$ in
$(k_x,k_y)$-plane for all considered $k_z=0,1,2$ as well as among
different components. The linear drift term also participates in
forming the final spectrum of $|\bar{b}_x|$ in the quasi-steady
turbulent state. It opposes the action of the nonlinear term: for
$k_y>0$ ($k_y<0$), ${\cal N}_x^{(b)}$, transfers modes to the left
(right), from the blue and dark blue region to the red and yellow
regions, while the drift transfers in the opposite direction. So,
the interplay of the drift, ${\cal I}_x^{(bu)}$ and ${\cal
N}_x^{(b)}$ yields the specific anisotropic spectra of $|\bar{b}_x|$
shown in the top row of this figure. Particularly noteworthy is the
role of the nonlinear term at $k_y=0, k_z=1,2$, because the drift
and the linear magnetic-kinetic exchange terms are proportional to
$k_y$ and hence vanish. As a result, axisymmetric modes with $k_y=0$
are energetically supported only by the nonlinear term. (At $k_y=0$,
although ${\cal N}_x^{(b)}$ is positive both at $k_z=1$ and $k_z=2$,
its values at $k_z=1$ are about an order of magnitude smaller than
those at $k_z=2$ and might not be well represented by light green
color in the bottom middle panel in Figure \ref{fig:spectral-bx}.)
So, $\bar{b}_x$, which is remarkably generated by the nonlinear
term, in turn, is a key factor in the production and distribution of
the energy-injecting Maxwell stress, ${\cal M}$, in Fourier space.
Indeed, note the correlation between the distributions of
$|\bar{b}_x|$ and ${\cal M}$ in $(k_x,k_y)$-plane depicted,
respectively, in the top row of Figure \ref{fig:spectral-bx} and in
the third row of Figure \ref{fig:spectral-by}.

From Figure \ref{fig:spectral-by} it is evident that, in fact, the
Maxwell stress, ${\cal M}$, which is positive in $(k_x,k_y)$-plane
and appreciable in the vital area, is the only source for the
quadratic form of the azimuthal field component, $|\bar{b}_y|^2$,
and hence for the turbulent magnetic energy, which is dominated by
this component. The linear exchange term, ${\cal I}_y^{(bu)}$,
appears to be much smaller with this stress term (and hence is not
shown in this figure). The nonlinear term, ${\cal N}^{(b)}_y$, is
negative in the vital area (blue regions in the bottom row of Figure
\ref{fig:spectral-by}), draining $|\bar{b}_y|^2$ there and
transferring it to large wavenumbers as well as among different
components. Thus, the sustenance of the magnetic energy is of linear
origin, due solely to the Maxwell stress that, in turn, is generated
from the radial field component. This stage constitutes the main
(linear) part of the sustenance scheme, which will be described in
the next subsection, and is actually a manifestation of the
azimuthal MRI.

The dynamics of the vertical field component $\bar{b}_z$ is shown in
\ref{fig:spectral-bz}. This components is smaller than $\bar{b}_x$
and $\bar{b}_y$. The linear exchange term, ${\cal I}_z^{(bu)}$, acts
as a source, supplying $\bar{b}_z$ from the vertical velocity
$\bar{u}_z$. The nonlinear term, ${\cal N}^{(b)}_z$, also realizes
the transverse cascade and scatters the modes in different areas of
$(k_x,k_y)$-plane (from the yellow and red to blue and dark blue
areas in the bottom row of Figure \ref{fig:spectral-bz}). However,
as it is seen from the related plot in Figure \ref{fig:integrated in
plane-ub}, the cumulative effect of ${\cal N}^{(b)}_z$ in
$(k_x,k_y)$-plane is positive and even prevails over the positive
cumulative contribution of ${\cal I}_z^{(bu)}$ in this plane at
every $k_z$. As it is clearly seen from Figure
\ref{fig:spectral-bz}, the linear drift term opposes the action of
the nonlinear term for $\bar{b}_z$, similar to that in the case of
$\bar{b}_x$.

Figure \ref{fig:spectral-ux} shows that the linear term ${\cal H}_x$
can be positive and act as a source for the radial velocity
$|\bar{u}_x|^2$ at the expense of the mean flow, while the nonlinear
term ${\cal N}^{(u)}_x$ is negative and drains it. The exchange
terms ${\cal I}_x^{(u\theta)}$, ${\cal I}_x^{(ub)}$ are also
negative, giving the energy of the radial velocity, respectively, to
$\bar{\theta}$ and $\bar{b}_x$, but their contributions are
negligible compared with ${\cal H}_x$ and ${\cal N}_x^{(u)}$ and
hence not shown in this figure. So, the sustenance of $|\bar{u}_x|$
is ensured by the interplay of the linear drift and ${\cal H}_x$
terms. Indeed, shifting the result of the action of ${\cal H}_x$ by
the linear drift to the right (left) for $k_y>0$ ($k_y<0$) gives the
spectrum of $|\bar{u}_x|$ presented in the top row this figure.

Figure \ref{fig:spectral-uy} shows that the dynamics of the
azimuthal velocity $\bar{u}_y$ is governed primarily by ${\cal
H}_y$, ${\cal I}_y^{(ub)}$ and ${\cal N}^{(u)}_y$. The action of
${\cal I}_y^{(u\theta)}$ is negligible compared with these terms, in
agreement with the corresponding plot of Figure \ref{fig:integrated
in plane-ub}, and is not shown in this figure. The contributions of
${\cal I}_y^{(ub)}$ and ${\cal N}^{(u)}_y$ can be positive and hence
these terms act as a source for $|\bar{u}_y|^2$. The distribution of
${\cal H}_y$ at $k_z=0$ is quite complex with alternating positive
and negative areas in $(k_x,k_y)$-plane, while it is negative for
$k_z=1,2$. A interplay between these three terms yields the spectrum
of $|\bar{u}_y|$ shown in the top row of Figure
\ref{fig:spectral-uy}. From this spectrum, the harmonic with $k_x=1,
k_y=k_z=0$ has the highest amplitude. Translating this result in
physical space, it implies that the turbulence forms quite powerful
azimuthal/zonal flow, which will be examined in more detail in the
next subsection.

Figure \ref{fig:spectral-uz} shows that the contribution of the
thermal ${\cal I}_z^{(u\theta)}$ in the dynamics of the quadratic
form of vertical velocity, $|\bar{u}_z|^2$, is mostly negative
(sink), but not so strong. The magnetic exchange term ${\cal
I}_z^{(ub)}$ also acts as a sink, but is much smaller than ${\cal
I}_z^{(u\theta)}$ and can be neglected. Of course, the role of the
linear drift term is standard and similar to those for other
components described above. The sustenance of $|\bar{u}_z|$ at
$k_z=0$ is ensured by the combination of the linear drift and the
positive nonlinear term ${\cal N}^{(u)}_z$, while at $k_z=1,2$ it is
maintained by the interplay of the linear drift and ${\cal H}_z$,
which provides a source, now the nonlinear term ${\cal N}_z^{(u)}$
acts as a sink.

\begin{figure}
\includegraphics[width=\columnwidth]{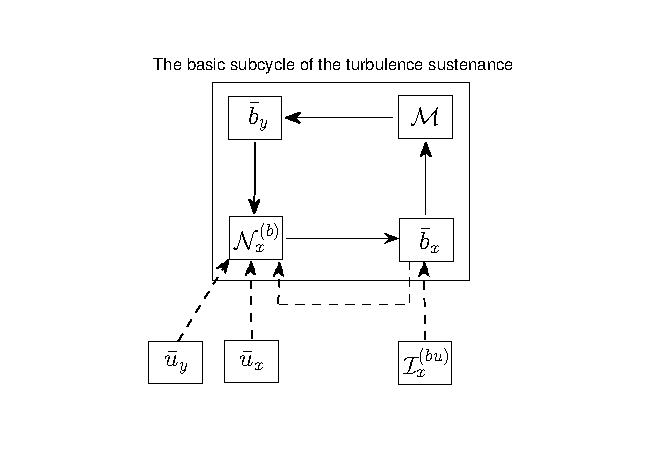}
\caption{Sketch of the basic subcycle of the sustaining process (the
solid arrows within the rectangle): (i) generation of $\bar{b}_x$ by
the nonlinearity ${\cal N}_x^{(b)}$, (ii) subsequent production of
${\cal M}$ from $\bar{b}_x$ and (iii) of the azimuthal field energy
$|\bar{b}_y|^2$ by ${\cal M}$ (the linear MRI stage) and finally
(iv) the nonlinear feedback -- contribution of $\bar{b}_y$ to ${\cal
N}_x^{(b)}$. The other contributions (dashed arrows) of ${\cal
I}_x^{(bu)}$ to the production of $\bar{b}_x$ and the feedback of
$\bar{u}_x, \bar{u}_y, \bar{b}_x$ to ${\cal N}_x^{b}$ ($\bar{u}_z$
and $\bar{b}_z$ are small and not shown), are a part of the overall
sustaining scheme, but outside the basic subcycle.}
\label{fig:sketch}
\end{figure}

\subsection{The basic subcycle of the turbulence sustenance}

As we already mentioned, the sustenance of the turbulence is the
result of a subtle intertwining of the anisotropic linear transient
growth and nonlinear transverse cascade processes, which have been
described in the previous section. The intertwined character of
these processes is too complex for a vivid schematization.
Nevertheless, based on the insight into the turbulence dynamics
gained from Figures \ref{fig:spectral-bx}-\ref{fig:spectral-uz}, we
can bring out the \emph{basic subcycle} of the sustenance that
clearly shows the equal importance of the linear and nonlinear
processes. The azimuthal and radial magnetic field components are
most energy-containing in this case. The basic subcycle of the
turbulence sustenance, which is concentrated in the vital area in
Fourier space, is sketched in Figure (\ref{fig:sketch}) (solid
arrows within a rectangle) and can be understood as follows. The
nonlinear term ${\cal N}_x^{(b)}$ contributes to the generation of
the radial field $\bar{b}_x$ through the transverse cascade process.
In other words, ${\cal N}_x^{(b)}$ provides a positive feedback for
the continual regeneration of the radial field, which, in turn, is a
seed/trigger for the linear growth of the MRI -- $\bar{b}_x$ creates
and amplifies the Maxwell stress, ${\cal M}$, due to the shear (via
linear term in Equation \ref{eq:App-byk} proportional to $q$). The
positive stress then increases the dominant azimuthal field energy
$|\bar{b}_y|^2/2$ at the expense of the mean flow, opposing the
negative nonlinear term ${\cal N}_y^{(b)}$ (and resistive
dissipation). Thus, this central energy gain process for turbulence,
as mentioned before, is of linear nature and a consequence of the
azimuthal MRI. The linearly generated $\bar{b}_y$ gives a dominant
contribution -- positive feedback -- to the nonlinear term ${\cal
N}_x^{(b)}$, closing the basic subcycle.

This is only a main part of the complete and more intricate
sustaining scheme that involves also the velocity components. In
this sketch, the dashed arrows denote the other, extrinsic to the
basic subcycle, processes. Namely, $\bar{b}_x$, together with the
nonlinear term, is fueled also by the linear exchange term, ${\cal
I}_x^{(bu)}$, which takes energy from the radial velocity
$\bar{u}_x$, while the azimuthal velocity $\bar{u}_y$ gets energy
from $\bar{b}_y$ via the linear exchange term ${\cal I}_y^{(ub)}$.
These are all linear processes, part of the MRI. (The vertical
velocity does not explicitly participate in this case.) All These
components of the velocity $\bar{u}_x$, $\bar{u}_y$, $\bar{u}_z$ and
the magnetic field $\bar{b}_x$, $\bar{b}_z$ then contribute to the
nonlinear feedback through the nonlinear term for the radial field,
${\cal N}_x^{(b)}$, which is the most important one in the
sustenance (see Equations \ref{eq:App-Nbi}), but still the
contribution of $\bar{b}_y$ in this nonlinear term is dominant. This
feedback process is essentially 3D: we verified that modes with
$|k_z|=1,2$ give the largest contribution to the horizontal integral
in the expression for the nonlinear term ${\cal N}_x^{(b)}$ (not
shown here).

It is appropriate here to give a comparative analysis of the
dynamical processes investigated in this paper and those underlying
sustained 3D MRI-dynamo cycles reported in \citet{Herault_etal11}
and \citet{Riols_etal15,Riols_etal17}, despite the fact that these
papers considered a magnetized Keplerian flow with different, zero
net vertical flux, configuration and different values of parameters
(smaller resolution, box aspect ratio, smaller Reynolds numbers)
than those adopted here. These apparently resulted in the resistive
processes penetrating into the vital area (in our terms) and
reducing a number of active modes to only first non-axisymmetric
ones (shearing waves) with the minimal azimuthal and vertical
wavenumbers, $k_y=2\pi/L_y, k_z=0,2\pi/L_z$, which undergo the
transient MRI due to the mean axisymmetric azimuthal (dynamo) field.
By contrast, the number of the active modes in our turbulent case is
more than hundred (Figure 7). Regardless of these differences, we
can trace the similarities in the sustenance cycles -- the energy
budget equations for these modes derived in those papers in fact
show that a similar scheme underlies the sustenance as in the
present case. The energy of the radial field $\bar{b}_x$ of new
leading non-axisymmetric modes is supplied by the joint action of
the induction term (i.e., ${\cal I}_x^{(bu)}$ in our notations) and
redistribution by the nonlinear term, however, a summation over
$k_x$ as used in those energy budget equations does not permit to
see how this nonlinear redistribution of modes over $k_x$ due to the
transverse cascade actually occurs in their analysis. As for the
energy of $\bar{b}_y$, it is amplified by the Maxwell stress during
the transient MRI phase (also called the $\Omega$-effect) and is
drained by the corresponding nonlinear term. Since in the turbulent
state considered here there are much more active modes, representing
various linear and nonlinear dynamical terms in $(k_x,k_y)$-plane
has a definite advantage over such low-mode-number models in that
gives a more general picture of nonlinear triad interactions among
all active modes. Such a comparison raises one more point for
thought: for a correct consideration of nonlinear triad
interactions, we gave preference to boxes symmetrical in
$(x,y)$-plane, while, all simulations in those papers are carried
out in azimuthally elongated boxes.

\begin{figure}
\includegraphics[width=\columnwidth]{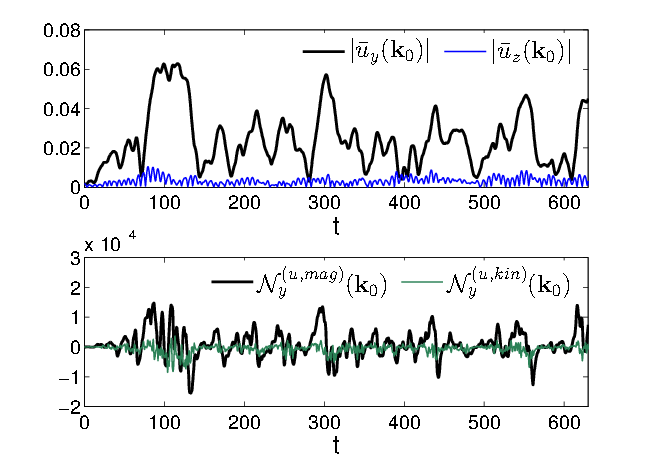}
\caption{Evolution of the large-scale mode ${\bf k}_0=(1,0,0)$,
which corresponds to the axisymmetric zonal flow. Shown are the
amplitudes of the azimuthal $|\bar{u}_y|$ (black) and the vertical
$|\bar{u}_z|$ (blue) velocities (upper panel) as well as the
magnetic ${\cal N}_y^{(u,mag)}$ (black) and hydrodynamic ${\cal
N}_y^{(u,kin)}$ (green) parts of the nonlinear term ${\cal
N}_y^{(u)}$ (lower panel). The dominant azimuthal velocity (i.e.,
zonal flow) is driven by the magnetic part of the nonlinear term and
is characterized by remarkably slower time variations.}
\label{fig:zonal flow}
\end{figure}

\subsection{Zonal flow}

Excitation of zonal flows by the MRI-turbulence was previously
observed by \citet{Johansen_etal09} and \citet{Bai_Stone14} in the
case of zero and nonzero net vertical magnetic flux, respectively.
We also observe it here in the case of the net azimuthal field. As
noted above, the mode corresponding to the zonal flow is
axisymmetric and vertically constant, $k_y=k_z=0$, with large scale
variation in the radial direction, $|k_x|=1$. The divergence-free
(incompressibility) condition (\ref{eq:App-divvk}) implies that the
radial velocity is zero, $\bar{u}_x=0$, for this mode and hence
${\cal H}_y=0$ at all times, also the magnetic exchange term is
identically zero at $k_y=0$, ${\cal I}_y^{(ub)}=0$. Therefore, a
source of the zonal flow can be only the nonlinear term ${\cal
N}_y^{(u)}$ in Equation (\ref{eq:uyk}). We can divide this term into
the magnetic, ${\cal N}^{(u,mag)}_y$, and hydrodynamic, ${\cal
N}^{(u,kin)}_y$, parts,
\begin{equation}\label{eq:nonlinzonal}
{\cal N}^{(u)}_y={\cal N}^{(u,mag)}_y + {\cal N}^{(u,kin)}_y.
\end{equation}
For the dominant mode ${\bf k_0}=(1,0,0)$, these two parts in
Equation (\ref{eq:nonlinzonal}) have the forms:
\begin{multline}\nonumber
{\cal N}^{(u,mag)}_y({\bf k}_0,t)\\=\frac{\rm
i}{2}\bar{u}^{\ast}_y({\bf k}_0,t)\int d^3{\bf k'}\bar{b}_y({\bf
k'},t)\bar{b}_x({\bf k}_0-{\bf k'},t) + c.c.,
\end{multline}
with the integrand composed of the turbulent magnetic stresses and
\begin{multline}\nonumber
{\cal N}^{(u,kin)}_y({\bf k}_0,t)\\= - \frac{\rm
i}{2}\bar{u}^{\ast}_y({\bf k}_0,t)\int d^3{\bf k'}\bar{u}_y({\bf
k'},t)\bar{u}_x({\bf k}_0-{\bf k'},t) + c.c.,
\end{multline}
with the integrand composed of the turbulent hydrodynamic stresses.
To understand the nature of the zonal flow, in Figure \ref{fig:zonal
flow} we present the time-development of the azimuthal and vertical
velocities as well as the driving nonlinear terms for this mode.
$|u_y({\bf k_0},t)|$ is characterized by remarkably longer timescale
(tens of orbits) variations and prevails over rapidly oscillating
$|u_z({\bf k_0},t)|$, i.e., the dominant harmonic indeed forms a
slowly varying in time axisymmetric zonal flow. Comparing the
time-development of $|\bar{u}_y({\bf k}_0,t)|$ with that of the
corresponding nonlinear terms in the lower panel of Figure
\ref{fig:zonal flow}, we clearly see that it is driven primarily by
the magnetic nonlinear term, ${\cal N}^{(u,mag)}_y({\bf k_0},t)$,
which physically describes the effect of the total azimuthal
magnetic tension (random forcing) exerted by all other smaller-scale
modes on the large-scale ${\bf k}_0$ mode, whereas ${\cal
N}^{(u,kin)}_y({\bf k_0},t)$, corresponding to the net effect of the
hydrodynamic stresses, is much smaller than the magnetic one. The
important role of the magnetic perturbations in launching and
maintaining the zonal flow is consistent with the findings of
\cite{Johansen_etal09}.

\begin{figure*}[t!]
\centering
\includegraphics[width=0.32\textwidth, height=0.2\textwidth]{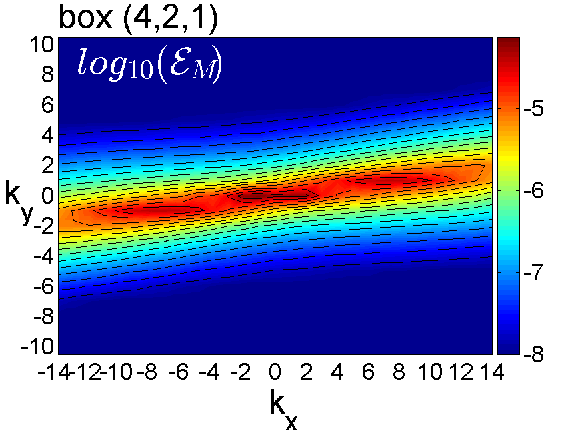}
\includegraphics[width=0.32\textwidth, height=0.2\textwidth]{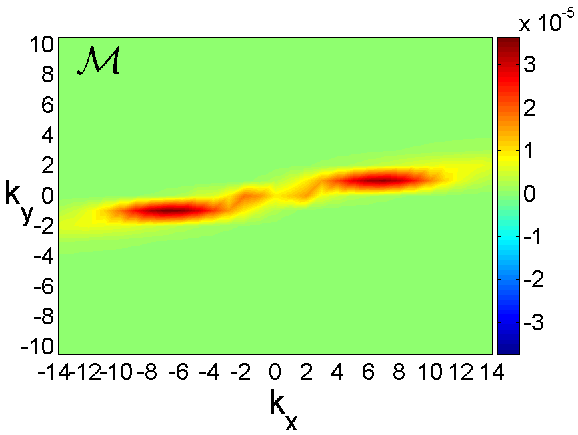}
\includegraphics[width=0.32\textwidth, height=0.2\textwidth]{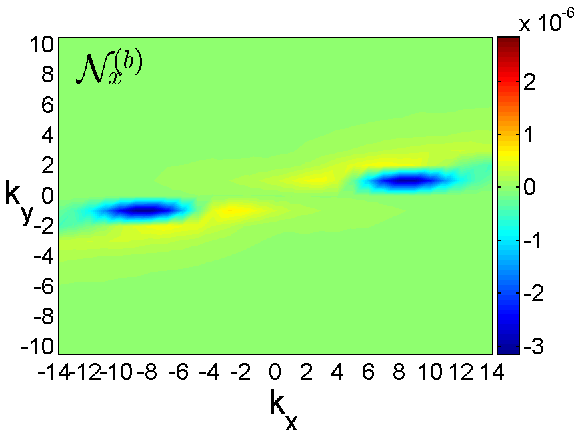}
\includegraphics[width=0.32\textwidth, height=0.2\textwidth]{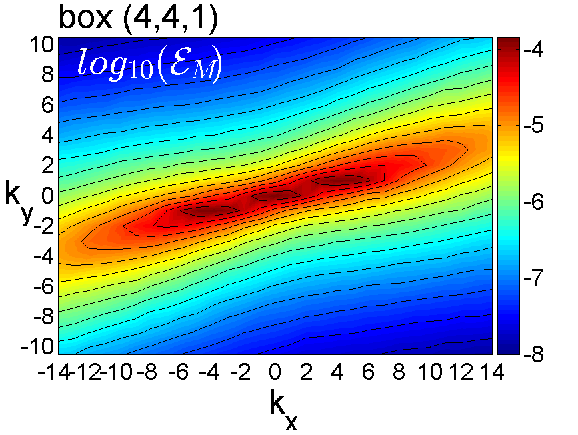}
\includegraphics[width=0.32\textwidth, height=0.2\textwidth]{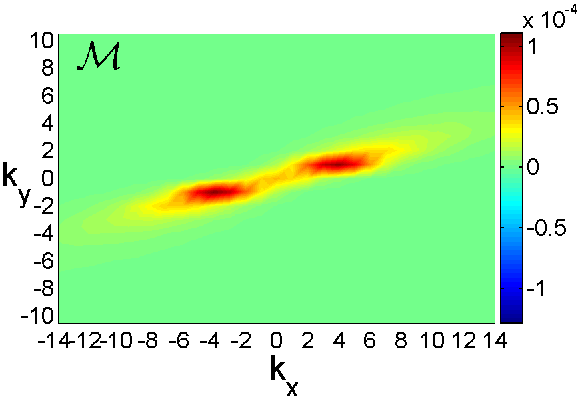}
\includegraphics[width=0.32\textwidth, height=0.2\textwidth]{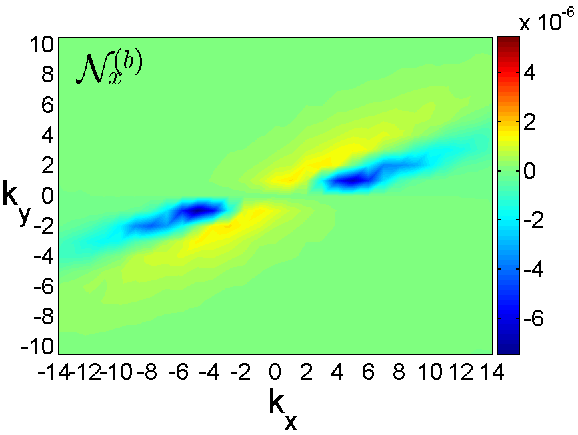}
\includegraphics[width=0.32\textwidth, height=0.2\textwidth]{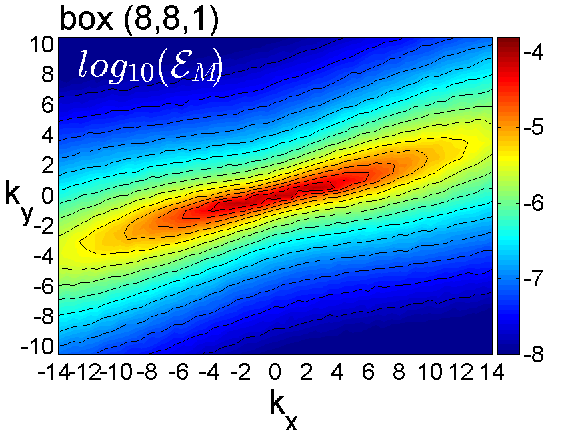}
\includegraphics[width=0.32\textwidth, height=0.2\textwidth]{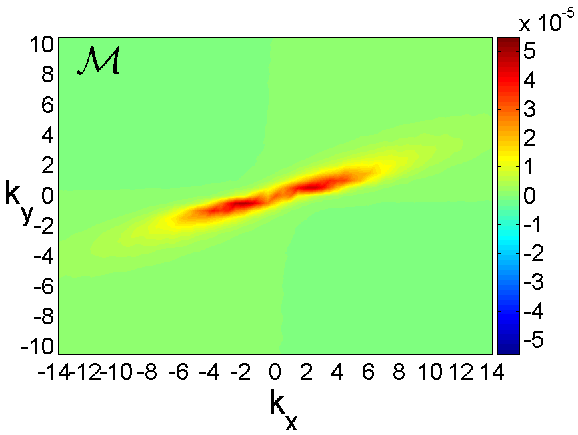}
\includegraphics[width=0.32\textwidth, height=0.2\textwidth]{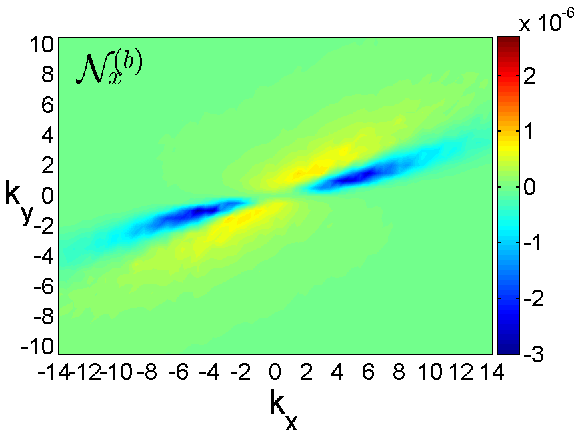}
\includegraphics[width=0.32\textwidth, height=0.2\textwidth]{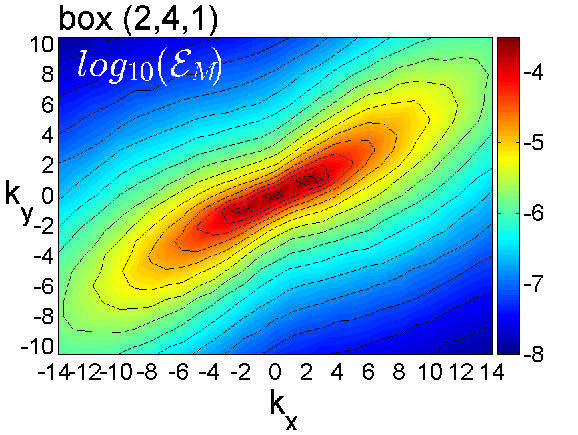}
\includegraphics[width=0.32\textwidth, height=0.2\textwidth]{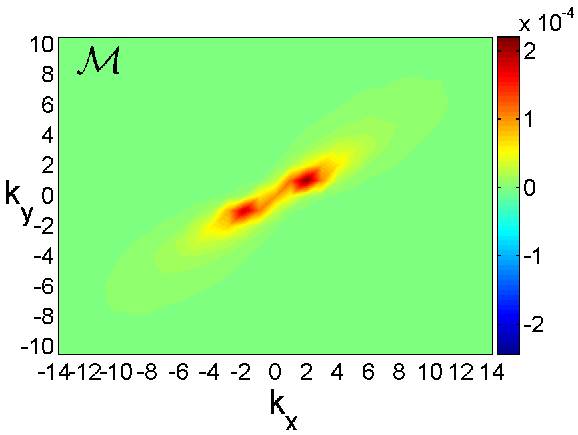}
\includegraphics[width=0.32\textwidth, height=0.2\textwidth]{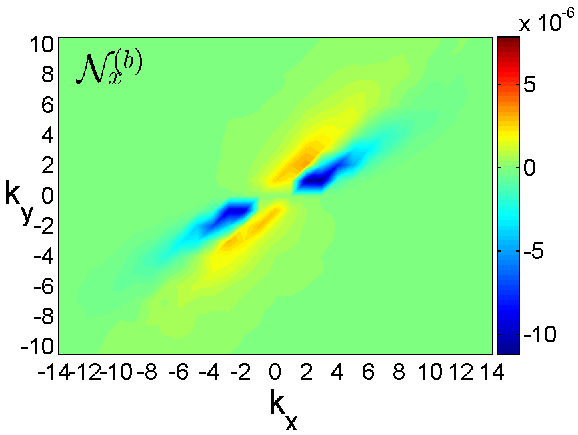}
\includegraphics[width=0.32\textwidth, height=0.2\textwidth]{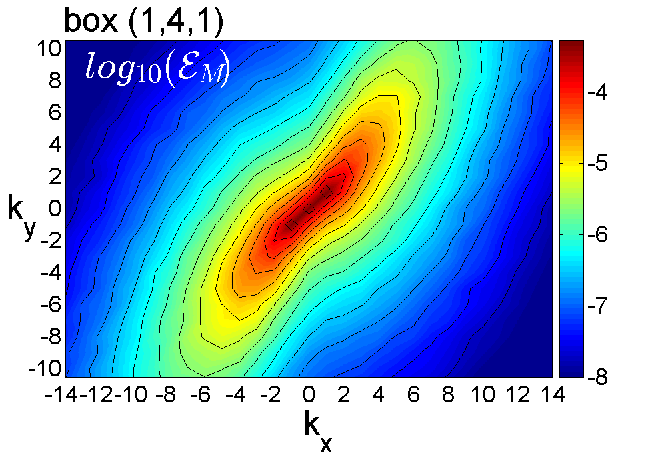}
\includegraphics[width=0.32\textwidth, height=0.2\textwidth]{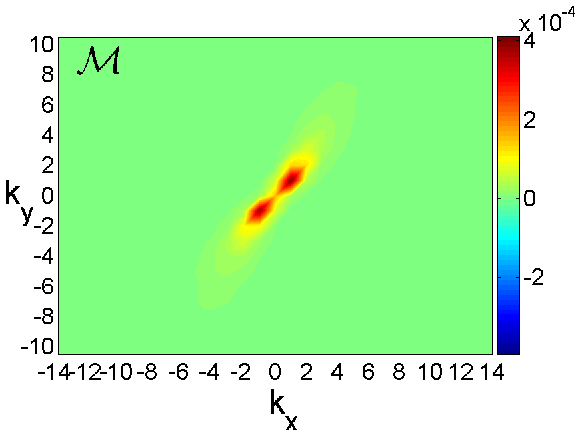}
\includegraphics[width=0.32\textwidth, height=0.2\textwidth]{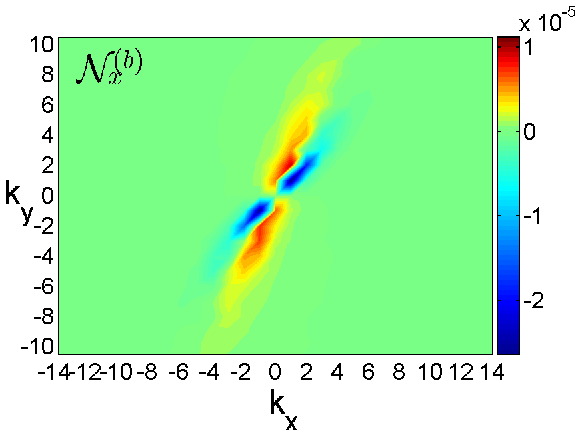}
\caption{Spectra of ${\cal E}_M$, ${\cal M}$ and ${\cal N}^{(b)}_x$
in $(k_x,k_y)$-plane at $k_z=1$ for all the boxes: $(4,2,1)$ (top
row), $(4,4,1)$ (second row), $(8,8,1)$ (third row), $(2,4,1)$
(fourth row) and $(1,4,1)$ (bottom row). In all the panels, the
general structure of these spectral terms is quite similar, that
indicates the universality and the robust character of the
turbulence sustaining scheme. At the same time, the symmetric boxes
$(4,4,1)$ and $(8,8,1)$ have similar spectral pictures with
identical inclinations, while in the asymmetric in $(x,y)$-plane
boxes, the spectral characteristics are somewhat deformed and have
different inclinations compared to the symmetric boxes.}
\label{fig:spectral-allboxes}
\end{figure*}

\subsection{Effect of the aspect ratio and the universality of the
turbulence sustenance scheme}

The main advantage of the box $(4,4,1)$ analyzed in the previous
subsection is that (i) it is symmetric in physical $(x,y)$-and
Fourier $(k_x,k_y)$-planes, where the effects of shear are most
important, (ii) the modes contained in this box densely cover the
vital area in $(k_x,k_y)$-plane and sufficiently comprise
effectively growing (optimal) harmonics (see the panel for the box
$(4,4,1)$ in Figure \ref{fig:nonmodalgrowthmodes}). In the three
asymmetric boxes -- $(1,4,1)$, $(2,4,1)$ and $(4,2,1)$ -- the modes
less densely cover the vital area (Figure
\ref{fig:nonmodalgrowthmodes}). As for the box $(8,8,1)$, as
mentioned above, the results qualitatively similar to the box
$(4,4,1)$ are expected. In this subsection, we examine how the box
aspect ratio influences the turbulence dynamics, and in particular,
the distribution of the linear and nonlinear process in Fourier
space.

A general temporal behavior of the volume-averaged energies,
stresses and rms values of the velocity and magnetic field
components is similar to that for the box $(4,4,1)$ represented in
Figure \ref{fig:time evolution} (see also Table 1) and we do not
show it here, but concentrate instead on the differences in Fourier
space. Figure \ref{fig:spectral-allboxes} juxtaposes the spectra of
the magnetic energy, Maxwell stress and the magnetic nonlinear term
${\cal N}_x^{(b)}$ for all the boxes. From this figure it is evident
that the skeleton of the balances of the various linear and
nonlinear processes and, in particular the basic subcycle,
underlying the sustenance of the azimuthal MRI-turbulence are
qualitatively the same in all the simulated boxes and quite robust
-- the variations in box sizes do not affect its effectiveness.
Changes in box aspect ratios lead to variation of the inclinations,
shapes and intensities of the energy spectra as well as the
distribution of linear and nonlinear dynamical terms in
$(k_x,k_y)$-plane. It is seen in Figure \ref{fig:spectral-allboxes}
that this variation is minimal between the symmetric in
$(x,y)$-plane boxes $(4,4,1)$ and $(8,8,1)$ -- they have similar
spectral characteristics with identical inclination angles -- but is
more remarkable among the asymmetric boxes, $(4,2,1)$, $(2,4,1)$,
$(1,4,1)$. Specifically, in the latter boxes, the spectral
characteristics are somewhat deformed and have different
inclinations compared to those in the symmetric boxes. The reason
for this is the reduction of the active modes' number/density along
the $k_x$- and $k_y$-axis in these boxes in contrast to the
symmetric ones (see Figure \ref{fig:nonmodalgrowthmodes}).

\section{Summary and discussion}

In this paper, we elucidated the essence of the sustenance of
MRI-driven turbulence in Keplerian disks threaded by a nonzero net
azimuthal field by means of a series of shearing box simulations and
analysis in 3D Fourier (\textbf{k}-)space. It is well known that in
the linear regime the MRI in the presence of a azimuthal field has a
transient nature and eventually decays without an appropriate
nonlinear feedback. We studied in detail the linear and nonlinear
dynamical processes and their interplay in Fourier space that ensure
such a feedback. Our first key finding is the pronounced anisotropy
of the nonlinear processes in \textbf{k}-space. This anisotropy is a
natural consequence of the anisotropy of linear processes due to the
shear and cannot be described in the framework of direct and inverse
cascades, commonly considered in the classical theory of HD and MHD
turbulence without shear, because the main activity of the nonlinear
processes is transfer of modes over wavevector orientation (angle)
in \textbf{k}-space, rather than along wavevector that corresponds
to direct/inverse cascades. This new type of nonlinear process --
\emph{the transverse cascade} -- plays a decisive role in the
long-term maintenance of the MRI-turbulence. Our second key result
is that the sustenance of the turbulence in this case is ensured as
a result of a subtle interplay of the linear transient MRI growth
and nonlinear transverse cascade. This interplay is intrinsically
quite complex. Nevertheless, one can isolate the basic subcycle of
the turbulence sustenance, which is as follows. The linear exchange
of energy between the magnetic field and the background flow,
realized by the Maxwell stress, ${\cal M}$, supplies only the
azimuthal field component $\bar{b}_y$. As for the radial field
$\bar{b}_x$, it is powered by the linear exchange ${\cal
I}_x^{(bu)}$ and the nonlinear ${\cal N}^{(b)}_x$ terms. So,
$\bar{b}_x$ and $\bar{b}_y$ have sources of different origin.
However, one should bear in mind that these processes are
intertwined with each other: the source of $\bar{b}_y$ (i.e., the
Maxwell stress, ${\cal M}$) is created by $\bar{b}_x$. In its turn,
the production of the nonlinear source of $\bar{b}_x$ (i.e., ${\cal
N}^{(b)}_x$) is largely due to $\bar{b}_y$. Similarly intertwined
are the dynamics of other spectral magnetic and kinematic
components. This sustaining dynamics of the turbulence is
concentrated mainly in a small wavenumber area of \textbf{k}-space,
i.e., involves large scale modes, and is appropriately called the
\emph{vital area}.

The spectra of the kinetic and magnetic energies that are
established in the turbulent state as a result of such interplay are
consequently also anisotropic and fundamentally differ from
classical Kolmogorov or IK spectra. So, the conventional
characterization of nonlinear MHD cascade processes in shear flows
in terms of direct and inverse cascades, which ignores the
shear-induced spectral anisotropy and the resulting important
transverse cascade process, is generally incomplete and misleading.
For this reason, we examined the dynamical processes in 3D Fourier
space in full without making the shell-averaging, which has been
commonly done in previous studies of MRI-turbulence and smears out
the anisotropy. We also showed that the turbulence is accompanied by
a large scale and slowly varying in time zonal (azimuthal) flow,
which is driven by the turbulent magnetic stresses.

The proposed scheme of the turbulence sustenance based on the
intertwined cooperated action of the linear and nonlinear processes
in the vital area is quite robust -- it is effective for different
aspect ratios of the simulation box. For all the box configurations
considered, $(4,4,1)$, $(1,4,1)$, $(2,4,1)$, $(4,2,1)$ and
$(8,8,1)$, the scheme is essentially universal, although there are
quantitative differences. The anisotropy of the box in
$(k_x,k_y)$-plane is superposed on the intrinsic shear-induced
anisotropy of the dynamical process and somewhat deforms the picture
of the turbulence, but the sustaining scheme is not changed. In any
case, an isotropic distribution of modes in $(k_x,k_y)$-plane seems
preferable for studying the own anisotropy of the shear flow system,
which is naturally achieved for equal radial and azimuthal sizes,
$L_x=L_y$, of the box.

In this paper, we considered a spectrally stable (i.e., without
purely exponential MRI) magnetized disk flow with an azimuthal
field, where the energy for turbulence can only be supplied via
linear transient growth of the MRI. Being associated with shear, it
seems obvious that the vital area and nonlinear transverse cascade
should be also present in disk flows with a nonzero net vertical
magnetic field, which can give rise to the exponentially growing MRI
\citep{Balbus_Hawley91,Goodman_Xu94,Pessah_Goodman09}. In this case,
besides purely exponentially growing axisymmetric (channel) modes,
energy supply and transport via (transient) growth of
non-axisymmetric ($k_y\neq 0$) modes are also important
\citep{Longaretti_Lesur10,Mamatsashvili_etal13,Squire_Bhattacharjee14}.
The latter, leading to anisotropic nonlinear dynamics
\citep{Murphy_Pessah15}, can inevitably effect the nonlinear
transverse cascade process. However, the presence of the purely
exponentially growing modes should somewhat alter the scheme of the
interplay of the dynamical processes that we studied here in the
case of the azimuthal field. We plan to explore this interplay also
in the case of vertical field MRI-turbulence, which will be
published elsewhere.

An interesting application of this approach -- analysis of
turbulence dynamics in Fourier space -- and a natural extension of
the present study would be understanding the nature of MRI
turbulence with zero net magnetic flux, where the classical linear
exponentially growing MRI is absent. This case has been studied in
several different configurations and there is much debate over the
nature of dynamo action, whether it is small-scale or large-scale
\citep{Lesur_Ogilvie08,Davis_etal10,Gressel10,Bodo_etal11,Bodo_etal12,Bodo_etal13,
Hirose_etal14,Shi_etal16}, and on the convergence with increasing
resolution/Reynolds number
\citep{Pessah_etal07,Fromang_Papaloizou07,Fromang10,Bodo_etal11,Bodo_etal14}.
A study of this kind will therefore be very helpful in the
resolution of these issues. In this regard, we would like to mention
recent high-resolution and high-Reynolds number simulations of
MRI-turbulence by \citet{Walker_etal16} and \citet{Zhdankin_etal17},
resolving larger wavenumbers outside the vital area -- inertial and
dissipation ranges. It was shown that the properties of turbulence
at these wavenumbers are insensitive to the specific nature of the
imposed large-scale magnetic field and are similar to those of
classical MHD turbulence without shear. In particular, the
characteristic energy spectra of the inertial range is close to the
IK spectrum, provided the energy of the large-scale azimuthal
magnetic field fluctuations is subtracted, while the small-scale
viscous and resistive dissipation characteristics are almost
unaffected by the presence of MRI. These studies, focusing on larger
wavenumbers, combined with our analysis, which focuses instead on
smaller wavenumbers that carry most of the energy and stress, should
be fruitful in shedding light on the dynamical picture of zero-net
flux MRI-turbulence.

\acknowledgments

This work is funded in part by the US Department of Energy under
grant DE-FG02-04ER54742 and the Space and Geophysics Laboratory at
the University of Texas at Austin. G.M. is supported by the Georg
Forster Research Fellowship from the Alexander von Humboldt
Foundation (Germany). We thank M. Pessah and O. Gressel for valuable
discussions and comments on the paper. We also thank the anonymous
referee for constructive comments that improved the presentation of
our results. The simulations were performed in Texas Advanced
Computing Center in Austin (TX) and on the high-performance Linux
cluster Hydra at the Helmholtz-Zentrum Dresden-Rossendorf (Germany).

\appendix

\section{Perturbation equations in physical space}

Equations governing the evolution of the velocity, total pressure
and magnetic field perturbations, ${\bf u}, p, {\bf b}$, about the
equilibrium Keplerian flow ${\bf U}_0=(0,-q\Omega x,0)$ with net
azimuthal field ${\bf B}_0=(0,B_{0y},0)$ are obtained from the basic
Equations (\ref{eq:mom})-(\ref{eq:divb}) and componentwise have the
form:
\begin{equation}\label{eq:App-ux}
\frac{Du_x}{Dt} = 2\Omega u_y- \frac{1}{\rho_0}\frac{\partial
p}{\partial x}+\frac{B_{0y}}{4\pi\rho_0}\frac{\partial b_x}{\partial
y}+\frac{\partial}{\partial
x}\left(\frac{b_x^2}{4\pi\rho_0}-u_x^2\right)+\frac{\partial}{\partial
y}\left(\frac{b_xb_y}{4\pi\rho_0}-u_xu_y\right)
+\frac{\partial}{\partial
z}\left(\frac{b_xb_z}{4\pi\rho_0}-u_xu_z\right)+\nu\nabla^2u_x,
\end{equation}
\begin{equation}\label{eq:App-uy}
\frac{Du_y}{Dt} = (q-2)\Omega u_x-\frac{1}{\rho_0}\frac{\partial
p}{\partial y}+\frac{B_{0y}}{4\pi\rho_0}\frac{\partial b_y}{\partial
y}+\\+\frac{\partial}{\partial
x}\left(\frac{b_xb_y}{4\pi\rho_0}-u_xu_y\right)+\frac{\partial}{\partial
y}\left(\frac{b_y^2}{4\pi\rho_0}-u_y^2\right)+\frac{\partial}{\partial
z}\left(\frac{b_zb_y}{4\pi\rho_0}-u_zu_y\right)+\nu\nabla^2u_y
\end{equation}
\begin{equation}\label{eq:App-uz}
\frac{Du_z}{Dt} = -\frac{1}{\rho_0}\frac{\partial p}{\partial z}-
N^2 \theta+\frac{B_{0y}}{4\pi\rho_0}\frac{\partial b_z}{\partial
y}+\\+\frac{\partial}{\partial
x}\left(\frac{b_xb_z}{4\pi\rho_0}-u_xu_z\right)+\frac{\partial}{\partial
y}\left(\frac{b_yb_z}{4\pi\rho_0}-u_yu_z\right)+\frac{\partial}{\partial
z}\left(\frac{b_z^2}{4\pi\rho_0}-u_z^2\right)+\nu\nabla^2u_z
\end{equation}
\begin{equation}\label{eq:App-theta}
\frac{D\theta}{Dt} = u_z -\frac{\partial}{\partial
x}(u_x\theta)-\frac{\partial}{\partial
y}(u_y\theta)-\frac{\partial}{\partial
z}(u_z\theta)+\chi\nabla^2\theta
\end{equation}
\begin{equation}\label{eq:App-bx}
\frac{Db_x}{Dt}= B_{0y}\frac{\partial u_x}{\partial
y}+\frac{\partial}{\partial y} \left(u_xb_y-u_yb_x
\right)-\frac{\partial}{\partial z}(u_zb_x-u_xb_z)+\eta\nabla^2b_x,
\end{equation}
\begin{equation}\label{eq:App-by}
\frac{Db_y}{Dt}= -q\Omega b_x + B_{0y}\frac{\partial u_y}{\partial
y}-\frac{\partial}{\partial x}
\left(u_xb_y-u_yb_x\right)+\frac{\partial}{\partial
z}(u_yb_z-u_zb_y)+\eta\nabla^2b_y,
\end{equation}
\begin{equation}\label{eq:App-bz}
\frac{Db_z}{Dt}= B_{0y}\frac{\partial u_z}{\partial
y}+\frac{\partial}{\partial x}
\left(u_zb_x-u_xb_z\right)-\frac{\partial}{\partial
y}(u_yb_z-u_zb_y)+\eta\nabla^2b_z,
\end{equation}
\begin{equation}\label{eq:App-divperu}
\frac{\partial u_x}{\partial x}+\frac{\partial u_y}{\partial
y}+\frac{\partial u_z}{\partial z}=0,
\end{equation}
\begin{equation}\label{eq:App-divperb}
\frac{\partial b_x}{\partial x}+\frac{\partial b_y}{\partial
y}+\frac{\partial b_z}{\partial z}=0,
\end{equation}
where $D/Dt=\partial/\partial t-q\Omega x\partial/\partial y$ is the
total derivative along the background flow.

\section{Derivation of spectral equations for quadratic terms}

Here we derive evolution equations for velocity, entropy and
magnetic field perturbations in Fourier space. Substituting
decomposition (\ref{eq:fourier}) into Equations
(\ref{eq:App-ux})-(\ref{eq:App-divperb}) and taking into account the
normalization made in the text, we arrive at the following equations
governing the dynamics of perturbation modes in Fourier space
\begin{equation}\label{eq:App-uxk}
\left(\frac{\partial}{\partial t}+qk_y\frac{\partial}{\partial
k_x}\right)\bar{u}_x=2\bar{u}_y-{\rm i}k_x\bar{p}+{\rm
i}k_yB_{0y}\bar{b}_x-\frac{k^2}{\rm Re}\bar{u}_x+{\rm i}k_x
N^{(u)}_{xx}+{\rm i}k_yN^{(u)}_{xy}+{\rm i}k_zN^{(u)}_{xz},
\end{equation}
\begin{equation}\label{eq:App-uyk}
\left(\frac{\partial}{\partial t}+qk_y\frac{\partial}{\partial
k_x}\right)\bar{u}_y=(q-2)\bar{u}_x-{\rm i}k_y\bar{p}+{\rm
i}k_yB_{0y}\bar{b}_y-\frac{k^2}{\rm Re}\bar{u}_y+{\rm
i}k_xN^{(u)}_{xy}+{\rm i}k_yN^{(u)}_{yy}+{\rm i}k_zN^{(u)}_{yz},
\end{equation}
\begin{equation}\label{eq:App-uzk}
\left(\frac{\partial}{\partial t}+qk_y\frac{\partial}{\partial
k_x}\right)\bar{u}_z=-{\rm i}k_z\bar{p}-N^2\bar{\theta}+{\rm
i}k_yB_{0y}\bar{b}_z-\frac{k^2}{\rm Re}\bar{u}_z+{\rm
i}k_xN^{(u)}_{xz}+{\rm i}k_yN^{(u)}_{yz}+{\rm i}k_zN^{(u)}_{zz},
\end{equation}
\begin{equation}\label{eq:App-thetak}
\left(\frac{\partial}{\partial t}+qk_y\frac{\partial}{\partial
k_x}\right)\bar{\theta}=\bar{u}_z-\frac{k^2}{\rm
Pe}\bar{\theta}+{\rm i}k_xN^{(\theta)}_{x}+{\rm
i}k_yN^{(\theta)}_{y}+{\rm i}k_zN^{(\theta)}_{z},
\end{equation}
\begin{equation}\label{eq:App-bxk}
\left(\frac{\partial}{\partial t}+qk_y\frac{\partial}{\partial
k_x}\right)\bar{b}_x={\rm i}k_yB_{0y}\bar{u}_x-\frac{k^2}{\rm
Rm}\bar{b}_x+{\rm i}k_y\bar{F}_z-{\rm i}k_z\bar{F}_y,
\end{equation}
\begin{equation}\label{eq:App-byk}
\left(\frac{\partial}{\partial t}+qk_y\frac{\partial}{\partial
k_x}\right)\bar{b}_y=-q\bar{b}_x+{\rm
i}k_yB_{0y}\bar{u}_y-\frac{k^2}{\rm Rm}\bar{b}_y+{\rm
i}k_z\bar{F}_x-{\rm i}k_x\bar{F}_z
\end{equation}
\begin{equation}\label{eq:App-bzk}
\left(\frac{\partial}{\partial t}+qk_y\frac{\partial}{\partial
k_x}\right)\bar{b}_z={\rm i}k_yB_{0y}\bar{u}_z-\frac{k^2}{\rm
Rm}\bar{b}_z+{\rm i}k_x\bar{F}_y-{\rm i}k_y\bar{F}_x
\end{equation}
\begin{equation}\label{eq:App-divvk}
k_x\bar{u}_x+k_y\bar{u}_y+k_z\bar{u}_z=0,
\end{equation}
\begin{equation}\label{eq:App-divbk}
k_x\bar{b}_x+k_y\bar{b}_y+k_z\bar{b}_z=0,
\end{equation}
where $k^2=k_x^2+k_y^2+k_z^2$ and $B_{0y}=\sqrt{2/\beta}$ is the
normalized background azimuthal field. These spectral equations
contain the linear as well as the nonlinear ($N^{(u)}_{ij}({\bf
k},t), N^{(\theta)}_i({\bf k},t), \bar{F}_i({\bf k},t)$,
$i,j=x,y,z$) terms that are the Fourier transforms of the
corresponding linear and nonlinear terms in the original Equations
(\ref{eq:App-ux})-(\ref{eq:App-divperb}). The latter are given by
convolutions
\begin{equation}\label{eq:App-Nuij}
N^{(u)}_{ij}({\bf k},t)=\int d^3{\bf k'}\left[\bar{b}_i({\bf
k'},t)\bar{b}_j({\bf k}-{\bf k'},t)-\bar{u}_i({\bf
k'},t)\bar{u}_j({\bf k}-{\bf k'},t)\right],
\end{equation}
\begin{equation}\label{eq:App-Nth}
N^{(\theta)}_{i}({\bf k},t)=-\int d^3{\bf k'}\bar{u}_i({\bf
k'},t)\bar{\theta}({\bf k}-{\bf k'},t)
\end{equation}
where $i,j=x,y,z$ and $ \bar{F}_x, \bar{F}_y, \bar{F}_z$ are the
fourier transforms of the respective components of the perturbed
electromotive force ${\bf F}={\bf u}\times {\bf b}$,
\begin{equation*}
\bar{F}_x({\bf k},t)=\int d^3{\bf k'}\left[\bar{u}_y({\bf
k'},t)\bar{b}_z({\bf k}-{\bf k'},t)-\bar{u}_z({\bf
k'},t)\bar{b}_y({\bf k}-{\bf k'},t)\right]
\end{equation*}
\begin{equation*}
\bar{F}_y({\bf k},t)=\int d^3{\bf k'}\left[\bar{u}_z({\bf
k'},t)\bar{b}_x({\bf k}-{\bf k'},t)-\bar{u}_x({\bf
k'},t)\bar{b}_z({\bf k}-{\bf k'},t)\right]
\end{equation*}
\begin{equation*}
\bar{F}_z({\bf k},t)=\int d^3{\bf k'}\left[\bar{u}_x({\bf
k'},t)\bar{b}_y({\bf k}-{\bf k'},t)-\bar{u}_y({\bf
k'},t)\bar{b}_x({\bf k}-{\bf k'},t)\right]
\end{equation*}
and describe the contribution from nonlinearity to the magnetic
field perturbations. In the case of classical forced MHD turbulence
without background shear flow, these nonlinear transfer terms in
${\bf k}$-space were also derived in \citet{Verma04}. From Equations
(\ref{eq:App-uxk})-(\ref{eq:App-uzk}) and the divergence-free
conditions (\ref{eq:App-divvk}) and (\ref{eq:App-divbk}) we can
eliminate pressure
\begin{equation}\label{eq:App-pk}
\bar{p}=2{\rm i}(1-q)\frac{k_y}{k^2}\bar{u}_x-2{\rm
i}\frac{k_x}{k^2}\bar{u}_y+{\rm
i}N^2\frac{k_z}{k^2}\bar{\theta}+\sum_{(i,j)=(x,y,z)}\frac{k_ik_j}{k^2}N^{(u)}_{ij}
\end{equation}
Substituting it back into Equations
(\ref{eq:App-uxk})-(\ref{eq:App-uzk}) we get
\begin{equation}\label{eq:App-uxk1}
\left(\frac{\partial}{\partial t}+qk_y\frac{\partial}{\partial
k_x}\right)\bar{u}_x=2\left(1-\frac{k_x^2}{k^2}\right)\bar{u}_y+
2(1-q)\frac{k_xk_y}{k^2}\bar{u}_x+
N^2\frac{k_xk_z}{k^2}\bar{\theta}+{\rm
i}k_yB_{0y}\bar{b}_x-\frac{k^2}{\rm Re}\bar{u}_x+Q_x,
\end{equation}
\begin{equation}\label{eq:App-uyk1}
\left(\frac{\partial}{\partial t}+qk_y\frac{\partial}{\partial
k_x}\right)\bar{u}_y=\left[q-2-2(q-1)\frac{k_y^2}{k^2}\right]\bar{u}_x-
2\frac{k_xk_y}{k^2}\bar{u}_y +N^2\frac{k_yk_z}{k^2}\bar{\theta}+{\rm
i}k_yB_{0y}\bar{b}_y-\frac{k^2}{\rm Re}\bar{u}_y+Q_y,
\end{equation}
\begin{equation}\label{eq:App-uzk1}
\left(\frac{\partial}{\partial t}+qk_y\frac{\partial}{\partial
k_x}\right)\bar{u}_z=2(1-q)\frac{k_yk_z}{k^2}\bar{u}_x-2\frac{k_xk_z}{k^2}\bar{u}_y-
N^2\left(1-\frac{k_z^2}{k^2}\right)\bar{\theta}+{\rm
i}k_yB_{0y}\bar{b}_z-\frac{k^2}{\rm Re}\bar{u}_z+Q_z,
\end{equation}
where
\begin{equation}\label{eq:App-Qi}
Q_i={\rm i}\sum_jk_jN^{(u)}_{ij}-{\rm
i}k_i\sum_{m,n}\frac{k_mk_n}{k^2}N^{(u)}_{mn}, ~~~~~ i,j,m,n=x,y,z.
\end{equation}

Multiplying Equations (\ref{eq:App-uxk1})-(\ref{eq:App-uzk1}),
respectively, by $\bar{u}_x^{\ast}$, $\bar{u}_y^{\ast}$,
$\bar{u}_z^{\ast}$, and adding up with their complex conjugates, we
obtain
\begin{equation}\label{eq:App-uxk2}
\frac{\partial}{\partial
t}\frac{|\bar{u}_x|^2}{2}=-qk_y\frac{\partial}{\partial k_x}
\frac{|\bar{u}_x|^2}{2} + {\cal H}_x + {\cal I}_x^{(u\theta)} +
{\cal I}_x^{(ub)}+{\cal D}_x^{(u)}+{\cal N}^{(u)}_x,
\end{equation}
\begin{equation}\label{eq:App-uyk2}
\frac{\partial}{\partial
t}\frac{|\bar{u}_y|^2}{2}=-qk_y\frac{\partial}{\partial k_x}
\frac{|\bar{u}_y|^2}{2}+{\cal H}_y + {\cal I}_y^{(u\theta)} + {\cal
I}_y^{(ub)}+{\cal D}_y^{(u)}+{\cal N}^{(u)}_y,
\end{equation}
\begin{equation}\label{eq:App-uzk2}
\frac{\partial}{\partial
t}\frac{|\bar{u}_z|^2}{2}=-qk_y\frac{\partial}{\partial k_x}
\frac{|\bar{u}_z|^2}{2} + {\cal H}_z + {\cal I}_z^{(u\theta)} +
{\cal I}_z^{(ub)}+{\cal D}_z^{(u)}+{\cal N}^{(u)}_z,
\end{equation}
where the terms of linear origin are
\begin{equation}\label{eq:App-Hx}
{\cal
H}_x=\left(1-\frac{k_x^2}{k^2}\right)(\bar{u}_x\bar{u}_y^{\ast}+\bar{u}_x^{\ast}\bar{u}_y)+2(1-q)\frac{k_xk_y}{k^2}|\bar{u}_x|^2,
\end{equation}
\begin{equation}\label{eq:App-Hy}
{\cal
H}_y=\frac{1}{2}\left[q-2-2(q-1)\frac{k_y^2}{k^2}\right](\bar{u}_x\bar{u}_y^{\ast}+\bar{u}_x^{\ast}\bar{u}_y)
- 2\frac{k_xk_y}{k^2}|\bar{u}_y|^2
\end{equation}
\begin{equation}\label{eq:App-Hz}
{\cal H}_z =
(1-q)\frac{k_yk_z}{k^2}(\bar{u}_x\bar{u}_z^{\ast}+\bar{u}_x^{\ast}\bar{u}_z)
-\frac{k_xk_z}{k^2}(\bar{u}_y\bar{u}_z^{\ast}+\bar{u}_y^{\ast}\bar{u}_z),
\end{equation}
\begin{equation}\label{eq:App-Iuthi}
{\cal I}_i^{(u\theta)}
=N^2\left(\frac{k_ik_z}{k^2}-\delta_{iz}\right)\frac{\bar{\theta}\bar{u}_i^{\ast}+\bar{\theta}^{\ast}\bar{u}_i}{2},
\end{equation}
\begin{equation}\label{eq:App-Iubi}
{\cal I}_i^{(ub)} = \frac{\rm
i}{2}k_yB_{0y}(\bar{u}_i^{\ast}\bar{b}_i -
\bar{u}_i\bar{b}_i^{\ast}),
\end{equation}
\begin{equation}\label{eq:App-Dui}
{\cal D}_i^{(u)}=-\frac{k^2}{\rm Re}|\bar{u}_i|^2,
\end{equation}
and the modified nonlinear transfer functions for the quadratic
forms of the velocity components are
\begin{equation}\label{eq:App-Nui}
{\cal
N}^{(u)}_i=\frac{1}{2}(\bar{u}_iQ^{\ast}_i+\bar{u}_i^{\ast}Q_i).
\end{equation}
Here $i=x,y,z$ and $\delta_{iz}$ is the Kronecker delta. It is
readily shown that the sum of ${\cal H}_i$ is equal to the Reynolds
stress spectrum multiplied by the shear parameter $q$, ${\cal
H}={\cal H}_x+{\cal H}_y+{\cal
H}_z=q(\bar{u}_x\bar{u}_y^{\ast}+\bar{u}_x^{\ast}\bar{u}_y)/2$

Similarly, multiplying Equation (\ref{eq:App-thetak}) by
$\bar{\theta}^{\ast}$ and adding up with its complex conjugate, we
get
\begin{equation}\label{eq:App-thk2}
\frac{\partial}{\partial
t}\frac{|\bar{\theta}|^2}{2}=-qk_y\frac{\partial}{\partial k_x}
\frac{|\bar{\theta}|^2}{2}+{\cal I}^{(\theta u)} + {\cal
D}^{(\theta)} + {\cal N}^{(\theta)},
\end{equation}
where the terms of linear origin are
\begin{equation}\label{eq:App-Ithu}
{\cal I}^{(\theta
u)}=\frac{1}{2}(\bar{u}_z\bar{\theta}^{\ast}+\bar{u}_z^{\ast}\bar{\theta}),
\end{equation}
\begin{equation}\label{eq:App-Dth} {\cal
D}^{(\theta)}=-\frac{k^2}{\rm Pe}|\bar{\theta}|^2
\end{equation}
and the modified nonlinear transfer function for the quadratic form
of the entropy is
\begin{equation}\label{eq:App-Nth1}
{\cal N}^{(\theta)}= \frac{\rm
i}{2}\bar{\theta}^{\ast}(k_xN^{(\theta)}_{x}+k_yN^{(\theta)}_{y}+k_zN^{(\theta)}_{z})+c.c.
\end{equation}

Multiplying Equations (\ref{eq:App-bxk})-(\ref{eq:App-bzk}),
respectively, by $\bar{b}_x^{\ast}$, $\bar{b}_y^{\ast}$,
$\bar{b}_z^{\ast}$, and adding up with their complex conjugates, we
obtain
\begin{equation}\label{eq:App-bxk2}
\frac{\partial}{\partial
t}\frac{|\bar{b}_x|^2}{2}=-qk_y\frac{\partial}{\partial k_x}
\frac{|\bar{b}_x|^2}{2} + {\cal I}_x^{(bu)}+{\cal D}_x^{(b)}+{\cal
N}^{(b)}_x
\end{equation}
\begin{equation}\label{eq:App-byk2}
\frac{\partial}{\partial
t}\frac{|\bar{b}_y|^2}{2}=-qk_y\frac{\partial}{\partial k_x}
\frac{|\bar{b}_y|^2}{2}+{\cal M}+{\cal I}_y^{(bu)}+{\cal
D}_y^{(b)}+{\cal N}^{(b)}_y
\end{equation}
\begin{equation}\label{eq:App-bzk2}
\frac{\partial}{\partial
t}\frac{|\bar{b}_z|^2}{2}=-qk_y\frac{\partial}{\partial k_x}
\frac{|\bar{b}_z|^2}{2}+{\cal I}_z^{(bu)}+{\cal D}_z^{(b)}+{\cal
N}^{(b)}_z,
\end{equation}
where ${\cal M}$ is the Maxwell stress spectrum multiplied by $q$,
\begin{equation}\label{eq:App-M}
{\cal
M}=-\frac{q}{2}(\bar{b}_x\bar{b}_y^{\ast}+\bar{b}_x^{\ast}\bar{b}_y),
\end{equation}
\begin{equation}\label{eq:App-Ibui}
{\cal I}_i^{(bu)}= - {\cal I}_i^{(ub)}=\frac{\rm
i}{2}k_yB_{0y}(\bar{u}_i\bar{b}_i^{\ast} -
\bar{u}_i^{\ast}\bar{b}_i)
\end{equation}
\begin{equation}\label{eq:App-Dbi}
{\cal D}_i^{(b)}=-\frac{k^2}{\rm Rm}|\bar{b}_i|^2
\end{equation}
and the modified nonlinear terms for the quadratic forms of the
magnetic field components are
\begin{equation}\label{eq:App-Nbi}
{\cal N}^{(b)}_x=\frac{\rm
i}{2}\bar{b}_x^{\ast}[k_y\bar{F}_z-k_z\bar{F}_y]+c.c.,~~~ {\cal
N}^{(b)}_y=\frac{\rm
i}{2}\bar{b}_y^{\ast}[k_z\bar{F}_x-k_x\bar{F}_z]+c.c.,~~~ {\cal
N}^{(b)}_z=\frac{\rm
i}{2}\bar{b}_z^{\ast}[k_x\bar{F}_y-k_y\bar{F}_x]+c.c.
\end{equation}

\bibliographystyle{aasjournal}
\bibliography{biblio}
\end{document}